\documentclass[aps,prb,groupedaddress,preprint,superscriptaddress,showkeys,showpacs]{revtex4-1}
\usepackage{amsfonts}
\usepackage{graphicx}
\usepackage{mathcomp}

\usepackage{graphicx}

\usepackage{savesym}
\usepackage{amsmath}
\savesymbol{iint}
\usepackage{wasysym}
\restoresymbol{WSS}{iint}

\begin{document}

\title{Theory of Multiwave Mixing within the Superconducting Kinetic-Inductance Traveling-Wave Amplifier}

\author{R. P. Erickson}
\author{D. P. Pappas}
\email[]{Electronic address: David.Pappas@NIST.gov}
\affiliation{National Institute of Standards and Technology, Boulder, Colorado 80305, USA}

\date{\today}

\begin{abstract}
We present a theory of parametric mixing within the coplanar waveguide (CPW) of a superconducting nonlinear kinetic-inductance traveling-wave (KIT) amplifier engineered with periodic dispersion loadings. This is done by first developing a metamaterial band theory of the dispersion-engineered KIT using a Floquet-Bloch construction and then applying it to the description of mixing of the nonlinear RF traveling waves. Our theory allows us to calculate signal gain vs. signal frequency in the presence of a frequency stop gap, based solely on loading design. We present results for both three-wave mixing (3WM), with applied DC bias, and four-wave mixing (4WM), without DC. Our theory predicts an intrinsic and deterministic origin to undulations of 4WM signal gain with signal frequency, apart from extrinsic sources, such as impedance mismatch, and shows that such undulations are absent from 3WM signal gain achievable with DC. Our theory is extensible to amplifiers based on Josephson junctions in a lumped LC transmission line (TWPA).
\end{abstract}


\maketitle

\section{Introduction}
Superconducting amplifiers with wide frequency bandwidth, high dynamic range, and low noise are used in both quantum computing\cite{Wallraff2004} and photon-detector\cite{DayNature2003} research. They have utility to measure large arrays of quantum-limited frequency-multiplexed microwave superconducting resonators, with recent strides made using nonlinear amplifiers based on Josephson junctions.\cite{CastellanosBeltran2008,Bergreal2010,Spietz2010,Hatridge2011,SLUG,Roch2012,Mutas2013,Macklin2015} In particular, a near-quantum-limited Josephson traveling-wave parametric amplifier (JTWPA) recently has been fabricated with a quantum efficiency of $75\%$ and a signal gain greater than 20 dB over a 3 GHz bandwidth.\cite{Macklin2015} Focus of the present discussion is the nonlinear kinetic-inductance traveling-wave (KIT) amplifier, first realized by Eom and co-workers.\cite{EomNature2012amplifier} In the coplanar waveguide (CPW) of the KIT, degenerate four-wave mixing (4WM) can occur between RF input pump and signal, resulting in signal amplification and generation of an idler product. This parametric mixing of traveling RF waveforms is analogous to 4WM realizable in the optical frequency regime,\cite{Agrawal2001,AghaOpticsExpress2009theoretical,*ChemboPRA2010modal,*ChemboPRL2010spectrum,
*HanssonArXiv2013dynamics,*GodeyArXiv2013stability} as in recent nonlinear resonance experiments involving strongly pumped, high-Q optical microcavities made from nonlinear media.\cite{KippenbergPhD2008thesis,*DelHayeNature2007comb,*DelhayPRL2011octavespanning,*FosterOpticsExpress2011SiBasedComb} In this 4WM-mode of operation the KIT possesses signal gain of $\sim 20$ dB, a bandwidth of $\sim 9$ GHz (centered about the pump tone), and a dynamic range of the order of $-100$ dB, comparable to microwave transistor amplifiers. Another promising mode of operation of the KIT is when a DC bias is applied: additional parametric three-wave mixing (3WM) may ensue, producing signal gain of $\sim 15$ dB over an exploitable bandwidth of $\sim 6$ GHz, centered about half the pump tone.\cite{Vissers2016} This latter 3WM gain is achieved with less pump input power than the 4WM mode due to DC biasing of the Kerr-like nonlinear kinetic inductance of the waveguide. Unwanted higher pump harmonics and shock waves, prevalent at higher pump powers, may be intentionally inhibited by proper CPW loading design, and additional dispersion loading may be engineered to customize the onset, magnitude, and bandwidth of the resulting signal gain. In this paper we present a quantifiable model of the dispersion engineering and parametric mixing of the KIT amplifier, taking into account DC biasing, showing how optimal amplifier design may be achieved.

\begin{figure}
\includegraphics[width=400pt, height=360pt]{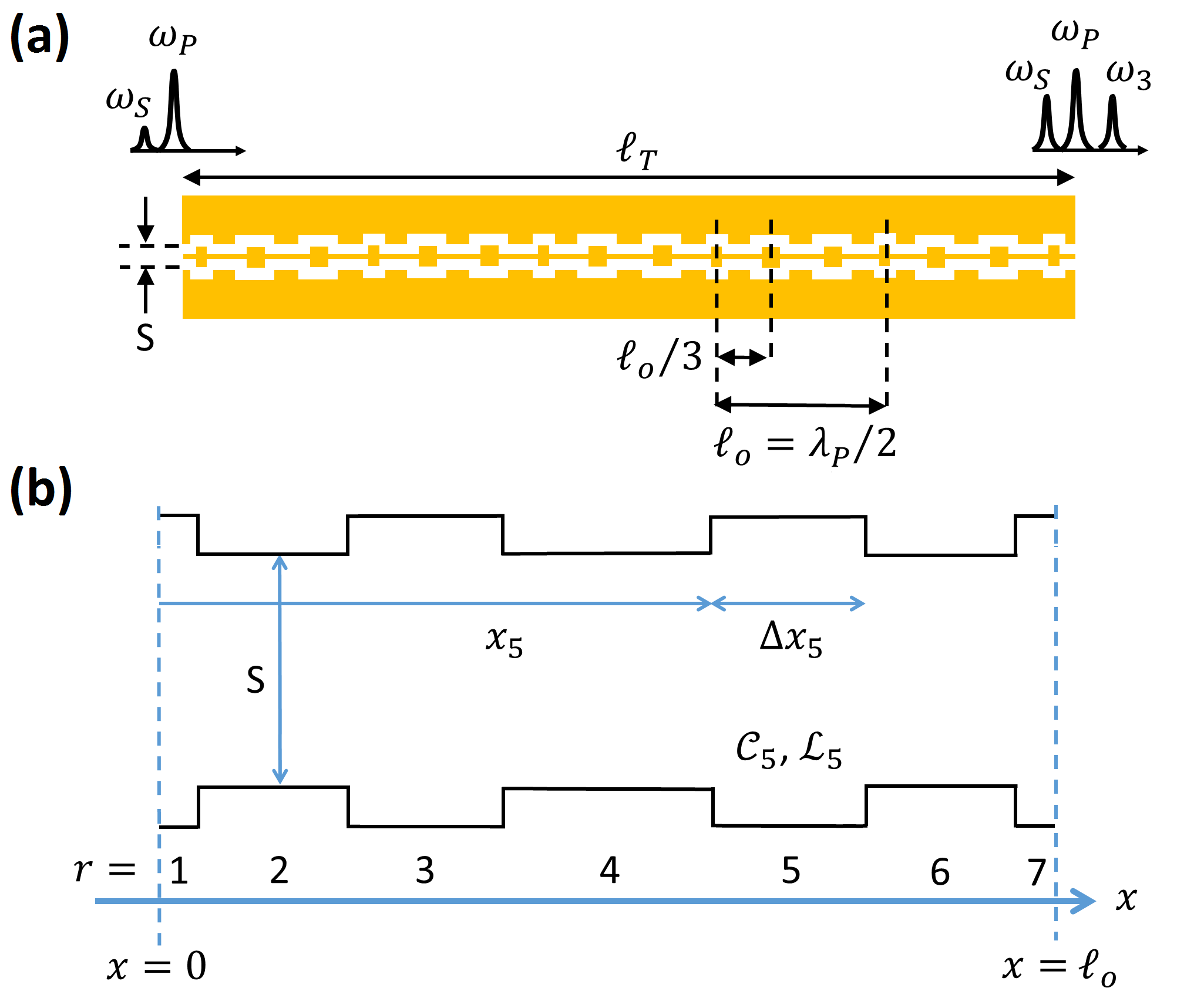}
\caption{\label{fig1} (a) Schematic of KIT CPW of length $\ell_T$ depicting example dispersion loadings of pattern length $\ell_o=\lambda_P/2$, with non-loading width $S$ and pump wavelength $\lambda_P$. With no DC applied, pump $\omega_P$ and signal $\omega_S$ tones are input and an additional 4WM idler $\omega_3$ is observed on output. (b) Corresponding periodic unit cell of length $\ell_o$ with $R=7$ regions, representing the loading pattern of (a). Showcased is loaded region $r=5$ at distance $x_5$ from start of cell, of length $\Delta x_5$, capacitance per unit length $\mathcal{C}_5$, and linear kinetic inductance per unit length $\mathcal{L}_5$.}
\end{figure}

KIT devices have been fabricated from superconducting TiN and NbTiN films on Si.\cite{DayNature2003,MazinAPL2006,*BarendsAPL2010} These materials are used to construct a CPW of a meter and more length, $\ell_T$, with width typically $S=2$ microns. As in Fig. \ref{fig1} (a), engineered loadings of repeat length $\ell_o$, where $\ell_T/\ell_o\sim 10^3$, are introduced that represent regions of increased width of the CPW--as much as 3 times wider than $S$. Loadings are designed with $l_o=\lambda_P/2$, where $\lambda_P$ is the wavelength of a propagating RF pump, to maximize destructive interference. If the pump propagates along the unloaded line with wavenumber $k$ and dispersion frequency $\Omega(k)=v_g k$, where $v_g=1/\sqrt{\mathcal{L}_o\mathcal{C}_o}$ is the group velocity and $\mathcal{L}_o$ and $\mathcal{C}_o$ are the unloaded inductance and capacitance per unit length, respectively, then one effect of the loadings is to open gaps in $\Omega(k)$ as a function of $k$. Loadings designed in this way are often referred to as frequency stops, and their corresponding gaps are known as stop gaps, since a tone will not propagate down the waveguide if its frequency falls within a gap. For example, if $\mathcal{L}_o=1.05$ pH/$\mu$m and $\mathcal{C}_o=0.540$ fF/$\mu$m in the unloaded line, such that $v_g=4.2\times 10^7$ m/s, then loadings placed at intervals of $\ell_o=2565$ $\mu$m introduce a first stop gap at frequency $v_g/\lambda_P=2v_g/\ell_o\cong 8.2$ GHz. Two additional loadings may be introduced between those of spacing $\ell_o$, such that the spacing between nearest loadings becomes $\ell_o/3$, as pictured in Fig. \ref{fig1} (a). The lengths of these additional loadings can be made greater than the initially described loadings, as suggested in the figure. This has the effect of broadening stop gaps at every third gap, starting with the third stop gap at $6v_g/\ell_o\cong 24.6$ GHz. The advantage of this design is that if a strong pump tone is placed just above or below the first stop gap, at $\sim 8$ GHz, then the higher third harmonic of the pump, which is prevalent in the KIT amplifier when operated in a 4WM nonlinear regime, can be suppressed, increasing efficiency of parametric amplification.

Geometries such as a double spiral and a meandering line have been used to create very long KIT CPWs, on chips of area $\sim 4$ cm\textsuperscript{2}. At low temperature, traveling RF waves, consisting of a strong pump of fixed frequency $\omega_P$ and a smaller-amplitude signal of adjustable frequency $\omega_S$, input to one port of the CPW, undergo degenerate 4WM along the direction $x$ of the CPW due to the Kerr-like nonlinear kinetic inductance per unit length, $L(x,t)$, of the underlying superconducting film, viz.
\begin{equation}	\label{kineticInductancePerLength}
L(x,t) = L_o(x) \left\{ 1 + { \left[ \frac{{I(x,t)}}{I_*} \right] }^2 \right\} ,
\end{equation}
where $L_o(x)$ is the linear kinetic inductance per unit length, made dependent on $x$ to account for engineered loadings, $I(x,t)$ is the total time-dependent electrical current of the mixing waveforms, and $I_*$ is a constant scaling factor.\cite{DayNature2003,EomNature2012amplifier} The CPW may be modeled as a straight LC ladder circuit where the total current $I(x,t)$ and voltage $V(x,t)$ satisfy the equations
\begin{gather}
\frac{\partial }{{\partial x}}I(x,t) + C(x)\frac{\partial }{{\partial t}}V(x,t) = 0 , 
\label{telegrapherEquation1} \\
\frac{\partial }{{\partial x}}V(x,t) + L(x, t)\frac{\partial }{{\partial t}}I(x,t) = 0 ,
\label{telegrapherEquation2}
\end{gather}
within the waveguide. The model assumes perfect impedance matching between end nodes of the CPW, although in practice high inductance of the thin lines can lead to mismatch, making it difficult to obtain a smooth transfer function, $S_{21}$, on output. The solution of Eqs. (\ref{telegrapherEquation1}) and (\ref{telegrapherEquation2}) for traveling-wave boundary conditions produces output from the second port of the amplifier that includes the amplified signal ($\omega_S$) and the pump ($\omega_P$), as well as a generated idler product of frequency $\omega_3$. In the spectral output, $\omega_S$ and $\omega_3$ are equidistant from $\omega_P$, as sketched in Fig. \ref{fig1} (a), in accordance with energy conservation, i.e., $\omega_S+\omega_3=2\omega_P$. The output is analogous to the products of degenerate 4WM that are encountered in nonlinear optical fibers.\cite{Agrawal2001} 

\subsection{Implications of a Periodic Loading Design}
In 4WM of pump, signal, and idler within nonlinear optical fibers, the electromagnetic fields are described by plane waves. Signal gain arises on output by satisfying the three criteria of (i) energy conservation, i.e., $\omega_S+\omega_3=2\omega_P$; (ii) linear momentum conservation, i.e., the respective plane-wave wavenumbers satisfy $k_S+k_3\cong 2k_P$; and (iii) overall phase matching of the constituent waveforms. In the last criterion, phase matching is tunable by adjusting the input power of the pump, which alters the extent of self-phase modulation of the waveforms, as well as the cross-phase modulation between them.\cite{Agrawal2001} The RF pump, signal, and idler currents that propagate along the CPW of the KIT amplifier also must obey these same criteria in order to achieve signal gain. However, due to the engineered periodic loadings, which necessitate $L_o(x+\ell_o)=L_o(x)$ and $C(x+\ell_o)=C(x)$ in Eqs. (\ref{telegrapherEquation1}) and (\ref{telegrapherEquation2}), the RF waveforms of the KIT amplifier cannot be described by plane waves. This has particular implication for the definitions of both momentum conservation and phase matching, and thus, parametric mixing as a whole, within the KIT.

A quantifiable theory of KIT operation, which addresses the magnitude and bandwidth of parametric signal gain without introduction of ad hoc fitting parameters, must account for the reduced translational symmetry imposed by periodic loadings. For example, in the linear limit, where $L(x,t)\cong L_o(x)$, with $\ell_T\gg \ell_o$, a solution of the voltage and current of Eqs. (\ref{telegrapherEquation1}) and (\ref{telegrapherEquation2}) is properly formed using Floquet-Bloch functions, viz.
\begin{equation}	\label{blochFunctions}
\left\{
\begin{array}{c}
V_k(x,t) \\
I_k(x,t)
\end{array}
\right\} =
\left\{
\begin{array}{c}
V_k(x) \\
I_k(x)
\end{array}
\right\}
e^{i k x - i{\Omega}(k) t} ,
\end{equation}
where $V_k(x)$ and $I_k(x)$ are Floquet-Bloch coefficients periodic in $\ell_o$. This construction introduces a Bloch wavenumber $k$, analogous to the wavenumber of a plane wave, but unique to the first of an infinite number of one-dimensional Brillouin zones. Any other wavenumber may be reduced to one within the first Brillouin zone by translation via a reciprocal lattice vector $2\pi n/\ell_o$, where $n$ is an integer.\cite{Kittel1976,*AshcroftMermin1976} In the KIT amplifier, momentum conservation between parametrically mixing waveforms is defined in terms of these Bloch wavenumbers, instead of their plane-wave counterparts. Additionally, the dispersion frequency $\Omega(k)$, as a function of $k$, forms one of a manifold of bands of dispersion frequencies that comprise the metamaterial band structure of the KIT amplifier. These engineered photonic bands are separated by the stop gaps we described earlier, and each band can have a distinctly different group velocity as a function of $k$. As we shall see from our theory, the parametrically mixing waveforms may be described as superpositions of these band states, which has important consequences for how overall phase matching is defined and achieved within the KIT.

Application of the Floquet-Bloch equation to the study of elementary excitations in the bulk of solid materials is well known, where the arrangement of atoms on a periodic lattice dictates solutions of the form of Eq. (\ref{blochFunctions}). Introductory texts, such as those of Ref. (\onlinecite{Kittel1976,*AshcroftMermin1976}), provide the reader with solutions for lattice vibrations (acoustic and optical phonons) and magnetic-moment precession (spin waves), to name a couple of examples. In particular, electronic states of the bulk formed in this way are the basis for determination of the electronic band structure of solids, and thus, account for the fundamental electronic properties of these materials. Similarly, the metamaterial bands engineered via loading design dictate the parametric behavior of the KIT amplifier, and resemble in principle the development of photonic crystals to manipulate light through the control of dispersion and formation of photonic band states.\cite{John1987,*Yablonovitch1989,*Leung1990,*Zhang1990,*Ho1990,*Lin1998}

Thus, the periodic variations in CPW width depicted in Fig. \ref{fig1}(a), designed to create frequency stops, also control, in a more general sense, the dispersion of RF traveling waves as they propagate along the KIT. A similar concept has been developed for Josephson junctions in a lumped LC transmission line (TWPA) using resonator-based dispersion engineering.\cite{OBrien2014,White2015} The engineered loadings are realized within our theory via the definitions we construct for both the periodic linear kinetic inductance per unit length, $L_o(x)$, and the periodic capacitance per unit length, $C(x)$, as these functions enter Eqs. (\ref{telegrapherEquation1}) and (\ref{telegrapherEquation2}). 

Figure \ref{fig1}(b) shows a schematic of a single loading pattern, or unit cell, of length $\ell_o$. The unit cell is representative of the repeated loadings of Fig. \ref{fig1}(a), and accounts for changes in inductance and capacitance attributable to variations in the width of the CPW. Specifically, within Fig \ref{fig1}(b), there are $R$ regions of different inductance and capacitance pairs, labeled by index $r=1,2,\dots,R$, and denoted by $\mathcal{L}_r$ and $\mathcal{C}_r$, respectively. To simplify matters, we confine our attention to unit cells of even symmetry, such that regions $r$ and $R-r+1$ of Fig. \ref{fig1}(b) possess the same loading sizes, and therefore, the same inductance and capacitance values. Thus, there exists a center region $(R+1)/2$ that may be defined as a non-loading region, with alternating loading and non-loading regions to either side, with the total number of regions $R$ always an odd number. We then model $C(x)$ and $L_o(x)$ of a unit cell ($0\le x<\ell_o$) as
\begin{equation} \label{loadingDesign}
\left\{
\begin{array}{c}
C(x) \\
L_o(x)
\end{array}
\right\} = 
\sum\limits_{r = 1}^R { 
\left\{
\begin{array}{c}
\mathcal{C}_r \\
\mathcal{L}_r
\end{array}
\right\}
\Theta \left( x - x_r \right) \left[ 1 - \Theta \left( x - x_r - \Delta x_r \right) \right] } ,
\end{equation}
where $x_r$ and $\Delta x_r$ are the starting position and length of region $r$, respectively, and $\Theta(x)$ is the conventional Heaviside step function. The definition in Eq. (\ref{loadingDesign}) may be extended to the entire length of the waveguide using $C(x+\ell_o)=C(x)$ and  $L_o(x+\ell_o)=L_o(x)$. To first approximation it is reasonable to model the waveguide as straight, with tens to thousands of repeated unit cells along the length.

\subsection{Parametric Multiwave Mixing within the KIT Amplifier}
\begin{figure}
\includegraphics[width=240pt, height=420pt]{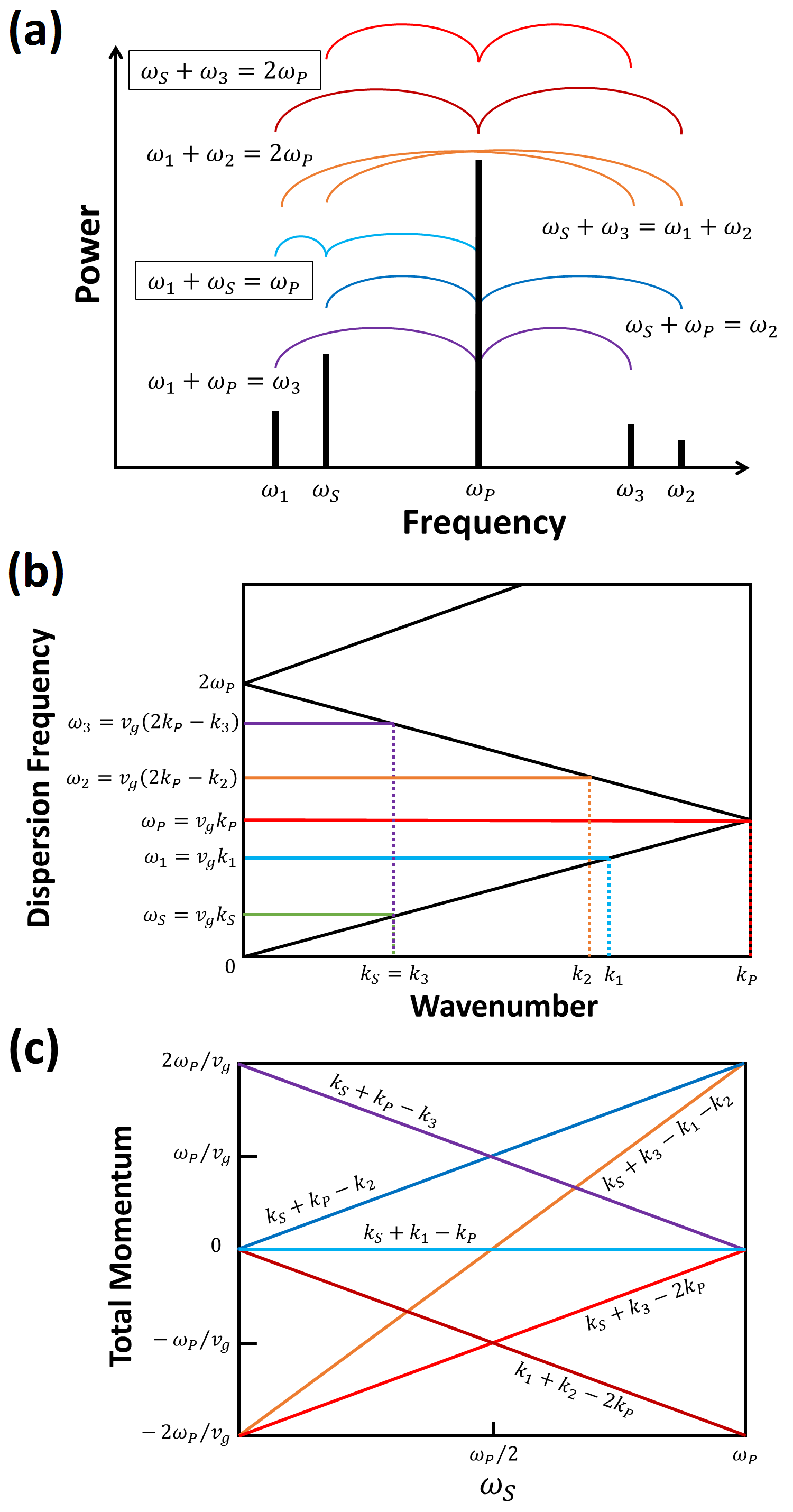}
\caption{\label{fig2} (a) Sketch of six parametric mixing processes of KIT with DC bias, involving pump ($\omega_P$), signal ($\omega_S$), and idler ($\omega_1$, $\omega_2$, $\omega_3$) frequencies. Degenerate 4WM (bright red) and topmost 3WM (cyan) produce broadband signal gain when $I_{DC}=0$ and $I_{DC}>0$, respectively. Additional processes are: 4WM (dark red and orange) and 3WM (blue and purple). (b) Simple depiction of the low-lying frequency-dispersion bands of the KIT amplifier approximated as a CPW with no engineered loadings, where $v_g$ is the magnitude of the approximate group velocity, assumed the same in each band. (c) Total momentum of the six parametric processes of (a) as a function of $\omega_S$ when $\omega_P$ is at the first stop gap, as obtained using the approximate band structure of (b).}
\end{figure}

As mentioned, application of a DC bias, $I_{DC}$, to the KIT amplifier can induce 3WM processes, as well as additional 4WM processes.\cite{Vissers2016} We refer to this scenario as parametric multiwave mixing. Specifically, with $I_{DC}>0$, mixing of a pump of frequency $\omega_P$ and a signal of frequency $\omega_S$ produces three idlers of frequencies $\omega_1$, $\omega_2$, and $\omega_3$. Figure \ref{fig2}(a) summarizes the six parametric scattering processes that occur with onset of $I_{DC}$. Only the degenerate 4WM process of $\omega_S+\omega_3=2\omega_P$ (bright red), when $I_{DC}=0$, and 3WM process of $\omega_S+\omega_1=\omega_P$ (cyan), when $I_{DC}>0$, contribute to broadband signal gain since only these processes achieve momentum conservation: $k_S+k_3\cong 2k_P$ and $k_S+k_1\cong k_P$, respectively, over a broad range of signal frequencies near the bottom of the amplifier dispersion-frequency manifold. 

To see this for the case of $I_{DC}>0$, one can make a simple estimate of the total momentum of each of the six scattering processes as a function of $\omega_S$, using the unloaded dispersion frequency $\Omega(k)=v_g k$. Conservation of energy for the six parametric processes is assumed as in Fig. \ref{fig2}(a). Recall that, within the loaded KIT, momentum conservation of parametrically mixing waveforms requires the use of Bloch wavenumbers. This necessitates folding $\Omega(k)$ into the first Brillouin zone so that every wavenumber $k$ is a reduced-zone wavenumber, as pictured in Fig. \ref{fig2}(b). In folding $\Omega(k)$, we place the pump of tone $\omega_P$ at the first stop gap, such that its corresponding wavenumber $k_P$ is representative of the zone edge, i.e., $\pi/\ell_o$, such that folding of $\Omega(k)$ is done with respect to $k_P$. Thus, for example, the 4WM idler of frequency $\omega_3$, matching to the second dispersion-frequency band, has the dispersion frequency $v_g \left( 2 k_P - k_3 \right)$, as illustrated in the figure. Because the simple approximate of Fig. \ref{fig2}(b) involves unloaded dispersion frequencies, there are technically no gaps in its manifold, and the magnitude of the group velocity is the same in each dispersion-frequency band. This is not the case in reality, but is useful for the present discussion.

For the case of the 3WM scattering process of $\omega_S+\omega_1=\omega_P$, we have $\omega_S\cong v_g k_S$, $\omega_1\cong v_g k_1$, and $\omega_P\cong v_g k_P$, such that $k_S+k_1-k_P\cong\left( \omega_S + \omega_1 - \omega_P \right) / v_g = 0$. Hence, assuming a pump tone placed near the first stop gap, momentum is conserved for this process across the range of signal frequencies $\omega_S$ that lie within the first dispersion-frequency band. As another example, consider the 4WM process of $\omega_S + \omega_3 = \omega_1 + \omega_2$, where we note $\omega_1\cong v_g k_1$ and $\omega_2\cong v_g \left( 2 k_P - k_2 \right)$. In particular, we have $k_2\cong 2k_P - \omega_2/v_g$, and also $k_3\cong 2k_P - \omega_3/v_g$. Thus, the total momentum in this case is $k_S+k_3-k_1-k_2\cong \left( \omega_S - \omega_3 - \omega_1 + \omega_2\right) / v_g$. From Fig. \ref{fig2}(a) we have $\omega_1=\omega_P-\omega_S$, $\omega_2=\omega_P+\omega_S$, and $\omega_3=2\omega_P-\omega_S$, so alternatively we may write the total momentum as $k_S+k_3-k_1-k_2\cong -2\omega_P/v_g + 4 \omega_S/v_g$, which represents a straight line as a function of $\omega_S$, with slope $4/v_g$ and intercept $-2\omega_P/v_g$. The total momentum of the other four processes may be approximated similarly. Figure \ref{fig2}(c) summaries the approximate total momentum of the six parametric scattering processes as a function of signal frequency. Apart from incidental momentum conservation of the 4WM process of $\omega_S + \omega_3 = \omega_1 + \omega_2$, at half the pump frequency, the 3WM process of $\omega_S+\omega_1=\omega_P$ is the source of signal amplification when $I_{DC}>0$, over the range of signal frequencies $\omega_S$ of the lowest-lying dispersion-frequency band. This exercise illustrates the importance of working within the reciprocal-lattice construct dictated by the underlying translational symmetry.

To understand the behavior of parametric multiwave mixing, let us define the current of the amplifier as $I(x,t)=I_{DC}+I_{RF}(x)$, where $I_{RF}(x,t)$ is the total current of all RF waveforms: pump, signal, and three idler products. Substituting this expression into Eq. (\ref{kineticInductancePerLength}) we have a nonlinear kinetic inductance given by
\begin{equation}	\label{kineticInductancePerLengthWithCurrent}
L(x,t) = L_o(x) \left\{ 1 + {\left[ \frac{I_{DC}}{I_*}\right] }^2 
+ 2 \left[ \frac{I_{DC}}{I_*} \right] \left[ \frac{{I_{RF}(x,t)}}{I_*} \right] + { \left[ \frac{{I_{RF}(x,t)}}{I_*} \right] }^2 \right\} ,
\end{equation}
where the term proportional to $I_{DC}\left/ I_* \right.$ is associated with the three 3WM processes of Fig. \ref{fig2}(a). The three 4WM processes correspond to the term involving the square of $I_{RF}(x,t)\left/ I_* \right.$. Noting the energy of kinetic inductance per unit length: $\mathcal{E}_{I}(x,t)=L(x,t)\, I_{RF}(x,t)^2/2$, as RF input power increases from zero with DC bias applied, it is the 3WM processes that activate first, generating the three idler products. Only after RF input power is increased further, beyond a threshold of amplitude $I_{RF}(x,t) \sim 2 I_{DC}$, do the three 4WM processes begin to dominate, with the two additional 4WM processes arising secondarily, after 3WM processes have generated the new idlers $\omega_1$ and $\omega_2$. When $I_{DC}>0$ and $\omega_P$ is situated just above the first stop gap, one finds the 3WM process of $\omega_S+\omega_1=\omega_P$ initiates broadband signal gain centered about half the pump frequency, i.e., $\omega_S\cong\omega_1\cong\omega_P/2$.\cite{Vissers2016} In this case, as sketched in Fig. \ref{fig2}(c), the total momentum $k_S+k_1-k_P$ (cyan) is essentially zero, i.e., $k_S+k_1\cong k_P$, for a range of signal frequencies $\omega_S$ about $\omega_P/2$. 

Because the energy of 3WM contributions is one integer exponent less in the RF current amplitude than 4WM contributions, it takes less pump input power to operate the amplifier in this 3WM mode. However, if the input power is increased too high, the amplitude of traveling-wave current will increase beyond the $I_{RF}(x,t) \sim 2 I_{DC}$ threshold, allowing 4WM to dominate and thus wash out the effect of 3WM broadband gain. In particular, incidental momentum conservation of the 4WM process of $\omega_S + \omega_3 = \omega_1 + \omega_2$, as approximated in Fig. \ref{fig2}(c), will become more prevalent for $\omega_S\cong \omega_P/2$. Hence, the 3WM mode of the amplifier is limited to a range of RF input powers. Similarly, if the waveguide length is made too long then the increased run length of the traveling waves will also lead to current amplitude exceeding the 4WM threshold. Thus, when operated in 3WM mode, the signal gain of the KIT does not exhibit exponential growth with waveguide length; to the contrary, the amplifier has a waveguide length limitation.

\subsection{Nonlinear Forward-Traveling Waves and the Floquet-Bloch Supermode}
Periodic loadings create stop gaps in the dispersion-frequency spectrum of the KIT amplifier, but they also modify the group velocity of RF traveling waves as they propagate along the CPW. In our theory we assume the dispersive propagation is monochromatic, such that, for example, an RF signal of frequency $\omega_S$ injected into the amplifier can be matched to a specific dispersion frequency of the amplifier, as in Fig. \ref{fig2}(b), i.e., one has $\omega_S=\Omega_{\alpha_S}(k_s)$, where $\Omega_{\alpha_S}(k_S)$ is the matching dispersion frequency, governed by the loading design, $k_S$ is the Bloch wavenumber of the signal as it propagates within the CPW, and $\alpha_S$ is the index of the matching band. If the the signal is injected at sufficiently low power, i.e., small amplitude, and the waveguide length is not too long, then the traveling wave will retain a linear form throughout the waveguide, with current and voltage satisfying the Floquet-Bloch condition of Eq. (\ref{blochFunctions}). Since the coefficients $V_k(x)$ and $I_k(x)$ of Eq. (\ref{blochFunctions}) are periodic in the unit cell length $\ell_o$, each may be expanded in a discrete Fourier series. Hence, in the linear limit of KIT operation, the voltage and current of the forward-traveling-wave signal assume the form
\begin{equation}	\label{linearRFTravelingWave}
\left\{
\begin{array}{c}
V_{k_S}(x,t) \\
I_{k_S}(x,t)
\end{array}
\right\} =
\sum\limits_{n=-\infty}^{\infty}
\left\{
\begin{array}{c}
V_n(k_S) \\
I_n(k_S)
\end{array}
\right\}
e^{i \left( k_S + 2\pi n / \ell_o \right) x - i\omega_S t} + \text{c.c.} ,
\end{equation}
where $V_n(k_S)$ and $I_n(k_S)$ are Fourier coefficients.

On the other hand, if the input power is sufficiently high, or the waveguide length is long enough, then $L(x,t)$ becomes nonlinear due to its dependence on the total current, and therefore Eq. (\ref{linearRFTravelingWave}) is no longer a viable solution of Eqs. (\ref{telegrapherEquation1}) and (\ref{telegrapherEquation2}). In this case, if the resulting nonlinear forward-traveling wave propagates adiabatically, then we may still assume the form of Eq. (\ref{linearRFTravelingWave}), except that the Fourier coefficients now take on a slowly varying dependence on $x$, i.e., $V_n(k_S)\rightarrow V_n(k_S,x)\equiv V^{(S)}_n(x)$ and $I_n(k_S)\rightarrow I_n(k_S,x)\equiv I^{(S)}_n(x)$. This nonlinear Floquet-Bloch forward-traveling-wave solution is then of the form
\begin{equation}	\label{nonlinearRFTravelingWave}
\left\{
\begin{array}{c}
V_{k_S}(x,t) \\
I_{k_S}(x,t)
\end{array}
\right\} =
\sum\limits_{n=-\infty}^{\infty}
\left\{
\begin{array}{c}
V^{(S)}_n(x) \\
I^{(S)}_n(x)
\end{array}
\right\}
e^{i \left( k_S + 2\pi n / \ell_o \right) x - i\omega_S t} + \text{c.c.} ,
\end{equation}
where the slowly varying coefficients satisfy the condition
\begin{equation}	\label{slowlyVaryingCondition}
\left| \frac{\partial^2 }{\partial x^2} 
\left\{
\begin{array}{c}
V^{(S)}_n(x) \\
I^{(S)}_n(x)
\end{array}
\right\}
\right|
\ll \left|   
\left( k_S + 2 \pi n \left/ \ell_o \right. \right)
\frac{\partial}{\partial x} 
\left\{
\begin{array}{c}
V^{(S)}_n(x) \\
I^{(S)}_n(x)
\end{array}
\right\}
\right| .
\end{equation}
Additionally, as we shall show in the development of the theory, the coefficients $V^{(S)}_n(x)$ and $I^{(S)}_n(x)$ can be further written as superpositions of the dispersion-frequency band states at $k_S$. In this later expansion, the nonlinear Floquet-Bloch forward-traveling-wave solution of Eq. (\ref{nonlinearRFTravelingWave}) may be referred to as a Floquet-Bloch supermode construction, and is not unlike the description of the plane-polarized electric field of nonlinear arrays of coupled optical waveguides.\cite{Mills1987,*Sipe1988} The nonlinear response of these latter optical superlattices admit transverse-propagating soliton and gap-soliton solutions.\cite{Chen1987-1,*Chen1987-2,*Christodoulides1988} These collective supermode excitations, referred to as Floquet-Bloch solitons, have been demonstrated experimentally in optical waveguide arrays.\cite{Mandelik2003} 

In the KIT amplifier, the parametric mixing of supermodes of pump, signal, and idler products makes for a complex description of overall phase matching, particularly as the band-superposition of each supermode evolves with increasing $x$. Each band-component waveform of a given supermode corresponds to its own characteristic group velocity, which compounds the description of self-phase and cross-phase modulation between these components. In our theory we are able to account for this complex overall phase matching and obtain a quantifiable result for the magnitude and bandwidth of signal gain of the KIT amplifier directly from the loading design.

In the preceding remarks we introduced several important concepts:
\begin{enumerate}
	\item[(i)] KIT amplifiers are engineered with periodic loadings to create frequency stops to inhibit higher pump harmonics, as well as to modify dispersion characteristics of RF traveling waves. The loading design is incorporated into our theory.
	\item[(ii)] Periodic loadings reduce the translational symmetry of the amplifier, which necessitates introduction of a band structure of dispersion frequencies and the Bloch wavenumber of a first Brillouin zone, as in Fig \ref{fig2}(b). In our theory the band structure contains the stop gaps of the loading design, as well as group velocity that may vary from Brillouin zone center to Brillouin zone edge, as well as from band to band.
	\item[(iii)] The criteria for parametric amplification is altered by the reduced translational symmetry: momentum conservation must be expressed in terms of Bloch wavenumbers and overall phase matching between parametrically mixing waveforms must include dispersion-frequency bands with varying group velocities. Our theory incorporates these modified criteria. 
	\item[(iv)] In our theory nonlinear, parametrically-mixing RF forward-traveling waves may be expressed as Floquet-Bloch supermodes constructed from slowly-varying superpositions of dispersion-frequency band states. As these dispersive forward-traveling waves propagate along the CPW, the evolution of the components of their respective superpositions defines the phase matching between them.
	\item[(v)] In this way our theory allows us to calculate the magnitude and bandwidth of parametric signal gain directly from the loading design, without need to introduce ad hoc fitting parameters.
\end{enumerate}
In what follows we first derive the metamaterial band theory of the KIT amplifier and use it as a basis for the theory of parametric multiwave mixing of nonlinear traveling waves. We then present results of calculations using a specific even-symmetry loading design for a KIT amplifier. The parameters of our model are those of Eq. (\ref{loadingDesign}), i.e., the set of $x_r$, $\Delta x_r$, $\mathcal{C}_r$, and $\mathcal{L}_r$, which determine the band structure. With band structure calculated, we show the dispersion of a single nonlinear forward-traveling wave as it propagates down the CPW of the KIT. We then present calculations of the signal gain of the KIT as a function of signal frequency, both without and with DC bias. We conclude with remarks about the extensiblility of our theory to other ladder-type, equivalent-circuit models of nonlinear traveling-wave parametric amplifiers.

\section{Theory}

\subsection{Band Theory of the KIT Amplifier}
We consider a KIT amplifier with engineered dispersion loadings as in Fig. \ref{fig1}(a), modeled as a straight LC ladder-type transmission line of total length $\ell_T$. The variable $x$ measures position along the length of the CPW. A unit cell of length $\ell_o$ of the loading design is sketched in Fig. \ref{fig1}(b) and expressed via Eq. (\ref{loadingDesign}). We first confine our attention to the linear limit of $L(x,t)\cong L_o(x)$, and with $\ell_T\gg\ell_o$ we look for band solutions of Eqs. (\ref{telegrapherEquation1}) and (\ref{telegrapherEquation2}) using the Floquet-Bloch construction of Eq. (\ref{blochFunctions}). The current and voltage defined in this way satisfy periodic boundary conditions where $k$ is the Bloch wavenumber and $\Omega(k)$ is the band frequency.

Like $L_o(x)$ and $C(x)$, the Bloch amplitudes ${V_k}(x)$, ${I_k}(x)$ of Eq. (\ref{blochFunctions}) are periodic in $x$ with periodicity $\ell_o$. Thus, we have four periodic functions that may be expanded in a discrete Fourier series, i.e., each may be transformed in the manner
\begin{gather}
f(x) = \sum\limits_{n = -\infty }^\infty 
f_n \, e^{ 2 \pi i n x \left/ \ell_o \right. } , 	\label{discreteFourierTransformPair1} \\
f_n = \frac{1}{{{\ell _o}}} {\int\limits_0}^{\ell _o} 
f(x) \, e^{ -2 \pi i n x \left/ \ell_o \right. } dx , 	\label{discreteFourierTransformPair2}
\end{gather}
where $f\in\left\{ V_k, I_k, C, L_o \right\}$. We also introduce discrete Fourier transforms $C^{-1}_n$ and  ${L_o}^{-1}_n$, which are elements of matrix inverses corresponding to $C_n$ and ${L_o}_n$, respectively. These may be written explicitly as
\begin{gather}
C_{n}^{-1} = \frac{1}{{{\ell _o}}} {\int\limits_0}^{\ell _o} \frac{1}{C(x)} \, e^{ -2 \pi i n x \left/ \ell_o \right. } dx , 	\label{matrixInverse1} \\
{L_o}_{n}^{-1} = \frac{1}{{{\ell _o}}} {\int\limits_0}^{\ell _o} { \frac{1}{L_o(x)} \, e^{ -2 \pi i n x \left/ \ell_o \right. } dx } , 	\label{matrixInverse2}
\end{gather}
from which one may easily show $\sum^{\infty}_{m=-\infty} C_{n-m}^{-1} C_{m-n'}=\delta_{n,n'}$ and $\sum^{\infty}_{m=-\infty} {L_o}_{n-m}^{-1} {L_o}_{m-n'}=\delta_{n,n'}$. 

If we now substitute Eq. (\ref{blochFunctions}) for $I(x,t)$ and $V(x,t)$ in Eqs. (\ref{telegrapherEquation1}) and (\ref{telegrapherEquation2}), make use of the discrete Fourier transform pair of Eqs. (\ref{discreteFourierTransformPair1}) and (\ref{discreteFourierTransformPair2}) for each of our four periodic functions, and apply the matrix inverses of Eqs. (\ref{matrixInverse1}) and (\ref{matrixInverse2}), we may decouple voltage and current in the transform space, obtaining the result
\begin{gather}
\sum\limits_{n'=-\infty}^{\infty} 
{D_{n,n'}^{\dag}}(k) \, V_{n'}(k) = {{\Omega}(k)}^2 \, V_n(k) , \\
\sum\limits_{n'=-\infty}^{\infty} 
{D_{n,n'}}(k) \, I_{n'}(k) = {{\Omega}(k)}^2 \, I_n(k) ,
\end{gather}
where we have introduced a non-Hermitian dispersion matrix with elements given by
\begin{equation}	\label{dispersionMatrix0}
D_{n,n'}(k) =
\sum\limits_{n''=-\infty}^{\infty}
{L_o}_{n-n''}^{-1} \, C_{n''-n'}^{-1} 
\left( k + 2\pi n'' \left/ \ell_o \right. \right) 
\left( k + 2\pi n' \left/ \ell_o \right. \right) .
\end{equation}
Diagonalization of the matrix of Eq. (\ref{dispersionMatrix0}) produces the metamaterial band structure of the KIT amplifier. 

Since the above dispersion matrix is non-Hermitian, we must introduce left, ${u_{n'}^{(\alpha)}}(k)$, and right, ${e_{n}^{(\alpha)}}(k)$, eigenvectors and express the diagonalization formally as
\begin{gather}
\sum\limits_{n'=-\infty}^{\infty} {u_{n'}^{(\alpha)}}(k) {D_{n',n}}(k) = 
{{\Omega_{\alpha}}(k)}^2 \, {u_{n}^{(\alpha)}}(k) , \label{linearEigenproblem1} \\
\sum\limits_{n'=-\infty}^{\infty} {D_{n,n'}}(k) \, {e_{n'}^{(\alpha)}}(k) = 
{{\Omega_{\alpha}}(k)}^2 \, {e_{n}^{(\alpha)}}(k) . \label{linearEigenproblem2}
\end{gather}
Here $\alpha$ refers to a band index, of which there are an infinite number at each value of $k$. For orthonormality and completeness we define
\begin{gather}
\sum\limits_{n=-\infty}^{\infty} {u_{n}^{(\alpha)}}(k) \, {e_{n}^{(\beta)}}(k) = \delta_{\alpha,\beta} , 	\label{orthonormalityCompleteness1} \\
\sum\limits_{\alpha} {e_{n}^{(\alpha)}}(k) \, {u_{n'}^{(\alpha)}}(k) = \delta_{n,n'} , 	\label{orthonormalityCompleteness2}
\end{gather}
which presupposes the existence of a non-unitary similarity transformation diagonalizing Eq. (\ref{dispersionMatrix0}). Several notable properties of the dispersion matrix, which are easily verified, are 
\begin{gather}	\label{eigenRelations}
D_{n,n'}(-k) = {{D_{-n,-n'}}(k)} , \;\; {\Omega_\alpha}(-k)={{\Omega_\alpha}(k)}^* , \\
u_n^{(\alpha)}(-k) = {{u_{-n}^{(\alpha)}}(k)}^*, \;\; {e_n^{(\alpha)}}(-k) = {{e_{-n}^{(\alpha)}}(k)}^* .
\end{gather}
These relations are consistent with reciprocity of linear waveform propagation in either direction of the waveguide.

Appendix \ref{appendixDispersionMatrix} shows how one may apply Eq. (\ref{loadingDesign}) to Eq. (\ref{dispersionMatrix0}) to obtain a useful formulation of $D_{n,n'}(k)$ for arbitrary loading design of even symmetry. The result is the real-valued dispersion matrix element of Eq. (\ref{arbitraryDispersionMatrix}), which we may write as 
\begin{multline}	\label{dispersionMatrix}
D_{n,n'}(k) = 
\frac{1}{ \mathcal{L}_{(R+1)/2} \mathcal{C}_{(R+1)/2} }
{ \Big( k + 2\pi n \left/ \ell_o \right. \Big) }^2 \delta_{n,n'} \\
+ \Bigg[ 
\mathcal{D}_{n-n'}(\left\{ \mathcal{L}_r \right\}, \left\{ \mathcal{C}_r \right\} ) \;
\Big( k + 2\pi n \left/ \ell_o \right. \Big) 
+ \mathcal{D}_{n-n'}(\left\{ \mathcal{C}_r \right\}, \left\{ \mathcal{L}_r \right\} ) \;
\Big( k + 2\pi n' \left/ \ell_o \right. \Big) \Bigg]
\Big( k + 2\pi n' \left/ \ell_o \right. \Big) ,
\end{multline}
where we have introduced coefficients
\begin{equation}	\label{dispersionMatrixCoefficient}
\mathcal{D}_n(\left\{ \mathcal{L}_r \right\}, \left\{ \mathcal{C}_r \right\} ) = 
\frac{1}{ 2\pi n } \sum\limits_{r=1}^{(R - 1)/2} 
\left( \frac{1}{\mathcal{L}_r} + \frac{1}{\mathcal{L}_{r+1}} \right)
\left( \frac{1}{\mathcal{C}_r} - \frac{1}{\mathcal{C}_{r+1}} \right)
\sin{ \left( 2\pi \sum\limits_{r'=1}^r \Delta x_{r'} \, n \left/ \ell_o \right. \right) } .
\end{equation}
Note that $\mathcal{D}_n(\left\{ \mathcal{C}_r \right\}, \left\{ \mathcal{L}_r \right\} )$ is just Eq. (\ref{dispersionMatrixCoefficient}) with $\mathcal{L}_r$, $\mathcal{L}_{r+1}$ interchanged with $\mathcal{C}_r$, $\mathcal{C}_{r+1}$. The generality of Eq. (\ref{dispersionMatrix}) allows us to explore the band structure of a wide range of loading designs of even symmetry. 

\begin{figure}
\includegraphics[width=400pt, height=440pt]{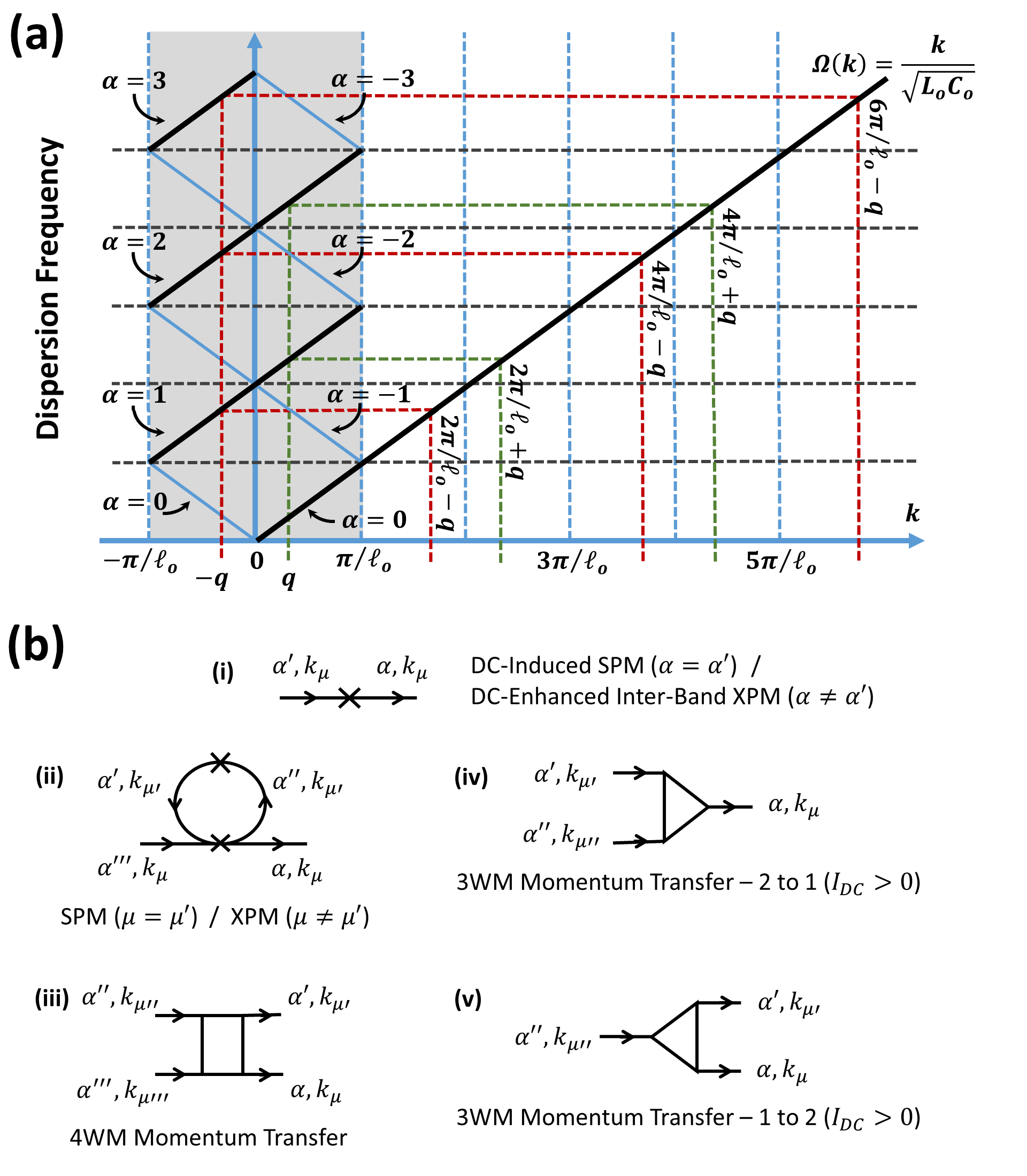}
\caption{\label{fig3} (a) Comparison of reduced-zone scheme ($-\pi/\ell_o<k<\pi/\ell_o$) and extended-zone scheme ($-\infty<k<\infty$) representations of KIT dispersion frequencies, as described in the text. (b) The five types of multichannel parametric scattering associated with multiwave mixing within the KIT amplifier, also discussed in the text.}
\end{figure}

For the remainder of our discussion we adopt a specific convention for labeling the bands $\alpha$ of the KIT amplifier. Figure \ref{fig3} provides a comparison of reduced-zone scheme ($-\pi/\ell_o<k<\pi/\ell_o$) and extended-zone scheme ($-\infty<k<\infty$) representations of KIT dispersion frequencies, using the no-load limit of constant $C(x)\equiv C_o$ and $L_o(x)\equiv L_o$ to illustrate how the extended-zone dispersion curve $\Omega(k)=k/\sqrt{L_oC_o}$ is mapped to bands. The dashed vertical blue lines represent Brillouin zone boundaries defined by reciprocal lattice vectors $G_\alpha=2i\pi\alpha/\ell_o$. The black (blue) line segments of the reduced zone, corresponding to specific positive (negative) indexes $\alpha$, map to the extended-zone dispersion curve of $k>0$ ($k<0$). Our convention is to label the bands $\alpha=\pm 1,\pm 2,\dots$ such that in the reduced zone we have the ascending order $\Omega_0(k)<\Omega_{-1}(k)<\Omega_{1}(k)<\Omega_{-2}(k)<\dots$, for $0<k<\pi/\ell_o$, and $\Omega_0(k)<\Omega_{1}(k)<\Omega_{-1}(k)<\Omega_{2}(k)<\dots$, for $-\pi/\ell_o<k<0$. In this way the mapping of extended to reduced zone follows as $\Omega(k+2\pi\alpha/\ell_o)=\Omega_\alpha(k)$, for any loading design. The red and green dashed lines show several examples of mapping, involving the wavenumbers $\pm q$ of the reduced zone. This labeling convention is useful to the understanding of the effect of a loading design on parametric mixing. 

\subsection{Theory of Multiwave Mixing within the KIT Amplifier}
If dispersion loadings of the KIT are engineered to inhibit formation of pump harmonics and shock waves then we may confine our solution of Eqs. (\ref{telegrapherEquation1}) and (\ref{telegrapherEquation2}) to forward-traveling RF waves, labeled by index $\mu$, consisting solely of the pump ($\mu=P$), signal ($\mu=S$), and idlers ($\mu=1,2,3$), with corresponding frequencies $\omega_\mu$, as depicted in Fig. \ref{fig2}(a). Inside the waveguide we approximate voltage and current of the $\mu$-th forward-traveling wave in terms of a Floquet-Bloch-like function, as in the example of Eq. (\ref{nonlinearRFTravelingWave}) for the case of $\mu=S$, again introducing a Bloch wavenumber, defined as $k_\mu$. Assuming monochromatic dispersion, $k_\mu$ is determined by matching frequency $\omega_\mu$ to a specific band $\alpha_\mu$ of the KIT, i.e., $\omega_\mu=\Omega_{\alpha_\mu}(k_\mu)$, as in Fig. \ref{fig2}(b). Like Eq. (\ref{nonlinearRFTravelingWave}), we allow the Fourier-like coefficients of each traveling wave to assume a slowly-varying dependence on $x$. Specifically, we may write
\begin{equation}	\label{travelingWaves}
\left\{
\begin{array}{c}
V^{(\mu)}_{k_\mu}(x,t) \\
I^{(\mu)}_{k_\mu}(x,t) 
\end{array}
\right\}
= \sum\limits_{n=-\infty}^{\infty} 
\left\{
\begin{array}{c}
V^{(\mu)}_n(x) \\
I^{(\mu)}_n(x)
\end{array}
\right\}
e^{i \left( k_\mu + 2\pi n \left/ \ell_o \right. \right) x - i\omega_\mu t} + \text{c.c.} , 
\end{equation}
where the assumption of a slowly-varying-amplitude may be expressed as
\begin{equation}	\label{slowlyVaryingApproximation}
\left| \frac{\partial^2 }{\partial x^2} 
\left\{
\begin{array}{c}
V^{(\mu)}_n(x) \\
I^{(\mu)}_n(x) 
\end{array}
\right\}
\right|
\ll \left|   
\left( k_\mu + 2 \pi n \left/ \ell_o \right. \right)
\frac{\partial}{\partial x} 
\left\{
\begin{array}{c}
V^{(\mu)}_n(x) \\
I^{(\mu)}_n(x) 
\end{array}
\right\}
\right| ,
\end{equation}
as in Eq. (\ref{slowlyVaryingCondition}). Allowing for a DC bias $I_{DC}$ the full solution, including all mixing waveforms and DC, may be expressed as
\begin{gather}
V(x,t) = \sum\limits_{\mu} V^{(\mu)}_{k_\mu}(x,t) + \text{c.c.} , \;\;  \label{fullSolution1} \\
I(x,t) = I_{DC} + \left[ \sum\limits_{\mu} I^{(\mu)}_{k_\mu}(x,t) + \text{c.c.} \right]  \label{fullSolution2}
\end{gather}

Equations (\ref{fullSolution1}) and (\ref{fullSolution2}) may be substituted into Eqs. (\ref{telegrapherEquation1}) and (\ref{telegrapherEquation2}) to obtain an expression of the current coefficients $I^{(\mu)}_{n}(x)$ decoupled from those of the voltage, $V^{(\mu)}_{n}(x)$. The steps of derivation are similar to those we outlined for the linear limit, with the caveat that we may also leverage the canonical transformation implied by Eqs. (\ref{orthonormalityCompleteness1}) and (\ref{orthonormalityCompleteness2}). For example, the current coefficients may be expanded as 
\begin{equation}	\label{currentExpansion}
I^{(\mu)}_{n}(x) = I_* \sum\limits_\alpha A^{(\mu)}_\alpha(x) \, e_n^{(\alpha)}(k_\mu) ,
\end{equation}
where the dimensionless current amplitudes $A^{(\mu)}_\alpha(x)$ also are slowly varying in $x$, in the manner of Eq. (\ref{slowlyVaryingApproximation}). Equation (\ref{currentExpansion}) expresses the fact that, although an initial boundary condition may be imposed on the slowly-varying amplitude at $x=0$, the effect of dispersion will cause the waveform to hybridize with other states of $k_\mu$ as it evolves in $x$ along the waveguide. These assumptions regarding the form of our solution hold as long as the traveling wave propagates adiabatically along the waveguide. 

With application of the canonical transformation, solution of the current $I(x,t)$, in particular, follows by solving an infinite set of coupled differential equations involving amplitudes $A^{(\mu)}_\alpha(x)$, expressed in terms of eigenvalues and vectors of our band theory. Appendix \ref{appendixNonlinearEquations} gives a general derivation of the coupled equations. Our focus will be restricted to dynamics of the six mixing processes of Fig. \ref{fig2}(a). These involve the phases
\begin{gather}
\Delta\beta^{(1)}_{3wm} = k_1 + k_S - k_P , \label{mismatch3wm1} \\
\Delta\beta^{(2)}_{3wm} = k_S + k_P - k_2 , \label{mismatch3wm2} \\
\Delta\beta^{(3)}_{3wm} = k_1 + k_P - k_3 , \label{mismatch3wm3} \\
\Delta\beta^{(1)}_{4wm} = k_S + k_3 - 2k_P , \label{mismatch4wm1} \\
\Delta\beta^{(2)}_{4wm} = k_1 + k_2 - 2k_P , \label{mismatch4wm2} \\
\Delta\beta^{(3)}_{4wm} = k_S + k_3 - k_1 - k_2 , \label{mismatch4wm3}
\end{gather}
which measure the extent of momentum conservation for each process, respectively. Hence, defining $A_{DC}=I_{DC}/I_*$, we have directly from Eq. (\ref{mwmEquations}) the coupled subset of pump, signal, and idler amplitude equations given by
\begin{multline}	\label{pump}
-i\frac{\partial }{\partial x} A^{(P)}_\alpha(x) \cong
\sum\limits_{\alpha'} f^{(P)}_{\alpha, \alpha'}(A_{DC}) \, A^{(P)}_{\alpha'}(x)
+ 2A_{DC} \sum\limits_{{\alpha}', {\alpha}''} \Bigg[
f^{(P,S,1)}_{\alpha, \alpha', \alpha''} \, A^{(S)}_{\alpha'}(x) \, A^{(1)}_{\alpha''}(x) \, e^{i \Delta\beta^{(1)}_{3wm} x} \\
+ {f^{(P,\bar{S},2)}_{\alpha, \alpha', \alpha''}} \, {A^{(S)}_{\alpha'}(x)}^* \, A^{(2)}_{\alpha''}(x) \, e^{-i \Delta\beta^{(2)}_{3wm} x} 
+ {f^{(P,\bar{1},3)}_{\alpha, \alpha', \alpha''}} \, {A^{(1)}_{\alpha'}(x)}^* \, A^{(3)}_{\alpha''}(x) \, e^{-i \Delta\beta^{(3)}_{3wm} x} \Bigg] \\
+ \sum\limits_{\alpha',\alpha'',\alpha'''} 
\Bigg\{ \Big[ f^{(P,\bar{P},P,P)}_{\alpha,\alpha',\alpha'',\alpha'''} \, {A^{(P)}_{\alpha'}(x)}^{*} A^{(P)}_{\alpha''}(x)
+ 2 f^{(P,\bar{S},S,P)}_{\alpha,\alpha',\alpha'',\alpha'''} \, {A^{(S)}_{\alpha'}(x)}^{*} A^{(S)}_{\alpha''}(x) \\
+ 2 f^{(P,\bar{1},1,P)}_{\alpha,\alpha',\alpha'',\alpha'''} \, {A^{(1)}_{\alpha'}(x)}^{*} A^{(1)}_{\alpha''}(x)
+ 2 f^{(P,\bar{2},2,P)}_{\alpha,\alpha',\alpha'',\alpha'''} \, {A^{(2)}_{\alpha'}(x)}^{*} A^{(2)}_{\alpha''}(x) \\
+ 2 f^{(P,\bar{3},3,P)}_{\alpha,\alpha',\alpha'',\alpha'''} \, {A^{(3)}_{\alpha'}(x)}^{*} A^{(3)}_{\alpha''}(x) \Big] A^{(P)}_{\alpha'''}(x) \\
+ 2 f^{(P,\bar{P},S,3)}_{\alpha,\alpha',\alpha'',\alpha'''} \, {A^{(P)}_{\alpha'}(x)}^* A^{(S)}_{\alpha''}(x) A^{(3)}_{\alpha'''}(x) \, 
e^{i\Delta\beta^{(1)}_{4wm} x} \\
+ 2 f^{(P,\bar{P},1,2)}_{\alpha,\alpha',\alpha'',\alpha'''} \, {A^{(P)}_{\alpha'}(x)}^* A^{(1)}_{\alpha''}(x) A^{(2)}_{\alpha'''}(x) \, 
e^{i\Delta\beta^{(2)}_{4wm} x} \Bigg\} ,
\end{multline}
\begin{multline}	\label{signal}
-i \frac{\partial }{\partial x} A^{(S)}_\alpha(x) \cong
\sum\limits_{\alpha'} f^{(S)}_{\alpha, \alpha'}(A_{DC}) \, A^{(S)}_{\alpha'}(x) \\
+ 2A_{DC} \sum\limits_{{\alpha}', {\alpha}''} \Bigg[
{f^{(S, \bar{1}, P)}_{\alpha, \alpha', \alpha''}} \, {A^{(1)}_{\alpha'}(x)}^* \, A^{(P)}_{\alpha''}(x) \, e^{-i \Delta\beta^{(1)}_{3wm} x} 
+ {f^{(S, \bar{P}, 2)}_{\alpha, \alpha', \alpha''}} \, {A^{(P)}_{\alpha'}(x)}^* \, A^{(2)}_{\alpha''}(x) \, e^{-i \Delta\beta^{(2)}_{3wm} x} \Bigg] \\
+ \sum\limits_{\alpha',\alpha'',\alpha'''} 
\Bigg\{ \Big[
f^{(S,\bar{S},S,S)}_{\alpha,\alpha',\alpha'',\alpha'''} \, {A^{(S)}_{\alpha'}(x)}^{*} A^{(S)}_{\alpha''}(x)
+ 2 f^{(S,\bar{P},P,S)}_{\alpha,\alpha',\alpha'',\alpha'''} \, {A^{(P)}_{\alpha'}(x)}^{*} A^{(P)}_{\alpha''}(x) \\
+ 2 f^{(S,\bar{1},1,S)}_{\alpha,\alpha',\alpha'',\alpha'''} \, {A^{(1)}_{\alpha'}(x)}^{*} A^{(1)}_{\alpha''}(x)
+ 2 f^{(S,\bar{2},2,S)}_{\alpha,\alpha',\alpha'',\alpha'''} \, {A^{(2)}_{\alpha'}(x)}^{*} A^{(2)}_{\alpha''}(x) \\
+ 2 f^{(S,\bar{3},3,S)}_{\alpha,\alpha',\alpha'',\alpha'''} \, {A^{(3)}_{\alpha'}(x)}^{*} A^{(3)}_{\alpha''}(x)
\Big] A^{(S)}_{\alpha'''}(x) \\
+ f^{(S,\bar{3},P,P)}_{\alpha,\alpha',\alpha'',\alpha'''} \, {A^{(3)}_{\alpha'}(x)}^* A^{(P)}_{\alpha''}(x) A^{(P)}_{\alpha'''}(x) \, 
e^{-i\Delta\beta^{(1)}_{4wm} x} \\
+ 2f^{(S,\bar{3},1,2)}_{\alpha,\alpha',\alpha'',\alpha'''} \, {A^{(3)}_{\alpha'}(x)}^* A^{(1)}_{\alpha''}(x) A^{(2)}_{\alpha'''}(x) \, 
e^{-i\Delta\beta^{(3)}_{4wm} x} \Bigg\} ,
\end{multline}
\begin{multline}	\label{idler1}
-i \frac{\partial }{\partial x} A^{(1)}_\alpha(x) \cong
\sum\limits_{\alpha'} f^{(1)}_{\alpha, \alpha'}(A_{DC}) \, A^{(1)}_{\alpha'}(x) \\
+ 2A_{DC} \sum\limits_{{\alpha}', {\alpha}''} \Bigg[
{f^{(1,\bar{S},P)}_{\alpha, \alpha', \alpha''}} \, {A^{(S)}_{\alpha'}(x)}^* \, A^{(P)}_{\alpha''}(x) \, e^{-i \Delta\beta^{(1)}_{3wm} x} 
+ {f^{(1,\bar{P},3)}_{\alpha, \alpha', \alpha''}} \, {A^{(P)}_{\alpha'}(x)}^* \, A^{(3)}_{\alpha''}(x) \, e^{-i \Delta\beta^{(3)}_{3wm} x} \Bigg] \\
\;\;\;\;\;
+ \sum\limits_{\alpha',\alpha'',\alpha'''} 
\Bigg\{ \Big[
f^{(1,\bar{1},1,1)}_{\alpha,\alpha',\alpha'',\alpha'''} \, {A^{(1)}_{\alpha'}(x)}^{*} A^{(1)}_{\alpha''}(x) 
+ 2 f^{(1,\bar{P},P,1)}_{\alpha,\alpha',\alpha'',\alpha'''} \, {A^{(P)}_{\alpha'}(x)}^{*} A^{(P)}_{\alpha''}(x)  \\
+ 2 f^{(1,\bar{S},S,1)}_{\alpha,\alpha',\alpha'',\alpha'''} \, {A^{(S)}_{\alpha'}(x)}^{*} A^{(S)}_{\alpha''}(x) 
+ 2 f^{(1,\bar{2},2,1)}_{\alpha,\alpha',\alpha'',\alpha'''} \, {A^{(2)}_{\alpha'}(x)}^{*} A^{(2)}_{\alpha''}(x) \\
+ 2 f^{(1,\bar{3},3,1)}_{\alpha,\alpha',\alpha'',\alpha'''} \, {A^{(3)}_{\alpha'}(x)}^{*} A^{(3)}_{\alpha''}(x)
\Big] A^{(1)}_{\alpha'''}(x) \\
+ f^{(1,\bar{2},P,P)}_{\alpha,\alpha',\alpha'',\alpha'''} \, {A^{(2)}_{\alpha'}(x)}^* A^{(P)}_{\alpha''}(x) A^{(P)}_{\alpha'''}(x) \, 
e^{-i\Delta\beta^{(2)}_{4wm} x} \\
+ 2 f^{(1,\bar{2},S,3)}_{\alpha,\alpha',\alpha'',\alpha'''} \, {A^{(2)}_{\alpha'}(x)}^* A^{(S)}_{\alpha''}(x) A^{(3)}_{\alpha'''}(x) \, 
e^{i\Delta\beta^{(3)}_{4wm} x} \Bigg\} ,
\end{multline}
\begin{multline}	\label{idler2}
-i \frac{\partial }{\partial x} A^{(2)}_\alpha(x) \cong
\sum\limits_{\alpha'} f^{(2)}_{\alpha, \alpha'}(A_{DC}) \, A^{(2)}_{\alpha'}(x)
+ 2A_{DC} \sum\limits_{{\alpha}', {\alpha}''} 
{f^{(2,P,S)}_{\alpha, \alpha', \alpha''}} \, A^{(P)}_{\alpha'}(x) \, A^{(S)}_{\alpha''}(x) \, e^{i \Delta\beta^{(2)}_{3wm} x} \\
+ \sum\limits_{\alpha',\alpha'',\alpha'''} 
\Bigg\{ \Big[
f^{(2,\bar{2},2,2)}_{\alpha,\alpha',\alpha'',\alpha'''} \, {A^{(2)}_{\alpha'}(x)}^{*} A^{(2)}_{\alpha''}(x) 
+ 2 f^{(2,\bar{P},P,2)}_{\alpha,\alpha',\alpha'',\alpha'''} \, {A^{(P)}_{\alpha'}(x)}^{*} A^{(P)}_{\alpha''}(x) \\
+ 2 f^{(2,\bar{S},S,2)}_{\alpha,\alpha',\alpha'',\alpha'''} \, {A^{(S)}_{\alpha'}(x)}^{*} A^{(S)}_{\alpha''}(x) 
+ 2 f^{(2,\bar{1},1,2)}_{\alpha,\alpha',\alpha'',\alpha'''} \, {A^{(1)}_{\alpha'}(x)}^{*} A^{(1)}_{\alpha''}(x) \\
+ 2 f^{(2,\bar{3},3,2)}_{\alpha,\alpha',\alpha'',\alpha'''} \, {A^{(3)}_{\alpha'}(x)}^{*} A^{(3)}_{\alpha''}(x)
\Big] A^{(2)}_{\alpha'''}(x) \\
+ f^{(2,\bar{1},P,P)}_{\alpha,\alpha',\alpha'',\alpha'''} \, {A^{(1)}_{\alpha'}(x)}^* A^{(P)}_{\alpha''}(x) A^{(P)}_{\alpha'''}(x) \, 
e^{-i\Delta\beta^{(2)}_{4wm} x} \\
+ 2 f^{(2,\bar{1},S,3)}_{\alpha,\alpha',\alpha'',\alpha'''} \, {A^{(1)}_{\alpha'}(x)}^* A^{(S)}_{\alpha''}(x) A^{(3)}_{\alpha'''}(x) \, 
e^{i\Delta\beta^{(3)}_{4wm} x} \Bigg\} ,
\end{multline}
\begin{multline}	\label{idler3}
-i \frac{\partial }{\partial x} A^{(3)}_\alpha(x) \cong
\sum\limits_{\alpha'} f^{(3)}_{\alpha, \alpha'}(A_{DC}) \, A^{(3)}_{\alpha'}(x)
+ 2A_{DC} \sum\limits_{{\alpha}', {\alpha}''}
{f^{(3,P,1)}_{\alpha, \alpha', \alpha''}} \, A^{(P)}_{\alpha'}(x) \, A^{(1)}_{\alpha''}(x) \, e^{i \Delta\beta^{(3)}_{3wm} x} \\
+ \sum\limits_{\alpha',\alpha'',\alpha'''} 
\Bigg\{ \Big[
f^{(3,\bar{3},3,3)}_{\alpha,\alpha',\alpha'',\alpha'''} \, {A^{(3)}_{\alpha'}(x)}^{*} A^{(3)}_{\alpha''}(x) 
+ 2 f^{(3,\bar{P},P,3)}_{\alpha,\alpha',\alpha'',\alpha'''} \, {A^{(P)}_{\alpha'}(x)}^{*} A^{(P)}_{\alpha''}(x) \\
+ 2 f^{(3,\bar{S},S,3)}_{\alpha,\alpha',\alpha'',\alpha'''} \, {A^{(S)}_{\alpha'}(x)}^{*} A^{(S)}_{\alpha''}(x) 
+ 2 f^{(3,\bar{1},1,3)}_{\alpha,\alpha',\alpha'',\alpha'''} \, {A^{(1)}_{\alpha'}(x)}^{*} A^{(1)}_{\alpha''}(x) \\
+ 2 f^{(3,\bar{2},2,3)}_{\alpha,\alpha',\alpha'',\alpha'''} \, {A^{(2)}_{\alpha'}(x)}^{*} A^{(2)}_{\alpha''}(x)
\Big] A^{(3)}_{\alpha'''}(x) \\
+ f^{(3,\bar{S},P,P)}_{\alpha,\alpha',\alpha'',\alpha'''} \, {A^{(S)}_{\alpha'}(x)}^* A^{(P)}_{\alpha''}(x) A^{(P)}_{\alpha'''}(x) \, 
e^{-i\Delta\beta^{(1)}_{4wm} x} \\
+ 2 f^{(3,\bar{S},1,2)}_{\alpha,\alpha',\alpha'',\alpha'''} \, {A^{(S)}_{\alpha'}(x)}^* A^{(1)}_{\alpha''}(x) A^{(2)}_{\alpha'''}(x) \, 
e^{-i\Delta\beta^{(3)}_{4wm} x} \Bigg\} .
\end{multline}
In the above equations we have defined $\bar{\mu}=-\mu$ and $k_{-\mu}=-k_\mu$, as introduced in Appendix \ref{appendixNonlinearEquations}. Boundary conditions corresponding to injected pump and signal may be expressed as $A^{(\mu)}_\alpha(0)=\bar{A}_\mu \delta_{\alpha,\alpha_\mu}$, where $\bar{A}_P$ and $\bar{A}_S$ are constants defined by the RF input power. For the idler products we set $\bar{A}_1=\bar{A}_2=\bar{A}_3=0$.

The mixing coefficients $f^{(\mu)}_{\alpha,\alpha'}(A_{DC})$, $f^{(\mu,\mu',\mu'')}_{\alpha,\alpha',\alpha''}$, and $f^{(\mu,\mu',\mu'',\mu''')}_{\alpha,\alpha',\alpha'',\alpha'''}$ of the above coupled equations are given by Eqs. (\ref{fHybridizationAppendix}) through (\ref{f4WMAppendix}), respectively. These are the hybridizing coefficients
\begin{equation}	\label{fHybridization}
f^{(\mu)}_{\alpha, \alpha'}(A_{DC}) =
\frac{1}{2} \, \tilde{\Lambda}_{\alpha,\alpha'}^{-1}(k_\mu)
\left[ {{\Omega_{\alpha_\mu}}(k_\mu)}^2 \, 
\left( 1 + A_{DC}^2 \right) - {{\Omega_{\alpha'}}(k_\mu)}^2 \right] ,
\end{equation}
the nonlinear 3WM coefficients
\begin{equation}	\label{f3WM}
f^{(\mu,\mu',\mu'')}_{\alpha, \alpha', \alpha''} =
\frac{1}{2} \, {{\Omega_{\alpha_\mu}}(k_\mu)}^2 \; \sum\limits_{ \substack{ n, n', n'' \\ ( n' + n'' = n ) } }
\sum\limits_{\beta} \tilde{\Lambda}_{\alpha,\beta}^{-1}(k_\mu) \,
{{u_n^{(\beta)}}(k_{\mu})} \; {e_{n'}^{(\alpha')}}(k_{\mu'}) \; {e_{n''}^{(\alpha'')}}(k_{\mu''}) ,
\end{equation}
and the nonlinear 4WM coefficients
\begin{equation}	\label{f4WM}
\begin{array}{l}
f^{(\mu,\mu',\mu'',\mu''')}_{\alpha, \alpha', \alpha'', \alpha'''} = 
\frac{1}{2} \, {{\Omega_{\alpha_\mu}}(k_\mu)}^2 \; \sum\limits_{ \substack{ n, n', n'', n''' \\ ( n' + n'' + n''' = n ) } }
\sum\limits_{\beta} \tilde{\Lambda}_{\alpha,\beta}^{-1}(k_\mu) \,
{{u_n^{(\beta)}}(k_{\mu})} \; {e_{n'}^{(\alpha')}}(k_{\mu'}) \; {e_{n''}^{(\alpha'')}}(k_{\mu''}) \; {e_{n'''}^{(\alpha''')}}(k_{\mu'''}) ,
\end{array}
\end{equation}
where, from Eq. (\ref{lambdaTildeAppendix}), we also have 
\begin{equation}	\label{lambdaTilde}
\tilde{\Lambda}_{\alpha,\alpha'}(k_\mu) =
\sum\limits_{n, n'} \Lambda_{n,n'}(k_\mu) \, 
{u_{n}^{(\alpha)}}(k_\mu) \, {e_{n'}^{(\alpha')}}(k_\mu) .
\end{equation}
Note that the mixing coefficients depend on a matrix of elements $\Lambda_{n,n'}(k)$ defined in Eq. (\ref{lambda}). Similar to the derivation in Appendix \ref{appendixDispersionMatrix} for the dispersion matrix $D_{n,n'}(k)$ of Eq. (\ref{dispersionMatrix}), we may write $\Lambda_{n,n'}(k)$ as
\begin{multline}	\label{lambdaMatrix}
\Lambda_{n,n'}(k) =
\frac{1}{ \mathcal{L}_{(R+1)/2} \mathcal{C}_{(R+1)/2} }
\Big( k + 2\pi n \left/ \ell_o \right. \Big) \delta_{n,n'} \\
+ \mathcal{D}_{n-n'}(\left\{ \mathcal{L}_r \right\}, \left\{ \mathcal{C}_r \right\} ) 
\Big[ k + \pi \left( n + n' \right) \left/ \ell_o \right. \Big] 
+ \mathcal{D}_{n-n'}(\left\{ \mathcal{C}_r \right\}, \left\{ \mathcal{L}_r \right\} )
\Big( k + 2\pi n' \left/ \ell_o \right. \Big) ,
\end{multline}
appropriate for a unit cell of even symmetry. 

As can be deduced from Eqs. (\ref{dispersionMatrix}) and (\ref{lambdaMatrix}), an important limiting form of the mixing coefficients is obtained when engineered loadings are absent from the waveguide; in this case we have
\begin{equation}	\label{noLoadingsFhyp}
f^{(\mu)}_{\alpha,\alpha'}(A_{DC}) = 
\frac{ { \left( k_\mu + 2\pi\alpha_{\mu} \left/ \ell_o \right. \right) }^2
\left( 1 + A_{DC}^2 \right)
- { \left( k_\mu + 2\pi\alpha \left/ \ell_o \right. \right) }^2 }
{ 2 \left( k_\mu + 2\pi\alpha \left/ \ell_o \right. \right) } \delta_{\alpha, \alpha'} ,
\end{equation}
\begin{equation}	\label{noLoadingsF3wmA}
f^{(\mu,\mu',\mu'')}_{\alpha,\alpha',\alpha''} = 
\frac{1}{2} \left( k_\mu + 2\pi\alpha \left/ \ell_o \right. \right) 
\delta_{\alpha, \alpha' + \alpha''} ,
\end{equation}
\begin{equation}	\label{noLoadingsF3wmB}
f^{(\mu,\bar{\mu}',\mu'')}_{\alpha,\alpha',\alpha''} = 
\frac{1}{2} \left( k_\mu + 2\pi\alpha \left/ \ell_o \right. \right) 
\delta_{\alpha + \alpha' , \alpha''} ,
\end{equation}
\begin{equation}	\label{noLoadingsF4wm}
f^{(\mu,\bar{\mu}',\mu'',\mu''')}_{\alpha,\alpha',\alpha'',\alpha'''} = 
\frac{1}{2} \left( k_\mu + 2\pi\alpha \left/ \ell_o \right. \right) 
\delta_{\alpha + \alpha' , \alpha'' + \alpha'''} .
\end{equation}
The limiting forms of the mixing coefficients are comparable to textbook formulations of plane-wave parametric mixing, provided comparisons are made in the limit $k_\mu \ell_o\ll 1$, as mentioned earlier. 

As a last topic of theory discussion we note the five types of parametric scattering processes implied by Eqs. (\ref{pump}) through (\ref{idler3}). Figure \ref{fig3}(b) illustrates these processes, where a photon is labeled by forward-traveling-wave subscript $\mu,\mu',\mu'',\mu'''\in\left\{ P, S, 1, 2, 3 \right\}$, with state described by both a Bloch-wavenumber momentum ($k_\mu,k_{\mu'},k_{\mu''}$, or $k_{\mu'''}$) and a band index ($\alpha$, $\alpha'$, $\alpha''$, or $\alpha'''$). For instance, diagram (i) expresses both the DC-induced self-phase modulation (SPM), corresponding to $\alpha=\alpha'$, and the DC-enhanced inter-band cross-phase modulation (XPM), corresponding to $\alpha\ne\alpha'$, that can occur between superposed components of a single traveling wave $\mu$. This process arises as a consequence of hybridization of the initial waveform into a supermode as it propagates down the waveguide. The process is mediated by mixing coefficient $f^{(\mu)}_{\alpha,\alpha'}(A_{DC})$ of Eq. (\ref{fHybridization}), whose strength may be augmented by application of a DC bias. Equation (\ref{noLoadingsFhyp}) is the limiting form of $f^{(\mu)}_{\alpha,\alpha'}(A_{DC})$ in the absence of loadings.

In Fig. \ref{fig3}(b) the 4WM one-loop diagram of (ii) contributes to SPM of a given traveling wave $\mu$, as well as XPM between mixing traveling waves, $\mu\ne\mu'$. These processes are mediated by a 4WM coefficient $f^{(\mu,\bar{\mu}',\mu',\mu)}_{\alpha,\alpha',\alpha'',\alpha'''}$ , as defined in Eq. (\ref{f4WM}). Diagram (iii) describes the diffusive momentum transfer of 4WM, as in the example of two pump photons ($2k_p$), a signal photon ($k_S$), and an idler-3 photon ($k_3$), where the phase satisfies $\Delta\beta^{(1)}_{4wm}\cong 0$. In particular, the phase condition of momentum conservation does not preclude photons of the collision from being in different band states. The viability of scattering from different bands ultimately depends on the extent of symmetry breaking of the loading design, which dictates the strength of the 4WM coefficient of Eq. (\ref{f4WM}) that mediates the process. Inclusion of bands in the description introduces multiple channels of scattering, which leads to a much richer description of 4WM signal gain in the KIT than in fiber-optic textbook treatments of 4WM.\cite{Agrawal2001} Note that Eq. (\ref{noLoadingsF4wm}) represents the limiting form of all the 4WM coefficients in the absence of loadings. In particular, diagrams (ii) and (iii) reduce to the parametric scattering processes of the textbook if $k_\mu \ell_o\ll 1$.  

\begin{table}
\caption{\label{tab0} Summary of important mathematical symbols employed in the formulation of the theory.}
\begin{ruledtabular}
\begin{tabular}{cl}
Symbol & Description \\
\hline
$A_{DC}$ & ratio of DC to scaling factor $I_*$ \\
$f_P$, $f_S$, $f_1$, $f_2$, $f_3$ & frequencies of traveling waves \\
$\omega_P$, $\omega_S$, $\omega_1$, $\omega_2$, $\omega_3$ & angular frequencies of traveling waves \\
$k_P$, $k_S$, $k_1$, $k_2$, $k_3$ & Bloch wavenumbers of traveling waves \\
$n$, $n'$, $n''$, $n'''$ & Fourier indexes \\
$\alpha, \alpha', \alpha'', \alpha'''$ & dispersion-frequency band indexes \\
$\mu,\mu',\mu'',\mu'''$ & traveling-wave indexes \\
$\alpha_P, \alpha_S, \alpha_1, \alpha_2, \alpha_3$ & index of band matching traveling wave \\
$D_{n,n'}(k)$ & non-Hermitian dispersion matrix \\
$\Omega_\alpha(k)$ & band dispersion frequency \\
$u_{n}^{(\alpha)}(k)$, $e_{n}^{(\alpha)}(k)$ & left and right eigenvectors, respectively \\
$\Delta\beta^{(1)}_{3wm}$, $\Delta\beta^{(2)}_{3wm}$, $\Delta\beta^{(3)}_{3wm}$ & 3WM phase factors \\
$\Delta\beta^{(1)}_{4wm}$, $\Delta\beta^{(2)}_{4wm}$, $\Delta\beta^{(3)}_{4wm}$ & 4WM phase factors \\
$A^{(\mu)}_\alpha(x)$ & dimensionless amplitude of traveling-wave current \\
$\Lambda_{n,n'}(k)$, $\tilde{\Lambda}_{\alpha,\alpha'}(k)$ & lambda and canonically-transformed lambda, respectively \\
$f^{(\mu)}_{\alpha, \alpha'}(A_{DC})$, $f^{(\mu,\mu',\mu'')}_{\alpha, \alpha', \alpha''}$, $f^{(\mu,\mu',\mu'',\mu''')}_{\alpha, \alpha', \alpha'', \alpha'''}$ & hybridization, 3WM, and 4WM coefficients, respectively \\
\end{tabular}
\end{ruledtabular}
\end{table}

Similarly, with onset of DC, diagrams (iv) and (v) of Fig. \ref{fig3}(b) are associated with diffusive 3WM momentum transfer. For example, when $\Delta\beta^{(1)}_{3wm}\cong 0$, diagram (iv) describes a signal photon ($k_S$) recombining with an idler-1 photon ($k_1$), as mediated by the 3WM coefficient $f^{(\mu,\mu',\mu'')}_{\alpha,\alpha',\alpha''}$ of Eq. (\ref{f3WM}), such that a pump photon ($k_P$) is created. In the absence of loadings this 3WM coefficient is given by Eq. (\ref{noLoadingsF3wmA}). Conversely, as in diagram (v), a pump photon ($k_P$) can scatter into a signal photon ($k_S$) and an idler-1 photon ($k_1$), mediated by a 3WM coefficient $f^{(\mu,\bar{\mu}',\mu'')}_{\alpha,\alpha',\alpha''}$, which has the no-loading limit given by Eq. (\ref{noLoadingsF3wmB}). As in the case of 4WM scattering these 3WM processes are multichannel contributors to 3WM signal gain, dependent on the loading design.

Table \ref{tab0} summarizes mathematical symbols used in our description of the theory. Several important observations arise from our theory:
\begin{enumerate}
	\item[(i)] A metamaterial band theory of the KIT amplifier follows from engineered periodic loadings. The band theory is obtained via diagonalization of the non-Hermitian dispersion matrix of Eq. (\ref{dispersionMatrix}).
	\item[(ii)] The band theory, via Eq. (\ref{currentExpansion}), serves as a basis for representing the current of forward-traveling waves as Floquet-Bloch supermodes.
	\item[(iii)] Dimensionless amplitudes $A^{(\mu)}_\alpha(x)$ of parametrically mixing supermodes satisfy a set of coupled, nonlinear first-order differential equations, not unlike those derived in simple treatments of parametric mixing.
	\item[(iv)] The mixing coefficients $f^{(\mu)}_{\alpha, \alpha'}(A_{DC})$, $f^{(\mu,\mu',\mu'')}_{\alpha, \alpha', \alpha''}$, and $f^{(\mu,\mu',\mu'',\mu''')}_{\alpha, \alpha', \alpha'', \alpha'''}$ that enter these coupled equations are calculated solely from knowledge of the KIT band theory, i.e., the engineered loading design.
\end{enumerate}
We now turn to a discussion of the results of calculations of forward-traveling waves and the signal gain that ensues from their parametric mixing.

\section{Results and Discussion}
We present numerical computations using an example loading design of even symmetry consisting of $R=7$ regions, as in Fig. \ref{fig1} (b). Model parameters for this design are shown in Table \ref{tab1}, where $\mathcal{L}_r$ and $\mathcal{C}_r$ of regions $r=1$ through $r=7$ are defined as in Eq. (\ref{loadingDesign}), and the unit cell corresponds to length $\ell_o = 2565$ $\mu$m. This particular design inhibits every third harmonic of a pump tone placed at $\sim 8$ GHz, in proximity to the first stop gap. While the design was optimized for 4WM we will use it throughout our discussion, for both 4WM and 3WM calculations, to maintain the closest comparisons between results.

In our approach to numerical solution we introduced a cutoff dimension $N_c$ for the matrices of $D_{n,n'}(k)$ and $\Lambda_{n,n'}(k)$ of Eqs. (\ref{dispersionMatrix}) and (\ref{lambdaMatrix}), respectively, such that $-N_c/2 \le n,n' \le N_c/2$. Consistent with this approximation, all sums over $n$, $n'$, $n''$, and $n'''$ in our formulation of mixing coefficients were evaluated using this same cutoff. This allowed us to numerically diagonalize the matrix of $D_{n,n'}(k)$, invert the matrix of $\tilde{\Lambda}_{\alpha,\alpha'}(k)$ of Eq. (\ref{lambdaTilde}), and evaluate the mixing coefficients of Eqs. (\ref{fHybridization}) through (\ref{f4WM}), for any $k$ of the first Brillouin zone. Calculations presented for the loading design of Table \ref{tab1} used $N_c=251$, which gives convergence to $\sim 0.1\%$ error in the ten lowest-lying dispersion-frequency bands. To evaluate the coupled amplitude equations of Eqs. (\ref{pump}) through (\ref{idler3}) we employed a Dormand-Prince eighth-order (853) adaptive-corrective Runge-Kutta method.\cite{NumericalRecipes2007} We introduced a cutoff value of $N_b$ for the number of lowest-lying bands used to solve these coupled equations, such that sums over band indexes $\alpha$, $\alpha'$, $\alpha''$, and $\alpha'''$ were restricted to a count of $N_b$. We then analyzed the convergence of our amplitude solutions as we increased $N_b$. For the loading design of Table \ref{tab1} we found $N_b=6$ produced decibel signal gain with less than $10\%$ error in most cases.\footnote{The numerical program that solves the coupled differential equations and obtains the signal gain is available from the authors upon request.} In a few instances, as we varied the signal frequency to obtain signal gain, we found the Dormand-Prince algorithm would not converge; in these cases, breaks appear in plots of signal gain as a function of signal frequency.

\begin{table}
\caption{\label{tab1} Model parameters of an example loading design of $R=7$ regions, like that of Fig. \ref{fig1} (b), with  unit cell of length $\ell_o = 2565$ $\mu$m, and $\mathcal{L}_r$ and $\mathcal{C}_r$ defined as in Eq. (\ref{loadingDesign}).}
\begin{ruledtabular}
\begin{tabular}{rrrrr}
$r$ & $x_r$ ($\mu$m) & $\Delta x_r$ ($\mu$m) & $\mathcal{L}_r$ (pH/$\mu$m) & $\mathcal{C}_r$ (fF/$\mu$m) \\
\hline
1 &      $0$ & $17.5$ & $2.10$ & $0.335$ \\
2 &   $17.5$ &  $805$ & $1.05$ & $0.540$ \\
3 &  $822.5$ &   $65$ & $2.10$ & $0.335$ \\
4 &  $887.5$ &  $790$ & $1.05$ & $0.540$ \\
5 & $1677.5$ &   $65$ & $2.10$ & $0.335$ \\
6 & $1742.5$ &  $805$ & $1.05$ & $0.540$ \\
7 & $2547.5$ & $17.5$ & $2.10$ & $0.335$ \\
\end{tabular}
\end{ruledtabular}
\end{table}

\begin{figure}
\includegraphics[width=240pt, height=340pt]{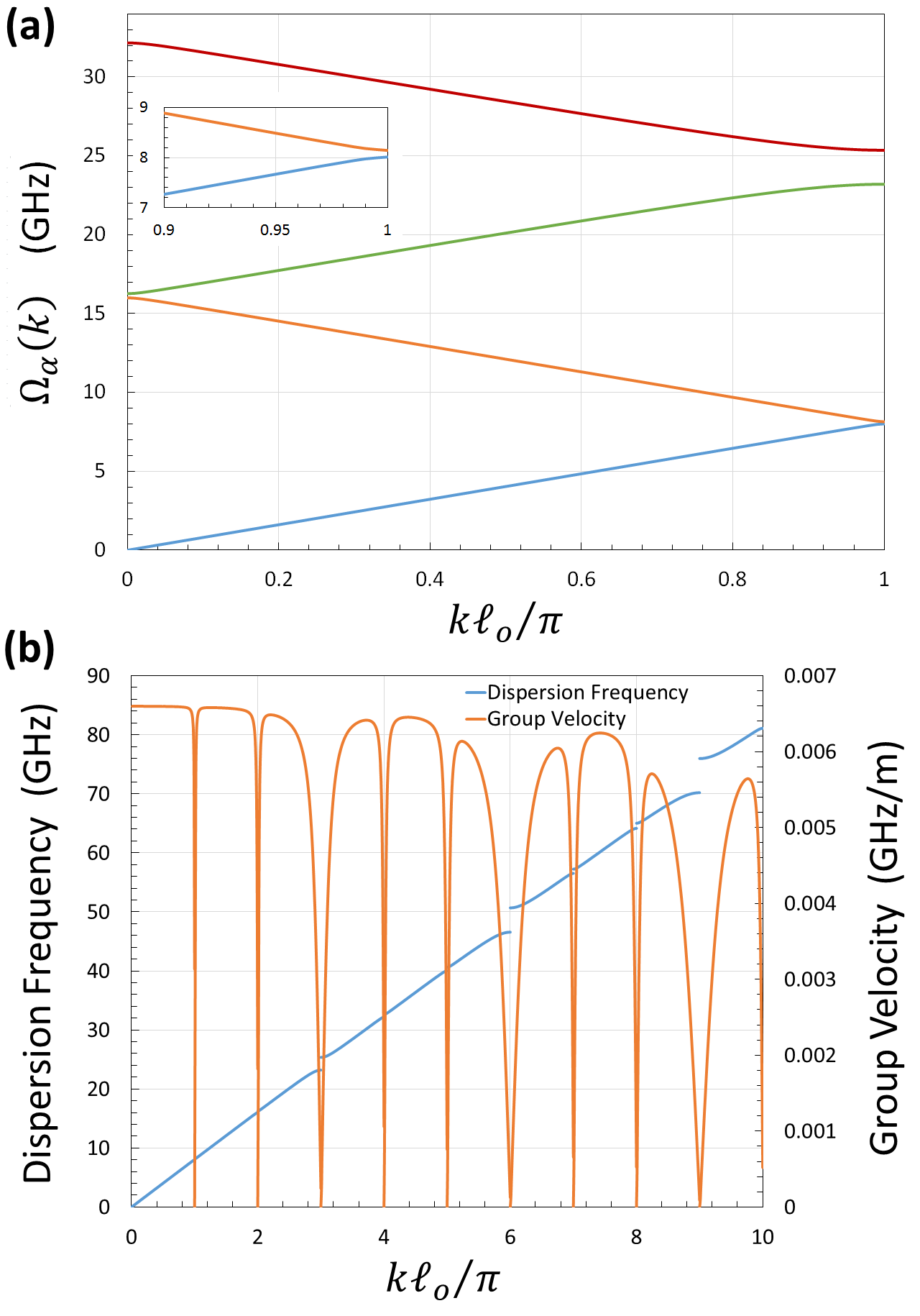}
\caption{\label{fig4} Metamaterial band structure of the example loading design of Table \ref{tab1}, as described in the text. (a) The lowest four bands of dispersion frequencies. The inset is a close-up view of the first stop gap of $130$ MHz size. (b) Group velocity as a function of wavenumber (orange), read from right vertical scale, with extended-zone dispersion frequencies overlayed (blue), read from left vertical scale.}
\end{figure}

Figure \ref{fig4}(a) is a reduced-zone-scheme plot of the four lowest-lying bands of dispersion frequencies $\Omega_0(k)<\Omega_{-1}(k)<\Omega_1(k)<\Omega_{-2}(k)$, denoted by the blue, orange, green, and red curves, respectively. The inset depicts the vicinity of the first stop gap at higher resolution. Table \ref{tab2} summarizes the lower bounds and sizes of the first 9 stop gaps of this band structure, those stop gaps that lie at the bottom of the dispersion-frequency manifold. 

\begin{table}
\caption{\label{tab2} List of lower bounds and sizes of the 9 lowest stop gaps, corresponding the the loading design of Table \ref{tab1}.}
\begin{ruledtabular}
\begin{tabular}{c|c|c}
Gap No. & Lower Bound (GHz) & Size (GHz) \\
\hline
1 & 8.013 & 0.1307 \\
2 & 16.00 & 0.2658 \\
3 & 23.20 & 2.153  \\
4 & 32.16 & 0.4735 \\
5 & 40.01 & 0.6274 \\
6 & 46.57 & 4.119  \\
7 & 56.53 & 0.7114 \\
8 & 64.15 & 0.8717 \\
9 & 70.19 & 5.780  \\
\end{tabular}
\end{ruledtabular}
\end{table}

Figure \ref{fig4}(b) shows the same dispersion frequencies plotted as a function of wavenumber in the extended-zone scheme (blue curve), with frequencies read from the left vertical scale. This plot is an unfolding in k-space of the reduced-zone-scheme representation of Fig \ref{fig4}(a), as in the sketch of Fig. \ref{fig3}(a). The locations of stop gaps in k-space are indicated by jump discontinuities in the dispersion-frequency curve. In Fig. \ref{fig4}(b) we also plot the group velocity as a function of wavenumber, with velocities read from the right vertical scale. The group velocities go to zero at every stop gap. For the lowest-lying bands the group velocity is nearly constant across the length of the band, although it diminishes on average from one band to the next as one traverses higher and higher bands. At higher bands the group velocity shows more variation across the length of a band as gap size invariably increases, particularly at every third stop gap. Every third stop gap shown in the figure was designed especially large to encompass 4WM harmonics of the pump frequency, as mentioned earlier, so at these locations we see a particularly strong variation of group velocity across the length of a given band.

Having solved for the metamaterial band structure of the KIT amplifier we now explore the nonlinear regime of its operation in the presence of these bands. This includes calculating the mixing coefficients $f^{(\mu)}_{\alpha,\alpha'}(A_{DC})$, $f^{(\mu,\mu',\mu'')}_{\alpha,\alpha',\alpha''}$, and $f^{(\mu,\mu',\mu'',\mu''')}_{\alpha,\alpha',\alpha'',\alpha'''}$ of Eqs. (\ref{fHybridization}) through (\ref{f4WM}) using this very band structure. We will first consider the effect of engineered loadings on a single pump tone as it propagates down the waveguide. We then turn to the results of calculations of signal gain, both with and without applied DC.

\subsection{Loading-Induced Dispersion of a Single Pump Tone}
We consider the effect of engineered frequency dispersion on a single forward-traveling pump tone injected into the CPW of Fig. \ref{fig1}. In this case Eq. (\ref{pump}) may be approximated as
\begin{equation}	\label{solePump}
-i\frac{\partial }{\partial x} A^{(P)}_\alpha(x) \cong
\sum\limits_{\alpha'} f^{(P)}_{\alpha, \alpha'}(A_{DC}) \, A^{(P)}_{\alpha'}(x) \\
+ \sum\limits_{\alpha',\alpha'',\alpha'''} 
f^{(P,\bar{P},P,P)}_{\alpha,\alpha',\alpha'',\alpha'''} \, {A^{(P)}_{\alpha'}(x)}^{*} A^{(P)}_{\alpha''}(x) \, A^{(P)}_{\alpha'''}(x) ,
\end{equation}
with initial boundary condition $A^{(P)}_\alpha(0)=\sqrt{P_o} \, \delta_{\alpha,\alpha_P}$, where $P_o$ is the relative input power of the RF tone. We also match frequency $\omega_P$ of the injected pump to a dispersion frequency $\Omega_{\alpha_P}(k_P)$ of Fig. \ref{fig4}, i.e., we set $\omega_P=\Omega_{\alpha_P}(k_P)$, where $k_P$ and $\alpha_P$ are the Bloch wavenumber and band index, respectively, of the matching dispersion frequency.

First note simple limiting forms of the solution of Eq. (\ref{solePump}). When nonlinearity is negligible then the solution is simply $A^{(P)}_\alpha(x)=\sqrt{P_o} \, \delta_{\alpha,\alpha_P}$. If we neglect off-diagonal components, i.e., we set $A^{(P)}_{\alpha\ne\alpha_P}(x)=0$, then we may approximate the solution of Eq. (\ref{solePump}) as
\begin{equation}	\label{pumpApproximation}
A^{(P)}_\alpha(x) \cong \sqrt{P}_o
\exp{ \left\{ i \left[ f^{(P)}_{\alpha_P, \alpha_P}(A_{DC}) 
+ f^{(P,\bar{P},P,P)}_{\alpha_P,\alpha_P,\alpha_P,\alpha_P} \, P_o 
\right] x \right\} } \; \delta_{\alpha,\alpha_P} .
\end{equation}
This approximation illustrates self modulation of the pump, where, in the absence of loadings, via Eqs. (\ref{noLoadingsFhyp}) and (\ref{noLoadingsF4wm}), we have
\begin{equation}	\label{pumpPhaseApproximation}
f^{(P)}_{\alpha_P, \alpha_P}(A_{DC}) 
+ f^{(P,\bar{P},P,P)}_{\alpha_P,\alpha_P,\alpha_P,\alpha_P} \, P_o =
\frac{1}{2} \left( k_P + 2\pi\alpha_P \left/ \ell_o \right. \right) 
\left( A_{DC}^2 + P_o \right) .
\end{equation}
The sum of $k_P + 2\pi\alpha_P \left/ \ell_o \right. $ is the extended-zone wavenumber, and the modulation is proportional to both the DC and RF input powers, resembling a plane-wave solution.

\begin{figure}
\includegraphics[width=440pt, height=400pt]{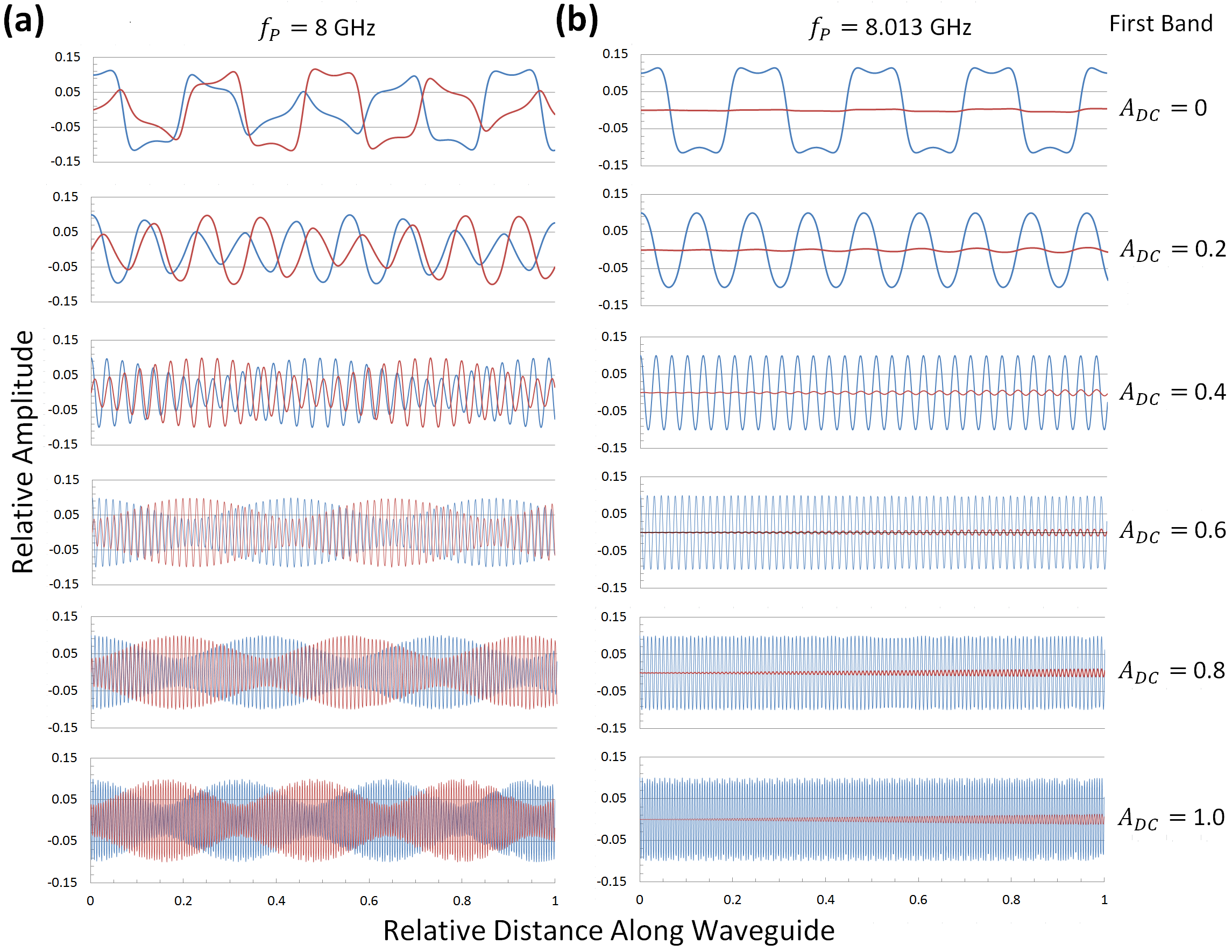}
\caption{\label{fig5} Real (blue) and imaginary (red) parts of relative amplitude $A^{(P)}_0(x)$ of first dispersion-frequency band as a function of relative position $x/\ell_T$ along KIT. Amplitude corresponds to pump frequencies (a) $f_P=8$ GHz and (b) $f_P=8.013$ GHz (bottom of first stop gap), for values of $A_{DC}=I_{DC}/I_*$ from zero to unity. Waveguide length is $\ell_T=2.0$ m and input power is set to $P_o=0.01$, i.e., $A^{(P)}_0(0)=0.1$.}
\end{figure}

Figure \ref{fig5} shows a more general numerical solution of Eq. (\ref{solePump}) using the Dormand-Prince eighth-order (853) adaptive-corrective Runge-Kutta method.\cite{NumericalRecipes2007} Calculations were performed with increasing DC bias, with pump frequency $f_P=\omega_P/2\pi$ set below the first stop gap, at (a) $f_P=8$ GHz ($k_P=1221$ $m^{-1}$, $\alpha_P=0$) and (b) $f_P=8.013$ GHz ($k_P=1225$ $m^{-1}$, $\alpha_P=0$), corresponding to the very bottom of the gap. The real (blue) and imaginary (red) parts of the relative amplitude of the lowest-lying band, $A^{(P)}_0(x)$, are shown as a function of $x/\ell_T$, where $\ell_T=2.0$ m is the length of the waveguide. The boundary condition was set to $A^{(P)}_0(x)=\sqrt{P_o}$, with relative input power at $P_o=0.01$, i.e., $A^{(P)}_0(0)=0.1$. 

Figure \ref{fig5}(a) depicts a jagged amplitude profile at $f_P=8$ GHz and zero DC bias. As DC is increased the profile shows a more sinusoidal-like modulation, with steep changes in amplitude as a function of $x/\ell_T$, and a pronounced envelope appearing by $A_{DC}=0.4$. The envelope exhibits a $90^\circ$ phase relationship between real and imaginary parts, as well as a period that decreases as $A_{DC}$ increases. In Fig. \ref{fig5}(b), where $f_P$ is increased by only $13$ MHz, to $f_P=8.013$ GHz, one sees distinct changes in the amplitude profile, owing to a sharp drop-off of group velocity near the gap. Recall from Fig. \ref{fig4}(b) that the group velocity rapidly decreases to zero near a stop gap. Thus, at $f_P=8.013$ GHz, where $f_P$ is at the bottom of the first stop gap, $A^{(P)}_0(x)$ as a function of $x$ corresponds to near-zero group velocity. Here, the imaginary part of the amplitude is also very nearly zero, everywhere in $x$. The real part of $A^{(P)}_0(x)$ is more square-like in shape at zero DC, but becomes more sinusoidal and of shorter period as $A_{DC}$ is increased. Though not shown, the results for $A^{(P)}_{-1}(x)$, corresponding to the second dispersion-frequency band, as a function of $x/\ell_T$, are similar to those of $A^{(P)}_0(x)$, except for an overall phase difference of $180^\circ$. At $P_o=0.01$, the maxima of higher-band amplitudes are no greater than $1\%$ of those of the first two bands, so the supermode traveling wave is well approximated by just the first two bands.

\begin{figure}
\includegraphics[width=440pt, height=400pt]{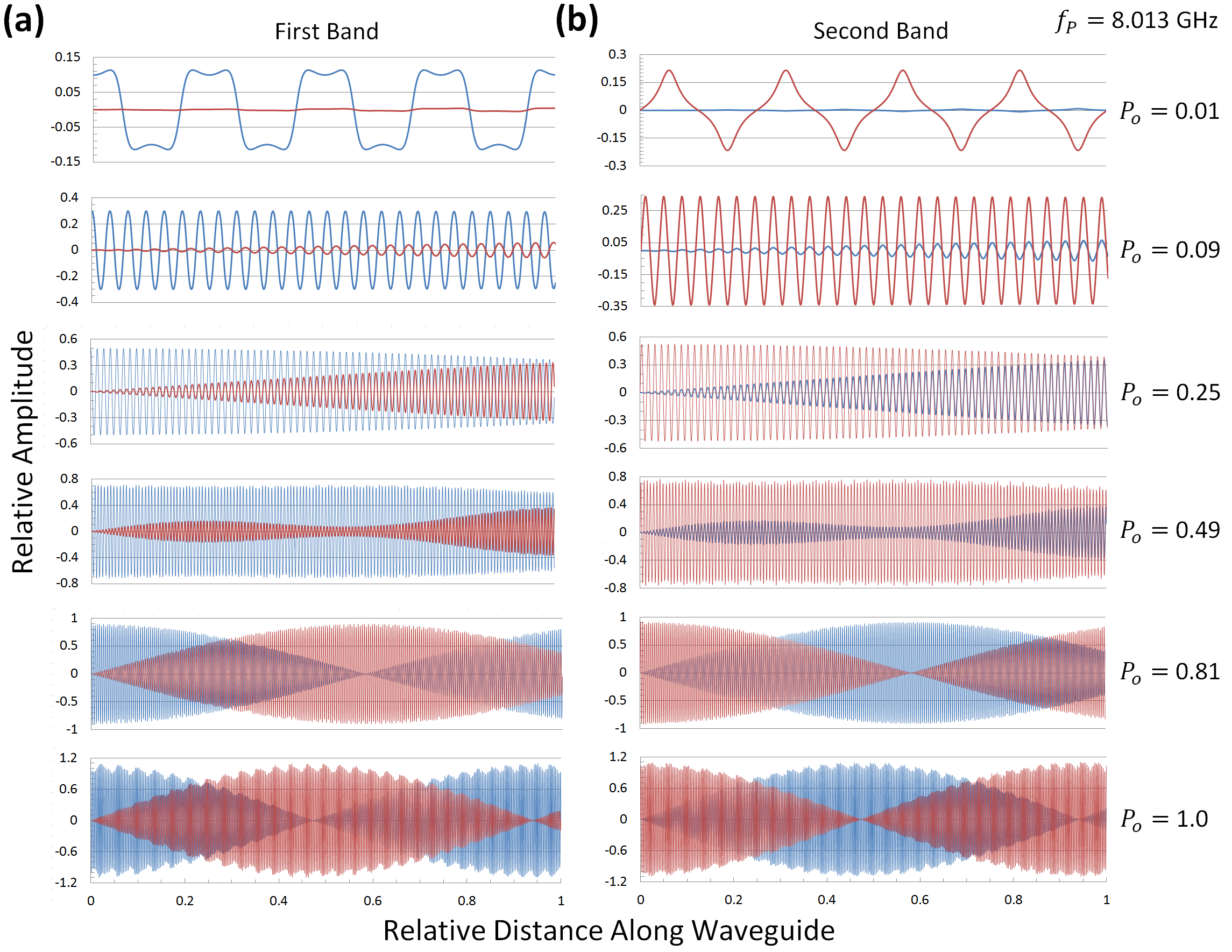}
\caption{\label{fig6} Real (blue) and imaginary (red) parts of relative amplitude correspond to (a) first band, $A^{(P)}_0(x)$, and (b) second band, $A^{(P)}_{-1}(x)$, as a function of relative distance $x/\ell_T$ along KIT, for incremented values of RF input power $P_o$. Waveguide length is $\ell_T=2.0$ m, pump frequency is $f_P=8.013$ GHz, at bottom of first stop gap, and no DC bias is applied.}
\end{figure}

Figure \ref{fig6}(a) shows plots of the real (blue) and imaginary (red) parts of amplitude $A^{(P)}_0(x)$ of the first band as a function of $x/\ell_T$ for incremented values of RF input power $P_o$. The DC bias is set to zero and the pump frequency is $f_P=8.013$ GHz, at the bottom of the first gap, and again the length of the waveguide is $\ell_T=2.0$ m. From Fig. \ref{fig6}(a) we see that the amplitude envelope, prevalent in the $f_P=8.0$ GHz results of Fig. \ref{fig5}(a), now also appears at $f_P=8.013$ GHz with increasing input power $P_o$, with envelope period decreasing as $P_o$ increases. As in Fig. \ref{fig5}(a), there is a $90^\circ$ phase difference between the real and imaginary parts of the amplitude. For comparison, Fig. \ref{fig6}(b) depicts a similar set of plots for the amplitude $A^{(P)}_{-1}(x)$ of the second band. The results of Fig. \ref{fig6}(b) show a $180^\circ$ phase difference from those of Fig. \ref{fig6}(a), such that the real and imaginary parts of the two amplitudes are more or less interchanged. However, at low power, $P_o=0.01$, the amplitude $A^{(P)}_{-1}(x)$ of the second band has a triangle-shaped profile, in contrast to the square-like profile of $A^{(P)}_0(x)$ of the first band, and its maximum amplitude is slightly greater in magnitude. Recall the jagged profile from the zero-bias result of Fig. \ref{fig5}(a), which appears to resemble a combination of square and triangle shapes as a function of $x$. While the maximum amplitudes of $A^{(P)}_0(x)$ and $A^{(P)}_{-1}(x)$ tend to differ somewhat at low input power, as $P_o$ is increased they both converge to $\sqrt{P_o}$.

In our calculations we varied the number of lowest-lying bands $N_b$ used to solve the coupled amplitude equations of Eq. (\ref{solePump}), considering values of $N_b=2$ to $N_b=8$. As mentioned, in our results for $P_o=0.01$, the maximum amplitude of the two lowest-lying bands was $\max{|A^{(P)}_0(x)|}\cong \max{|A^{(P)}_{-1}(x)|}\cong\sqrt{P_o}$, with the maxima of amplitudes of higher bands no greater than $1\%$ of $\sqrt{P_o}$. Thus, at low input powers, the first two bands were sufficient to describe the supermode traveling wave. However, as $P_o$ was increased toward unity, we found amplitudes of the higher-lying bands became more significant, indicating the need to incorporate more and more bands into the supermode solution. Generally speaking, for the loading design of Table \ref{tab1}, band-matching of injected frequencies to one of the first two bands, and input powers less than unity, we found $N_b=6$ to be sufficient to describe the supermode traveling waves.

We have shown results of calculations of amplitude for a single nonlinear forward-traveling wave as it propagates along the waveguide of the KIT in the presence of the engineered loadings of Table \ref{tab1}. This was accomplished as a function of both applied DC ($A_{DC}$) and input RF power ($P_o$). We next discuss the parametric signal gain of the amplifier when no DC is applied.

\subsection{Four-Wave Mixing Signal Gain of the KIT Amplifier}
For the loading design of Table \ref{tab1}, and a waveguide of length $\ell_T=2$ m, we calculated the 4WM signal gain produced in the KIT amplifier with zero DC. In this case a strong pump, of frequency $f_P=\omega_P/2\pi=8.013$ GHz and initial amplitude $A^{(P)}_\alpha(0)=\bar{A}^{(P)}\, \delta_{\alpha,\alpha_P}$, and a small signal of variable frequency $f_S=\omega_S/2\pi$ and initial amplitude $A^{(S)}_\alpha(0)=\bar{A}^{(S)}\, \delta_{\alpha,\alpha_S}$, were injected into the waveguide. The resulting 4WM idler product, of frequency $f_3=2f_P-f_S$, was defined with initial boundary condition $A^{(3)}_\alpha(0)=0$. We specifically solved the zero-DC forms of Eqs. (\ref{pump}), (\ref{signal}), and (\ref{idler3}) for the relative current amplitudes $A^{(P)}_\alpha(x)$, $A^{(S)}_\alpha(x)$, and $A^{(3)}_\alpha(x)$ using the same Dormand-Prince eighth-order (853) adaptive-corrective Runge-Kutta method\cite{NumericalRecipes2007} we employed for the single traveling wave. In our calculations we set $\bar{A}^{(P)}=\sqrt{P_o}$ and $\bar{A}^{(S)}=10^{-3}$, where $P_o$ serves as a relative measure of the RF input power. 

\begin{figure}
\includegraphics[width=240pt, height=300pt]{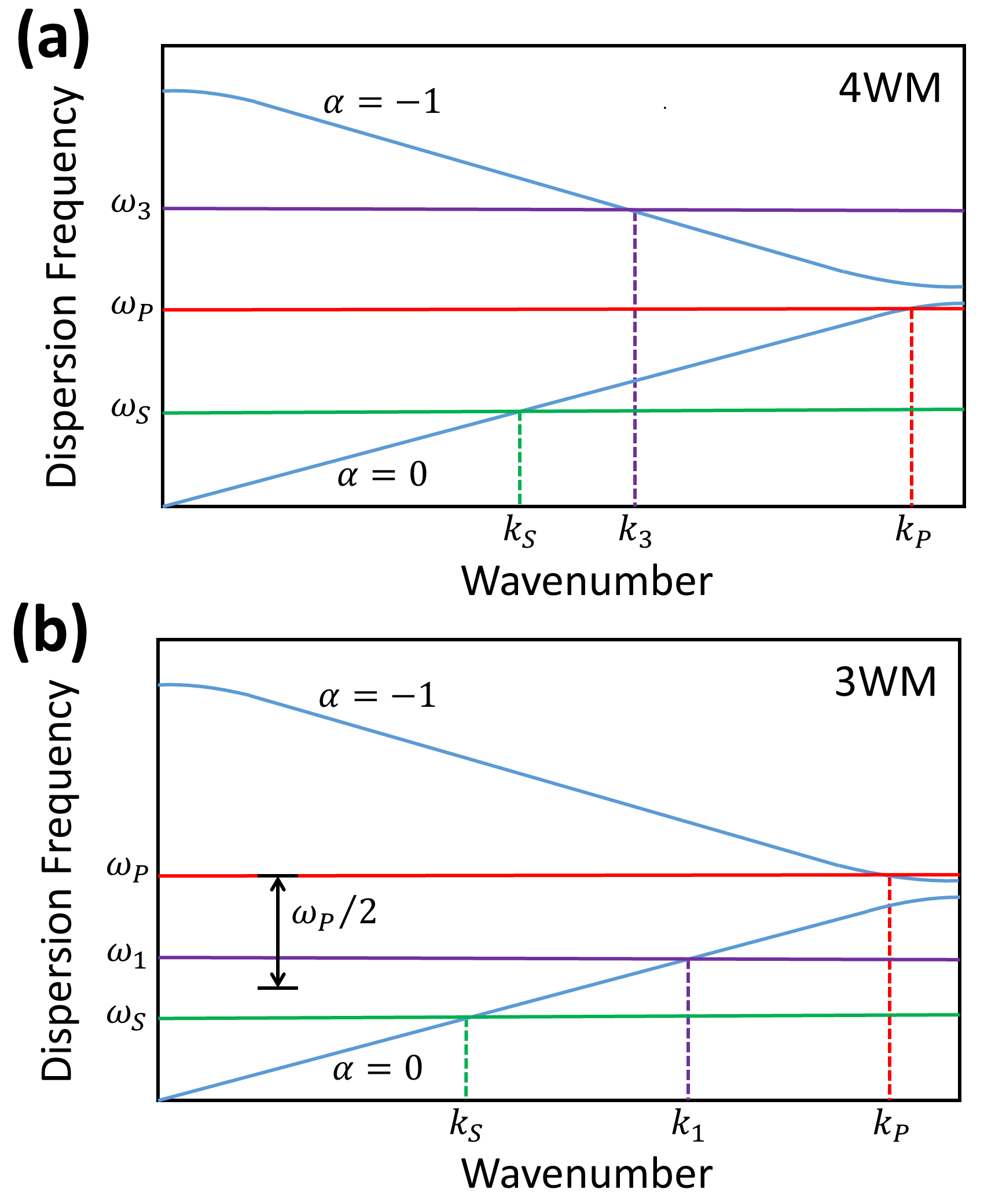}
\caption{\label{fig7} Sketch showing traveling-wave pump frequency $\omega_P$ (red) and signal frequency $\omega_S$ (green) matched to the KIT amplifier band structure upon injection, assuming $\omega_S<\omega_P$. In the case of (a) the degenerate-4WM idler frequency $\omega_3$ (purple) is shown matched to the second band ($\alpha=-1$). Similarly, in (b) the idler of frequency $\omega_1$ (purple), responsible for 3WM broadband gain, is shown matched to the first band ($\alpha=0$).}
\end{figure}

The pump frequency was fixed to the bottom of the first stop gap, with $\alpha_P=0$ and $k_P=1225$ $m^{-1}$. The frequency $\omega_S$ of the injected signal was varied and the gain was calculated at each value of $\omega_S$. As $\omega_S$ was incremented it was matched to either the first band ($\alpha_S=0$) or the second band ($\alpha_S=-1$), depending on whether $\omega_S$ was below or above the stop gap. If $\omega_S$ fell within the gap then calculation of signal gain was skipped, with gain reported as zero, since $k_S$ is indeterminate in this case. The frequency of the 4WM idler, $\omega_3=2\omega_P-\omega_S$, was matched to the band structure in a manner similar to the pump and signal, with $k_3$ obtained from the matching band of index $\alpha_3$. Figure \ref{fig7}(a) is a sketch of a typical scenario of band matching of the pump ($\omega_P$), signal ($\omega_S$), and 4WM idler ($\omega_3$) frequencies. In the figure the pump frequency is just below the gap, and the signal and idler straddle the gap, equidistant from the pump ($\omega_S+\omega_3=2\omega_P$), matched to opposing bands. 

\begin{figure}
\includegraphics[width=440pt, height=300pt]{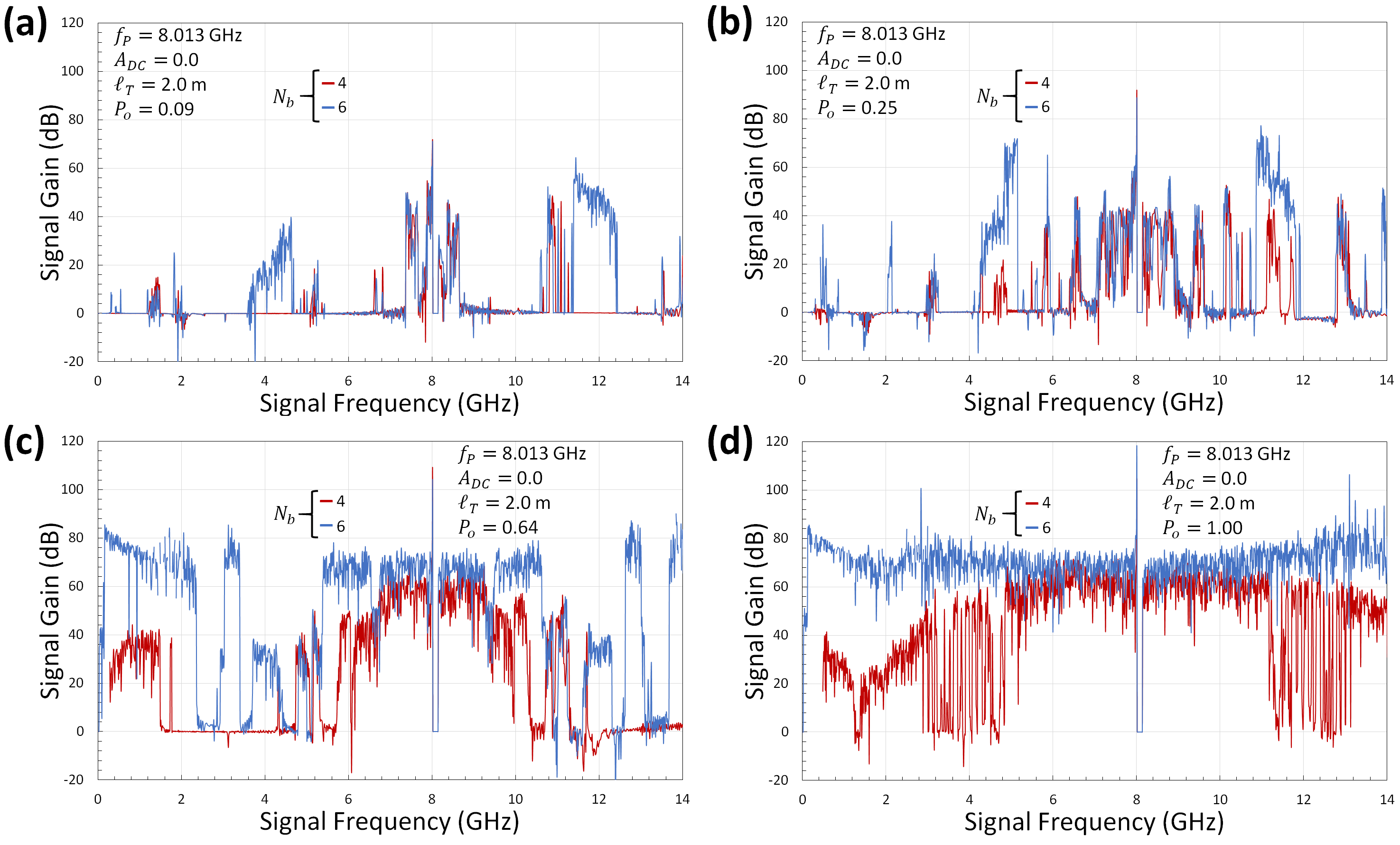}
\caption{\label{fig8} Calculations of 4WM signal gain as a function of signal frequency for engineered loadings of Table \ref{tab1}, with pump frequency $f_P=8.013$ GHz set just below first stop gap and waveguide length set to $\ell_T=2$ m. Relative RF input power $P_o$ was increased from panel to panel to illustrate approach to full amplification, with (a) $P_o=0.09$, (b) $P_o=0.25$, (c) $P_o=0.64$, and (d) $P_o=1.00$. Calculations were performed using $N_b=4$ (red) and $N_b=6$ (blue) lowest-lying dispersion-frequency bands, as indicated.}
\end{figure}

Using Eqs. (\ref{travelingWaves}) and (\ref{currentExpansion}) to form the root-mean-square (RMS) current, the signal gain may be expressed as
\begin{equation}	\label{signalGain}
G_S(\ell_T) = {\left| \frac{ \sum\limits_{\alpha} \sum\limits_{n=-\infty}^{\infty} A^{(S)}_\alpha(\ell_T) \, e_n^{(\alpha)}(k_S) }
{ A^{(S)}_{\alpha_S}(0) \sum\limits_{n=-\infty}^{\infty} e_n^{(\alpha_S)}(k_S) } \right| }^2 .
\end{equation}
Figure \ref{fig8} shows calculations of $G_S(\ell_T)$ as a function of $f_S=\omega_S/2\pi$, for several values of relative input power $P_o$. Results were reported for both $N_b=4$ (red curve) and $N_b=6$ (blue curve) to illustrate convergence characteristics. 

In either case of $N_b$, Fig. \ref{fig8} shows how the amplifier becomes fully lit as $P_o$ is increased to unity. At lower powers, panels (a) and (b), features prevalent for both $N_b=4$ and $N_b=6$, particularly near the pump frequency ($8.013$ GHz), show little difference in magnitude, indicating that, at these signal frequencies, the traveling-wave supermodes are well represented by the lowest-lying bands. In contrast, at higher powers, panels (c) and (d), these features show some increase in magnitude from $N_b=4$ to $N_b=6$, indicative of the importance of higher-lying bands at these higher powers. For all curves of $N_b=6$ we see additional strong features in the signal-gain profile, which tend to occur at signal frequencies of $\sim 4$ GHz and $\sim 12$ GHz. 
These are sideband contributions corresponding to participation of higher-lying, supermode band states in the parametric scattering, which are additional channels of amplification. These are particularly evident in the sidebands of panel (d) where we see very large undulations in the profile of $N_b=4$, but not for $N_b=6$. For $N_b=4$, amplification is essentially either on or off in the sidebands as $f_S$ increases; for $N_b=6$, there are more channels that can contribute, so there is never a value of $f_S$ where the amplification is zero. As $P_o$ increases, Fig. \ref{fig8} shows that these sidebands coalesce with the center band to form one long broadband signal-gain profile, as in panel (d), $N_b=6$.

What is most striking about the 4WM signal gain depicted in Fig. \ref{fig8} is the intrinsic undulation of the gain with increasing $f_S$. This occurs no matter the number of low-lying bands incorporated into the calculation. Its origin is in the engineered dispersion and the fact that in 4WM the signal and idler are matched to different dispersion-frequency bands. As we saw in Fig. \ref{fig4}(b), the group velocity differs between bands, even in the centers of two lowest-lying bands, where the group velocity is nearly constant. Thus, as $f_S$ is increased, the momentum-conserving criterion $k_S+k_3\cong 2k_P$ cannot be maintained robustly because $k_S$ and $k_3$ change at different rates. This creates the effect of undulation in the signal gain as $f_S$ is varied. The effect is compounded by inclusion of additional parametric scattering channels of supermode band states since each has its own phase-matching criterion.

Experimental results reported in Fig. 3(a) of Ref. (\onlinecite{Vissers2016}), for an unbiased NbTiN KIT amplifier of loading design similar to Table \ref{tab1}, show a 4WM signal gain of $\sim 20$ dB, which is much smaller than that calculated in Fig. \ref{fig8}. The differences can be attributed to (i) the lack of loss, i.e., intrinsic $Q$, modeling in Eqs. (\ref{telegrapherEquation1}) and (\ref{telegrapherEquation2}), and (ii) the sensitivity of the signal gain to the size of the first stop gap. In the later case, the theoretical results correspond to a stop gap of $130$ MHz whereas the stop gap of the results presented in Fig. 3(a) of Ref. (\onlinecite{Vissers2016}) is about 3 times larger. We are presently engaged in follow-on research to understand these dependencies more clearly. In particular, for ripple observed in the 4WM signal gain of the KIT, there is a need to understand the extent of reflections at the impedance transformers as these operate at high input powers. What is clear from the present theoretical analysis is that operation of the KIT amplifier without DC to produce 4WM signal gain is prone to intrinsic spectral gain fluctuations that reduce its efficacy as an amplification device. As we shall see next, when a DC bias is applied to the KIT, multiple channels of parametric scattering once again come into play but the resulting signal amplification differs greatly as a function of signal frequency from what we have presented in Fig. \ref{fig8}.

\subsection{Multiwave Mixing of the KIT Amplifier with Application of Direct Current}
Recall that if a DC bias is applied to the KIT amplifier then, in addition to the degenerate 4WM process of the zero-DC case, mixing of injected pump and signal results in five new parametric processes, as sketched in Fig. \ref{fig2}(a). In this case momentum conservation of the 3WM process involving pump ($f_P=\omega_P/2\pi$), signal ($f_S=\omega_S/2\pi$), and idler ($f_1=\omega_1/2\pi$) becomes possible over an extended range of signal frequencies centered about $f_P/2$, where one has $\omega_S+\omega_1=\omega_P$ and $k_S+k_1\cong k_P$. In what follows we consider the signal gain attributable to this 3WM process.

\subsubsection{Runge-Kutta Results with Application of a DC Bias}
As we did for the zero-DC 4WM signal gain, we calculated the 3WM signal gain for the loading design of Table \ref{tab1}, setting the DC to one tenth of the nonlinear scaling factor $I_*$, i.e., $A_{DC}=I_{DC}/I_*=0.1$. Also, we typically set the length of the waveguide to $\ell_T=2$ m and the pump frequency to $f_P=8.144$ GHz, at the top of the first stop gap. However, we also explored other values for $\ell_T$ and $f_P$, particularly contrasting $f_P$ above the gap with $f_P$ below the gap. With DC applied we solved the full set of coupled amplitude equations, Eqs. (\ref{pump}) through (\ref{idler3}), using the same Runge Kutta method as before, with calculations involving $N_b=4$ and $N_b=6$ of the lowest-lying bands. As in the case of zero applied DC, the initial boundary conditions assigned to pump and signal amplitudes were $A^{(P)}_\alpha(0)=\bar{A}^{(P)}\, \delta_{\alpha,\alpha_P}$ and $A^{(S)}_\alpha(0)=\bar{A}^{(S)}\, \delta_{\alpha,\alpha_S}$, with $\bar{A}^{(P)}=\sqrt{P_o}$ and $\bar{A}^{(S)}=10^{-3}$. The idler products, of frequency $f_1=f_P-f_S$, $f_2=f_P+f_S$, and $f_3=2f_P-f_S$ were defined with initial amplitudes of zero. Again, $P_o$ serves as a relative measure of the RF input power, which we varied as we calculated signal gain as a function of $f_S$ using Eq. (\ref{signalGain}). Figure \ref{fig7}(b) is a sketch exemplifying the matching of pump, signal, and 3WM idler frequencies to the band structure.

\begin{figure}
\includegraphics[width=440pt, height=300pt]{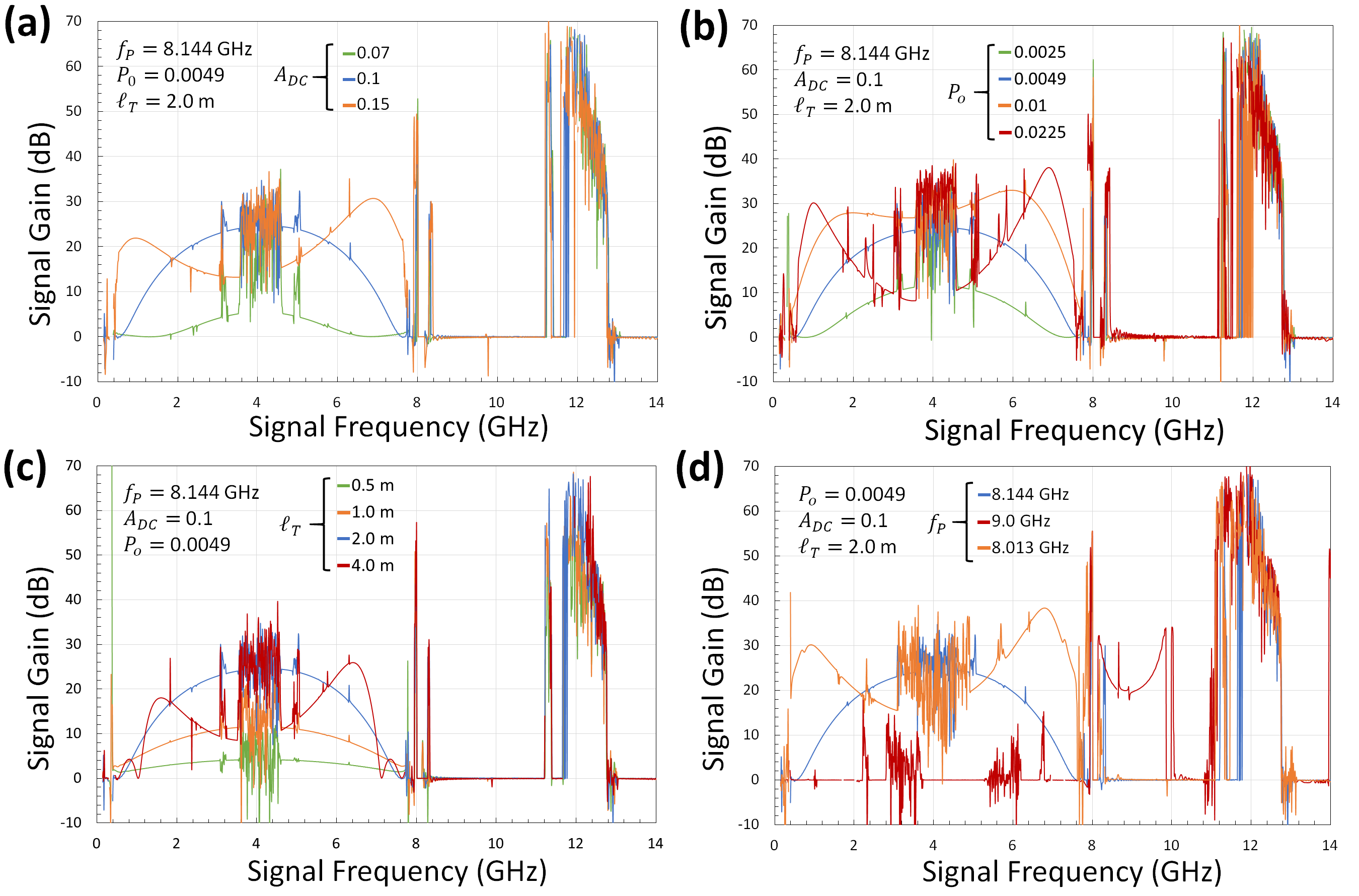}
\caption{\label{fig9} Calculations of 3WM signal gain for the loading design of Table \ref{tab1}, as described in the text, as a function of signal frequency $f_S$, for several values of (a) relative DC bias $A_{DC}$, (b) relative input power $P_o$, (c) waveguide length $l_T$, and (d) pump frequency $f_P$, with other parameters as labeled. In all panels optimal signal gain is the blue curve corresponding to $A_{DC}=0.1$, $P_o=0.0049$, $l_T=2.0$ m, and $f_P=8.144$ GHz.}
\end{figure}

In Fig. \ref{fig9} we present calculations of 3WM signal gain $G_S(\ell_T)$, again computed via Eq. \ref{signalGain}, as a function of $f_S$ for different values of pump frequency $f_P$, relative RF input power $P_o$, waveguilde length $\ell_T$, and relative DC bias $A_{DC}$. In all of the panels the blue curve is the optimal signal gain corresponding to pump frequency just above the gap ($f_P=8.144$ GHz), with $P_o=0.0049$, $\ell_T=2$ m, and $A_{DC}=0.01$. All curves show the behavior of multiwave mixing as a function of $f_S$: a smooth broadband 3WM contribution centered about $f_P/2$ and a 4WM contribution concentrated at both $\sim 4$ GHz and $\sim 12$ GHz, equidistant from $f_P$. The 4WM features show the characteristic undulation with $f_S$ that we saw in Fig. \ref{fig8}. In contrast, the 3WM contributions show smooth progression with $f_S$ because the frequencies of signal and 3WM idler match to the same lowest-lying dispersion-frequency band, as in the sketch of Fig \ref{fig7}(b). Moreover, group velocity of this first band, as shown in Fig \ref{fig4}(b), is nearly constant throughout the width of the band.

In Fig. \ref{fig9}(a) we varied $A_{DC}$ to show how the 3WM signal gain collapses beyond $A_{DC}\cong 0.1$. As $A_{DC}$ increases, the propensity for higher amplitude response increases as the traveling wave propagates down the waveguide. Initially, this translates into higher 3WM signal gain, resulting in an optimal gain at $A_{DC}\cong 0.1$. However, recalling our introductory remarks that 4WM will begin to dominate 3WM beyond amplitude threshold $I_{RF}(x,t)\sim 2 I_{DC}$, as $A_{DC}$ is increased past $A_{DC}\cong 0.1$ the amplitude response will eventually exceed this threshold. The collapse begins in the central region of the 3WM band because this where 4WM is strongest--recall from Fig. \ref{fig2}(c) that the 4WM process of $\omega_S+\omega_3=\omega_1+\omega_2$ is momentum conserving when the signal frequency is near half the pump frequency. The broadband 3WM signal gain is therefore confined to a finite range of applied DC. 

In Fig. \ref{fig9}(b) we instead varied $P_o$ about the optimal gain scenario. Similar to panel (a), as $P_o$ is increased from zero we see a corresponding increase in broadband 3WM signal gain, but eventually the gain profile collapses from the center of the band when the RF input power exceeds $P_o\cong 0.0049$. As in (a), increasing $P_o$ increases the amplitude response, leading to an initial increase in gain. However, beyond $P_o\cong 0.0049$ the amplitude response again exceeds the threshold of 4WM dominance. Thus, broadband 3WM signal gain exists within a range of RF input powers.

In Fig. \ref{fig9}(c) we next varied the length of the waveguide $\ell_T$. As $\ell_T$ is increased the run length of the mixing traveling waves correspondingly increases leading to increased amplitude response on exit from the waveguide. As in (a) and (b), there is an initial signal gain, but again the amplitude threshold is exceeded beyond $\ell_T\cong 2$ m and the broadband profile collapses. In contrast to broadband 4WM signal gain, which tends to grow exponentially with $\ell_T$, the 3WM signal gain has a limiting waveguide length.

In Fig. \ref{fig9}(d) we varied the pump frequency $f_P$, where the optimal 3WM signal gain (blue curve) occurs when $f_P$ is placed just above the first stop gap, at $f_P=8.144$ GHz. If pump frequency is moved to a position below the gap then the condition of ideal 3WM phase matching diminishes, with gain profile collapsing from the center of the band. If pump frequency is moved above $f_P\cong 8.144$ GHz, to $f_P=9.0$ GHz, then the 3WM signal gain all but disappears. However, at $f_P=9.0$ GHz (red curve), we see the emergence of a broadband 4WM gain profile about this value of pump frequency, albeit depressed within its center. Note the smoothness of 4WM signal gain in this particular case; this occurs because the pump, signal, and 4WM idler frequencies are all now matched to the same second dispersion-frequency band.

\subsubsection{Semi-Analytical Approximation of 3WM Signal Gain}
To facilitate understanding of the underlying physics of broadband 3WM signal gain in the presence of engineered dispersion we derived a two-band approximation in Appendix \ref{appendixApproximation3wmGain}. The approximation applies to an undepleted pump, with $f_P$ near the first stop gap, and with $P_o$ sufficiently small for broadband 3WM to occur. It tends to underestimate the magnitude and breadth of gain, particularly if $f_P$ is positioned away from the gap, but it provides insight into the underlying mechanism of the parametric gain.

In the approximation we assume amplitudes of signal and idler traveling waves can be represented via the lowest dispersion-frequency band, i.e., $\alpha=0$, and we allow $f_P$ to be positioned to either side of the first stop gap, treating the pump traveling wave as a superposition involving the first two bands, i.e., $\alpha=0,-1$. From Eqs. (\ref{approxPump1}) and (\ref{approxPump2}), the amplitudes of the two components of the pump traveling wave are then of the form
\begin{equation}
A^{(P)}_0(x) \cong A^{(P)}_{0,+}(P_o,A_{DC}) \, e^{i\phi^{(P)}_+(A_{DC}) x} + A^{(P)}_{0,-}(P_o,A_{DC}) \,e^{i\phi^{(P)}_-(A_{DC}) x} ,
\end{equation}
\begin{equation}
A^{(P)}_{-1}(x) \cong A^{(P)}_{-1,+}(P_o,A_{DC}) \, e^{i\phi^{(P)}_+(A_{DC}) x} + A^{(P)}_{-1,-}(P_o,A_{DC}) \,e^{i\phi^{(P)}_-(A_{DC}) x} ,
\end{equation}
with canonical phase coefficients given by Eq. (\ref{coefficients}) as
\begin{multline}	\label{pumpPhaseRelation}
\phi^{(P)}_\pm(A_{DC}) = 
\frac{1}{2} \, \left[ f^{(P)}_{0,0}(A_{DC}) + f^{(P)}_{-1,-1}(A_{DC}) \right] \\
\pm \frac{1}{2} \sqrt{ { \left[ f^{(P)}_{0,0}(A_{DC}) - f^{(P)}_{-1,-1}(A_{DC}) \right] }^2 + 4 f^{(P)}_{0,-1}(A_{DC}) \, f^{(P)}_{-1,0}(A_{DC}) } .
\end{multline}
The canonical phase coefficients embody both self-phase modulation (SPM) of pump components, via $f^{(P)}_{0,0}(A_{DC})$ and $f^{(P)}_{-1,-1}(A_{DC})$, and cross-phase modulation (XPM) between components, via $f^{(P)}_{0,-1}(A_{DC})$ and $f^{(P)}_{-1,0}(A_{DC})$. The mixing coefficients $f^{(P)}_{0,-1}(A_{DC})$ and $f^{(P)}_{-1,0}(A_{DC})$ convey the strength of hybridization of the supermode traveling wave between the two bands as it propagates down the waveguide.

Our approximation therefore simplifies the supermode representation of the pump traveling wave to just the two most significant components, adequate for $P_o\ll 1$. Because the pump is comprised of these two component waveforms, there are then two criteria for phase matching of pump with signal and idler. From Eq. (\ref{canonicalPhase}) the relevant phase relationships may be defined as
\begin{equation}	\label{phaseRelations}
\varphi_\pm(A_{DC})=\phi^{(P)}_\pm(A_{DC})-f^{(S)}_{0,0}(A_{DC})-f^{(1)}_{0,0}(A_{DC})-\Delta\beta^{(1)}_{3wm} ,
\end{equation}
each of which depends on (i) momentum conservation, in terms of Bloch wavenumbers, (ii) SPM of both signal and idler, matched to the same lowest-lying band, and (iii) a canonical pump phase coefficient, $\phi^{(P)}_-(A_{DC})$ or $\phi^{(P)}_+(A_{DC})$, which contains both SPM and inter-band XPM of the pump components. 

As either of $\varphi_\pm(A_{DC})$ tends to zero a corresponding channel of signal gain may arise. From Eq. (\ref{appendixGainCoefficients}), the gain coefficient of each of the two channels may be expressed as
\begin{equation}	\label{twoChannelGainCoefficients}
g_\pm(P_o,A_{DC}) = 
\sqrt{ \tilde{f}^{(S,\bar{1},P)}_\pm(P_o,A_{DC}) \, \tilde{f}^{(1,\bar{S},P)}_\pm(P_o,A_{DC}) - \frac{1}{4} {\varphi_\pm(A_{DC})}^2 } ,
\end{equation}
where, via Eq. (\ref{canonicalMixingCoefficients}), we defined canonical nonlinear mixing coefficients
\begin{equation}
\tilde{f}^{(\mu,\bar{\nu},P)}_{m}(P_o,A_{DC}) = 2A_{DC} \sum\limits_{\alpha} {f^{(\mu,\bar{\nu},P)}_{0,0,\alpha}} \, A^{(P)}_{\alpha,m}(P_o,A_{DC}) ,
\end{equation}
which are real-valued for the loading design of Table \ref{tab1}. The resulting two-channel signal gain may be approximated from Eq. (\ref{appendixSignalGain}) as
\begin{multline}	\label{twoChannelSignalGain}
G_S(P_o,A_{DC},\ell_T)\cong \\
\Bigg| {\mathcal{A}_+(P_o,A_{DC})}^* \Big[ \cosh{g_+(P_o,A_{DC})\, \ell_T} 
- \frac{i\varphi_+(A_{DC})}{2g_+(P_o,A_{DC})} \, \sinh{g_+(P_o,A_{DC})\, \ell_T} \Big] e^{i \varphi_+(A_{DC})\, \ell_T / 2} \\
+ {\mathcal{A}_-(P_o,A_{DC})}^* \Big[ \cosh{g_-(P_o,A_{DC})\, \ell_T} 
- \frac{i\varphi_-(A_{DC})}{2g_-(P_o,A_{DC})} \, \sinh{g_-(P_o,A_{DC})\, \ell_T} \Big] e^{i \varphi_-(A_{DC})\, \ell_T / 2} \Bigg| ^2 ,
\end{multline}
where, via Eq. (\ref{psuedoAmplitudes}), we have defined pseudo dimensionless amplitudes
\begin{equation}
\mathcal{A}_\pm(P_o,A_{DC}) = \frac{ \sum\limits_{m=\pm} \tilde{f}^{(1,\bar{S},P)}_{m}(P_o,A_{DC}) - i\tilde{f}^{(1,\bar{S},P)}_{ \mp }(P_o,A_{DC}) }
{ \tilde{f}^{(1,\bar{S},P)}_\pm(P_o,A_{DC}) - \tilde{f}^{(1,\bar{S},P)}_\mp(P_o,A_{DC}) } .
\end{equation}
Note that if the two channels of Eq. (\ref{twoChannelSignalGain}) were ideally separable we could express them as
\begin{multline}	\label{twoChannelSignalGainComponents}
G^{(\pm)}_S(P_o,A_{DC},\ell_T) = \\
{ \left| \mathcal{A}_\pm(P_o,A_{DC}) \right| }^2
{ \left| \cosh{g_\pm(P_o,A_{DC})\, \ell_T} - \frac{i\varphi_\pm(A_{DC})}{2g_\pm(P_o,A_{DC})} \, \sinh{g_\pm(P_o,A_{DC})\, \ell_T} \right| }^2 .
\end{multline}

\begin{figure}
\includegraphics[width=240pt, height=280pt]{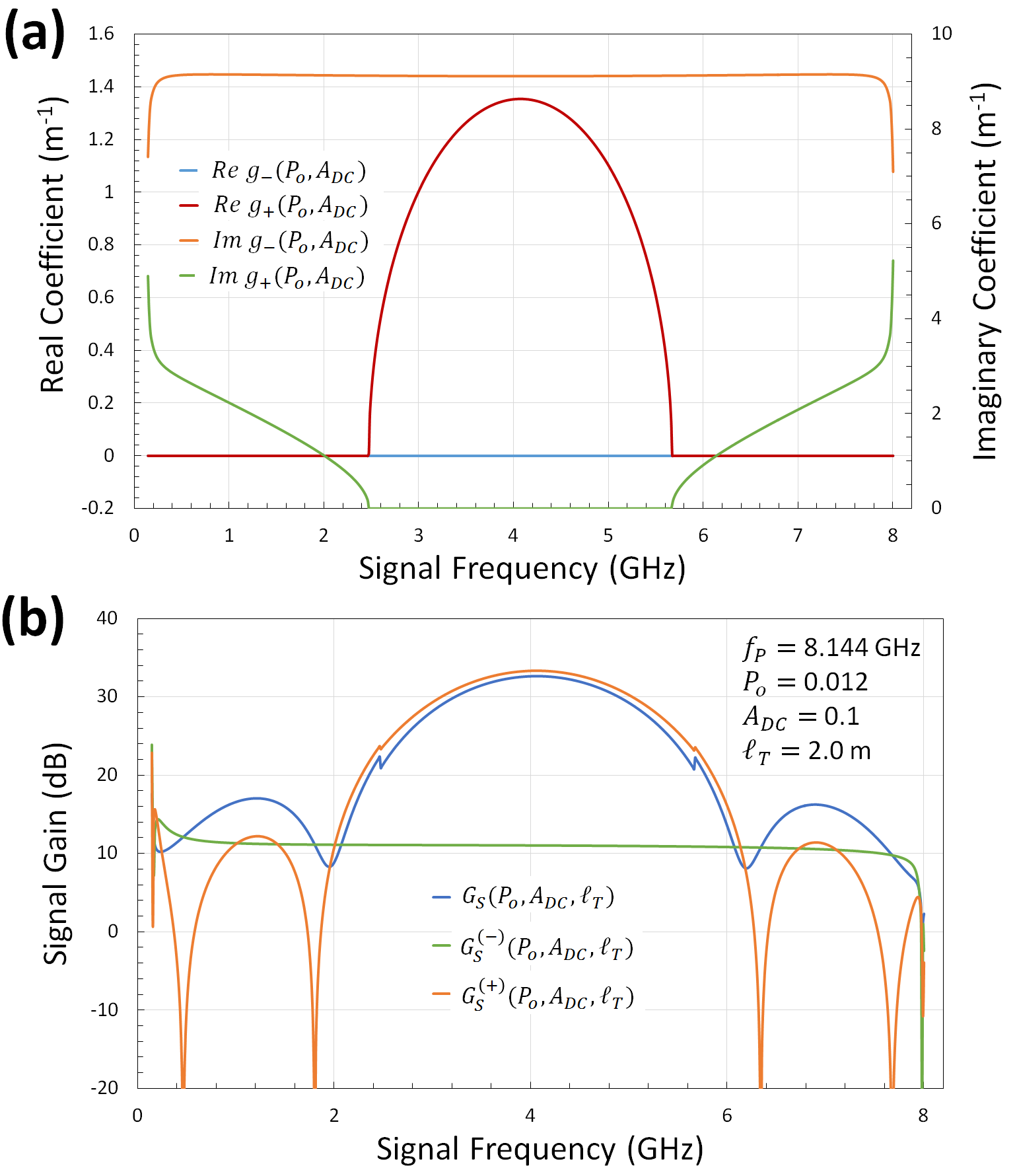}
\caption{\label{fig10} (a) Real and imaginary parts of approximate gain coefficients of Eq. (\ref{twoChannelGainCoefficients}) as a function of signal frequency, as described in the text. (b) Approximate 3WM signal gain of Eqs. (\ref{twoChannelSignalGain}) and (\ref{twoChannelSignalGainComponents}) as a function of signal frequency. All results employed mixing coefficients calculated from the loading design of Table \ref{tab1}.}
\end{figure}

If hybridization is very weak, i.e., $f^{(P)}_{0,-1}(A_{DC})\cong 0$ and $f^{(P)}_{-1,0}(A_{DC})\cong 0$, then $\phi^{(P)}_+(A_{DC})\cong f^{(P)}_{0,0}(A_{DC})$ and $\phi^{(P)}_-(A_{DC})\cong f^{(P)}_{-1,-1}(A_{DC})$. As hybridization increases, such as by structurally altering the loading design or by simply increasing the DC dynamically, the splitting between $\phi^{(P)}_+(A_{DC})$ and $\phi^{(P)}_-(A_{DC})$ also increases, such that for the loading design of Table \ref{tab1}, we find $\phi^{(P)}_+(A_{DC})$ ($\phi^{(P)}_-(A_{DC})$) becomes larger (smaller) and eventually strongly positive (negative). Since the terms $f^{(S)}_{0,0}(A_{DC})+f^{(1)}_{0,0}(A_{DC})+\Delta\beta^{(1)}_{3wm}$ tend to be collectively positive in Eq. (\ref{phaseRelations}), for the loadings of Table \ref{tab1}, we find $\varphi_+(A_{DC})$ tends to tune to zero as the hybridization increases, whereas $\varphi_-(A_{DC})$ becomes increasing detuned from zero, becoming more and more negative. Figure \ref{fig10}(a) shows the resulting impact this has on the gain coefficients of Eq. (\ref{twoChannelGainCoefficients}) when these are computed using the band structure, for the case of $P_o=0.012$, $A_{DC}=0.1$, $\ell_T=2$ m, and $f_P=8.014$ GHz. Here $g_+(P_o,A_{DC})$ as a function of $f_S$ implies a channel exhibiting broadband signal gain whereas $g_-(P_o,A_{DC})$ corresponds to a channel that is essentially inert.

The underlying physics can be explained as follows. In the supermode representation the injected traveling wave is hybridized by the dispersive property of the loadings, made increasing so by applied DC, for example. A pump traveling wave with frequency just below the gap takes on an additional component corresponding to the excited band state at $k_p$ of the second dispersion-frequency band, with occupancy of this excited state at $x$ proportional to ${\left| A^{(P)}_{-1}(x) \right|}^2$. One then has XPM induced between the two superposed components of the wave, which in turn creates a strong, broadband condition of phase matching, i.e., $\varphi_+(A_{DC})\cong 0$. This condition of phase matching persists whether the pump frequency is just below or just above the gap, with optimal phase matching occurring just above the gap. 

In contrast, if the pump frequency is set well below the first stop gap, such that the pump traveling wave can be described by just one component, with amplitude $A^{(P)}_0(x) \cong A^{(P)}_{0}(P_o,A_{DC}) \, e^{i f^{(P)}_{0,0}(A_{DC})\, x}$, then in analogy with Eq. (\ref{phaseRelations}), the phase matching condition in this case is  
\begin{equation}
\varphi(A_{DC})=f^{(P)}_{0,0}(A_{DC})-f^{(S)}_{0,0}(A_{DC})-f^{(1)}_{0,0}(A_{DC})-\Delta\beta^{(1)}_{3wm} ,
\end{equation}
which is the no-hybridization limit of $\varphi_+(A_{DC})$. For the loading design of Table \ref{tab1} one finds the condition $\varphi(A_{DC})\cong 0$ is never definitively established, i.e., strong inter-band XPM is required to create the phase matching condition, with additional tuning supplied by a DC bias. This explains why the signal gain diminishes in Fig. \ref{fig9}(d) as $f_P$ is decreased below the stop gap. The engineered dispersion, via the stop gap, is truly necessary to establish the broadband 3WM signal gain, borne out in the details of the band structure.

Figure \ref{fig10}(b) illustrates 3WM signal gain computed from the approximation of Eqs. (\ref{twoChannelSignalGain}) and (\ref{twoChannelSignalGainComponents}) using the band structure of the loading design of Table \ref{fig1}. The figure shows the dominance of the $+$ channel in defining $G_S(P_o,A_{DC},\ell_T)$. The results are comparable to the optimal signal gain presented in Fig. \ref{fig9} (blue curve), although we had to set the RF input power $P_o\cong 0.012$ to a value about twice as large as that of Fig. \ref{fig9} in order to obtain the comparison.

In Fig. \ref{fig9} we calculated a maximum 3WM signal gain of $\sim 30$ dB. This is larger than the experimental results of Fig. 3(b) of Ref. (\onlinecite{Vissers2016}), where a 3WM signal gain of $\sim 10$ dB was measured. As in our earlier comments regarding 4WM, the higher theoretical gain may be attributable to (i) the lack of a finite intrinsic $Q$ in our model and (ii) the stop gap size and position--in the results of Fig. 3(b) of Ref. (\onlinecite{Vissers2016}) the loading design differs markedly from Table \ref{tab1}, particularly the design of the first stop gap at $\sim 16$ GHz rather than $\sim 8$ GHz. Nevertheless, as we saw in Fig. \ref{fig9}, operation of the KIT with a DC bias increases its efficacy as an amplification device because the 3WM signal gain so produced is a much smoother function of signal frequency. Given a loading design more optimal for 4WM, we also saw in Fig. \ref{fig9} that 4WM features exist in the center of the 3WM signal band. Using our theoretical framework as a guide, we are presently engaged in follow-on research to obtain a loading design that maximizes the 3WM gain and bandwidth, and is also free of the 4WM features seen in Fig. \ref{fig9}. Important highlights from the results of our numerical calculations are:
\begin{enumerate}
	\item[(i)] Dispersion-frequency bands of the KIT amplifier may be calculated from knowledge of loading design, which creates stop gaps between bands and group velocities that differ between bands, as shown in Fig. \ref{fig4}.
	\item[(ii)] A forward-traveling current of the KIT amplifier may be represented by a superposition of these dispersion-frequency band states, as in Eq. (\ref{currentExpansion}), where the slowly-varying dimensionless amplitudes $A^{(P)}_\alpha(x)$ of the superposition are solutions of the coupled nonlinear first-order differential equations of Eq. (\ref{solePump}). Initial boundary conditions match the frequency of the traveling wave to a particular band state, but as the traveling wave propagates along the CPW it hybridizes into adjacent band states. The superposition may be approximated by a number $N_b$ of the lowest-lying bands, but as input power increases hybridization becomes stronger, requiring $N_b$ to be increased to retain numerical accuracy.
	\item[(iii)] Onset, magnitude, and bandwidth of both 4WM and 3WM signal gain is obtained solely from knowledge of loading design, as in Figs. \ref{fig8} and \ref{fig9}.
	\item[(iv)] Apart from external sources, such as impedance mismatch, intrinsic undulations in 4WM signal gain as a function of signal frequency are characteristic of matching of signal and idler frequencies to bands on either side of the first stop gap, as pictured in Fig. \ref{fig7}(a). Because signal and idler frequencies lie in different bands, they have different group velocities as a function of changing wavenumber, as calculated in Fig. \ref{fig4}(b), which means momentum conservation and overall phase-matching cannot be robustly maintained as signal frequency is varied.
	\item[(v)] In contrast to 4WM signal gain, 3WM signal gain is a relatively smooth function of signal frequency because signal and idler are matched to the same lowest-lying band, as in Fig. \ref{fig7}(b).
\end{enumerate}

\section{Concluding Remarks}
We presented a theoretical framework that describes the operation of a KIT amplifier designed with periodic loadings placed along the length of its CPW. Within this framework we first developed a metamaterial band theory of the dispersion engineering using Floquet-Bloch waves as a basis. From the band theory we constructed nonlinear forward-traveling waves as Floquet-Bloch supermodes built up from band eigenstates, showing how one defines and computes the slowly-varying amplitudes of these traveling waves. We then applied the supermode representation to a formulation of parametric mixing of an injected pump and signal, showing how one can calculate signal gain as a function of signal frequency, both with and without application of a DC bias. Our framework can be applied to other equivalent-circuit models of nonlinear traveling-wave parametric amplifiers that leverage periodic engineered dispersion to generate signal amplification, such as TWPAs based on Josephson junctions in lumped LC transmission lines.

In our analysis of a model KIT amplifier using the framework we came to several important conclusions. One is that, in the absence of a DC bias, the 4WM signal gain exhibits intrinsic undulations as a function of signal frequency. These occur apart from any extrinsic sources that may be encountered in real KIT amplifiers, such as ripples attributable to impedance mismatch between end nodes of the CPW. The cause of intrinsic undulations is the fact that injected signal and 4WM idler product are matched to different dispersion-frequency bands, as in Fig \ref{fig7}(a). As signal frequency is varied the criterion of phase matching cannot be maintained robustly because group velocity differs in the two bands, as demonstrated in Fig \ref{fig4}(b). The problem is exacerbated by the presence of multiple channels of parametric scattering involving excited-band-state constituents of the supermode traveling waves. The contributions of excited band states becomes more prevalent the higher the RF input power and the longer the waveguide length because these factors increase the sensitivity of traveling waves to the dispersive property of engineered loadings. Each new channel of parametric scattering, resulting from the symmetry-breaking of the loading design, corresponds to a distinct phase matching criterion and channel of signal gain. Thus, with variation of signal frequency, each phase-matching criterion exhibits either strengthening or weakening, which can lead to large non-monotonic variations in signal gain.

Intrinsic undulations are a general consequence of 4WM in the presence of engineered dispersion because, on average, group velocity cannot remain invariant from band to band as one progresses upward through the dispersion-frequency manifold. The issue may be mitigated by (i) the choice of nonlinear traveling-wave parametric amplifier, (ii) a design of loadings that minimizes the size of stop gaps, (iii) reduction of waveguide length, (iv) minimization of required RF input power, or (v) application of a DC bias, as described below. One can model the amplifier using the present framework to establish the extent of intrinsic undulations and then compare these to Fourier analysis of the length scales in the signal gain of the actual amplifier. Assuming traveling waves of frequencies adjacent to the first stop gap, if intrinsic undulations are mitigated to a single channel of contribution then the length scale of undulations should be the order of the stop-gap size. In the case of the KIT amplifier of Table \ref{tab1}, with large stop gaps at every third multiple of the pump frequency, we find very many fine undulations due to the effect of a number of contributing channels.

A second conclusion of our analysis of the KIT amplifier is that we can mitigate the issue of intrinsic signal-gain undulation by applying a DC bias, thereby leveraging 3WM signal gain on either side of the center of the resulting band. Incidental 4WM features present in the center of the 3WM band may be addressed by any of the methods mentioned above, although removal of these incidental features is not essential to operation of the amplifier. The 3WM signal gain is achievable at an order of magnitude less RF input power than 4WM signal amplification, albeit with reduced gain when compared to 4WM. Because the injected signal and 3WM idler are matched to the same first dispersion-frequency band, as in Fig \ref{fig7}(b), phase matching remains robust as signal frequency is varied, with the result that 3WM signal gain is a very smooth function of signal frequency. Caveats to utilization of 3WM signal amplification include the limited range of input powers and waveguide lengths permitted, due to the continual presence of 4WM processes, which tend to dominate at higher signal amplitude.

A last point of our analysis is that we showed how dispersion engineering generates 3WM signal gain. In our presentation of results we developed an analytical approximation of the 3WM signal, given by the two-channel form of Eq. (\ref{twoChannelSignalGain}). We showed how phase matching arises in the $+$ channel of this approximation via $\varphi_+(A_{DC})\cong 0$, where $\varphi_+(A_{DC})$ is given by Eqs. (\ref{pumpPhaseRelation}) and (\ref{phaseRelations}). From $\phi^{(P)}_+(A_{DC})$ of Eq. (\ref{pumpPhaseRelation}) we saw how inter-band XPM of the pump, i.e, cross-phase modulation between components of the dispersion-hybridized supermode traveling wave, create the necessary condition of phase matching, leading to the gain coefficient and signal gain depicted in Fig \ref{fig10}. For the engineered dispersion of Table \ref{tab1}, the phase matching condition was shown to be particularly tuned when the pump frequency was just above the stop gap.

Photonic metamaterials of the micro scale and smaller that exhibit strongly-correlated behavior continue to be an interesting and exploitable realm of quantum physics.\cite{Liu2016} Nonlinear traveling-wave parametric amplifiers, like the present KIT amplifier, represent one such category of device. In these devices one must account for the physical role of dispersion engineering to fully understand and quantify their behavior. In the case of periodic loading designs, the photonic band structure is essential to this comprehension, both within the waveguide and at its nodes, just as electronic band structure is essential to understanding many basic electromagnetic properties of solids, both in the bulk and at material interfaces. We hope that our theoretical framework can help to build better amplifiers and also stimulate new avenues of research.

\begin{acknowledgments}
This work was supported by the Army Research Office and the Laboratory for Physical Sciences under EAO221146, EAO241777, and the NIST Quantum Initiative. RPE acknowledges grant 60NANB14D024 from the US Department of Commerce, NIST. We have benefited greatly from discussions with Peter Day, Tom Ohki, Hsiang-sheng Ku, and Mustafa Bal. This work is property of the US Government and not subject to copyright.
\end{acknowledgments}

\appendix

\section{Summation Relations}	\label{appendixSummations}
Sums that arise in the metamaterial band theory of the KIT amplifier are
\begin{gather}
S_{n,n'}^{(1)}(\alpha,\beta)=
\sum\limits_{n''=-\infty}^\infty
\frac{ \sin{\pi\alpha \left( n - n'' \right) } \;
\sin{\pi\beta \left( n'' - n' \right) } }
{\pi^2 \left( n - n'' \right) \left( n'' - n' \right) } , \label{appendixSum1Def} \\
S_{n,n'}^{(2)}(\alpha,\beta)=
\sum\limits_{n''=-\infty}^\infty
n'' \, \frac{ \sin{\pi\alpha \left( n - n'' \right) } \;
\sin{\pi\beta \left( n'' - n' \right) } }
{\pi^2 \left( n - n'' \right) \left( n'' - n' \right) } , \label{appendixSum2Def}
\end{gather}
where we assume $0<\alpha,\beta<1$. Alternatively, these may be expressed as
\begin{equation}
S_{n,n'}^{(1)}(\alpha,\beta)=
\int_0^\alpha d\alpha' \int_0^\beta d\beta'
\sum\limits_{n''=-\infty}^\infty
\cos{\pi\alpha' \left( n - n'' \right) } \;
\cos{\pi\beta' \left( n'' - n' \right) } ,
\end{equation}
\begin{equation}
S_{n,n'}^{(2)}(\alpha,\beta)=
-\frac{i}{\pi}\lim_{\phi\rightarrow 0} \frac{\partial}{\partial\phi}
\int_0^\alpha d\alpha' \int_0^\beta d\beta'
\sum\limits_{n''=-\infty}^\infty
e^{i\pi\phi n''}
\cos{\pi\alpha' \left( n - n'' \right) } \;
\cos{\pi\beta' \left( n'' - n' \right) } .
\end{equation}

The cosine functions that appear in the above integral expressions may be written in the form $\cos{x}=(e^{ix}+e^{-ix})/2$, such that their respective sums may be evaluated using the definition of the delta function given by $\delta(x)=(1\left/ 2\right. )\sum\limits_{n=0}^\infty e^{\pm i\pi x n}$. In this way the sums are removed from the above integrals and we now have
\begin{equation}	\label{sum1}
S_{n,n'}^{(1)}(\alpha,\beta)=
\int_0^\alpha d\alpha' \int_0^\beta d\beta'
\left[ \delta(\alpha' + \beta')
\cos{\pi \left( \alpha' n + \beta' n' \right) } 
+\delta(\alpha' - \beta')
\cos{\pi \left( \alpha' n - \beta' n' \right) } \right] ,
\end{equation}
\begin{multline}	\label{sum2}
S_{n,n'}^{(2)}(\alpha,\beta)=
-\frac{i}{2\pi}\lim_{\phi\rightarrow 0} \frac{\partial}{\partial\phi} 
\int_0^\alpha d\alpha' \int_0^\beta d\beta'
\bigg[ \delta(\phi - \alpha' - \beta')
e^{i\pi \left( \alpha' n + \beta' n' \right) } \\
+ \delta(\phi + \alpha' + \beta') 
e^{-i\pi \left( \alpha' n + \beta' n' \right) }
+ \delta(\phi - \alpha' + \beta')
e^{i\pi \left( \alpha' n - \beta' n' \right) } 
+\delta(\phi + \alpha' - \beta')
e^{-i\pi \left( \alpha' n - \beta' n' \right) } \bigg] .
\end{multline}

In Eq. (\ref{sum1}) the first term in square brackets evaluates to zero since $\alpha,\beta>0$ while evaluation of the second term depends on the relative magnitudes of $\alpha$ and $\beta$. The integrated result is then
\begin{equation}	\label{appendixSum1}
S_{n,n'}^{(1)}(\alpha,\beta)=
\frac{1}{\pi \left( n - n' \right)} \left[
\Theta(\beta - \alpha) \sin{\pi \alpha \left( n - n' \right) } 
+ \Theta(\alpha - \beta) \sin{\pi \beta \left( n - n' \right) } \right] ,
\end{equation}
where $\Theta(x)$ is a step function given by
\begin{equation}	\label{stepFuncton}
\Theta(x) = \left\{
\begin{array}{lll}
0 & ; & x < 0 \\
\frac{1}{2} & ; & x = 0 \\
1 & ; & x > 0
\end{array}
\right. .
\end{equation}
Similarly, in Eq. (\ref{sum2}), weighing also the magnitude of $\phi$, we have the intermediate step
\begin{multline}
S_{n,n'}^{(2)}(\alpha,\beta)=
\frac{1}{2\pi^2 \left( n - n' \right) }
\lim_{\phi\rightarrow 0} \frac{\partial}{\partial\phi} \Bigg\{
\left[ \Theta(\phi - \alpha - \beta) - \Theta(\phi - \alpha + \beta) \right] 
e^{ i\pi \left[ \alpha \left( n - n' \right) + \phi n' \right] } \\
+ \left[ \Theta(-\phi - \alpha + \beta) - \Theta(-\phi - \alpha - \beta) \right] 
e^{ -i\pi \left[ \alpha \left( n - n' \right) - \phi n' \right] } \\
+ \left[ \Theta(-\phi - \alpha - \beta) - \Theta(-\phi + \alpha - \beta) \right] 
e^{ i\pi \left[ \beta \left( n - n' \right) + \phi n \right] } \\
+ \left[ \Theta(\phi + \alpha - \beta) - \Theta(\phi - \alpha - \beta) \right] 
e^{ -i\pi \left[ \beta \left( n - n' \right) - \phi n \right] } \Bigg\} .
\end{multline}
Differentiating with respect to $\phi$ then taking the limit as $\phi\rightarrow 0$ we arrive at
\begin{equation}	\label{appendixSum2}
S_{n,n'}^{(2)}(\alpha,\beta)=
\frac{1}{\pi \left( n - n' \right)} \left[
\Theta(\beta - \alpha) \; n' \; \sin{\pi \alpha \left( n - n' \right) } 
+ \Theta(\alpha - \beta) \; n \; \sin{\pi \beta \left( n - n' \right) } \right] .
\end{equation}

\section{Dispersion Matrix of Arbitrary Loading Design of Even Symmetry}	\label{appendixDispersionMatrix}
Assume a unit cell of alternating regions of loadings and non-loadings, symmetric about the center of the cell, as in Fig. \ref{fig1}(b). The total number of regions is $R$, an odd whole number, with region $(R+1)/2$ at the center of the cell. Pairs of regions $r$ and $R-r+1$, on either side of the center region, are identical in length $\Delta x_r$ and have the same capacitance per unit length $\mathcal{C}_r$ and linear inductance per unit length $\mathcal{L}_r$. 

Substituting Eq. (\ref{loadingDesign}) into the Fourier coefficients of Eqs. (\ref{matrixInverse1}) and (\ref{matrixInverse2}) we initially find
\begin{equation}
\left\{
\begin{array}{c}
C_n^{-1} \\
{L_o}_n^{-1}
\end{array}
\right\} =
\frac{i}{2\pi n}
\sum\limits_{r=1}^{R}
\left\{
\begin{array}{c}
\frac{1}{\mathcal{C}_r} \\
\frac{1}{\mathcal{L}_r}
\end{array}
\right\}
\left(
e^{-2\pi i n \Delta x_r \left/ \ell_o \right. } - 1
\right) 
e^{-2\pi i n x_r \left/ \ell_o \right. } .
\end{equation}
Then exploiting the symmetry of the unit cell, as described above, we obtain after some algebra the real-valued coefficients
\begin{multline}
\left\{
\begin{array}{c}
C_n^{-1} \\
{L_o}_n^{-1}
\end{array}
\right\} =
\frac{ { \left( -1 \right) }^n }{\pi n}
\left\{
\begin{array}{c}
\frac{1}{\mathcal{C}_{(R+1)/2}} \\
\frac{1}{\mathcal{L}_{(R+1)/2}}
\end{array}
\right\}
\sin{ \Big( \pi \Delta x_{(R+1)/2} \, n \left/ \ell_o \right. \Big) } \\
+ \frac{1}{\pi n} \sum\limits_{r=1}^{(R-1)/2}
\left\{
\begin{array}{c}
\frac{1}{\mathcal{C}_r} \\
\frac{1}{\mathcal{L}_r}
\end{array}
\right\}
\Bigg[
\sin{ \left( 2\pi \sum\limits_{r'=1}^r \Delta x_{r'} \, n \left/ \ell_o \right. \right) }
- \sin{ \left( 2\pi \sum\limits_{r'=1}^{r-1} \Delta x_{r'} \, n \left/ \ell_o \right. \right) }
\Bigg] .
\end{multline}
The sums of this expression may be further simplified to
\begin{equation}	\label{appendixInversesEvaluated}
\left\{
\begin{array}{c}
C_n^{-1} \\
{L_o}_n^{-1}
\end{array}
\right\} = 
\left\{
\begin{array}{c}
\frac{1}{\mathcal{C}_{(R+1)/2}} \\
\frac{1}{\mathcal{L}_{(R+1)/2}}
\end{array}
\right\}
\delta_{n,0}
+ \frac{1}{\pi n} \sum\limits_{r=1}^{(R - 1)/2}
\left\{
\begin{array}{c}
\frac{1}{\mathcal{C}_r} - \frac{1}{\mathcal{C}_{r+1}} \\
\frac{1}{\mathcal{L}_r} - \frac{1}{\mathcal{L}_{r+1}}
\end{array}
\right\}
\sin{ \left( 2\pi \sum\limits_{r'=1}^r \Delta x_{r'} \, n \left/ \ell_o \right. \right) } .
\end{equation}

With the aid of Appendix \ref{appendixSummations}, if we then substitute Eq. (\ref{appendixInversesEvaluated}) into Eq. (\ref{dispersionMatrix0}), and apply the sums $S_{n,n'}^{(1)}(\alpha,\beta)$ and $S_{n,n'}^{(2)}(\alpha,\beta)$ defined in Eqs. (\ref{appendixSum1Def}) and (\ref{appendixSum2Def}), we obtain
\begin{multline}
D_{n,n'}(k) = 
\frac{1}{ \mathcal{L}_{(R+1)/2} \mathcal{C}_{(R+1)/2} } 
{ \left( k + 2\pi n \left/ \ell_o \right. \right) }^2 \delta_{n,n'}
\\
+ \frac{1}{\pi \left( n - n' \right) } 
\sum\limits_{r=1}^{(R - 1)/2} 
\Bigg[
\frac{1}{\mathcal{L}_{(R+1)/2}} 
\Big( \frac{1}{\mathcal{C}_r} - \frac{1}{\mathcal{C}_{r+1}} \Big)
\Big( k + 2\pi n \left/ \ell_o \right. \Big) \\
+ \Big( \frac{1}{\mathcal{L}_r} - \frac{1}{\mathcal{L}_{r+1}} \Big)
\frac{1}{\mathcal{C}_{(R+1)/2}} 
\Big( k + 2\pi n' \left/ \ell_o \right. \Big)
\Bigg]
\sin{ \left( 2\pi \sum\limits_{r'=1}^r \Delta x_{r'} \, \left( n - n' \right) 
\left/ \ell_o \right. \right) }
\left( k + 2\pi n' \left/ \ell_o \right. \right)
\\
+ \sum\limits_{r_1=1}^{(R - 1)/2} \sum\limits_{r_2=1}^{(R - 1)/2}
\Big( \frac{1}{\mathcal{L}_{r_1}} - \frac{1}{\mathcal{L}_{r_1+1}} \Big)
\Big( \frac{1}{\mathcal{C}_{r_2}} - \frac{1}{\mathcal{C}_{r_2+1}} \Big) \\
\times\; \Bigg[ k \, S_{n,n'}^{(1)} \left( 
2 \sum\limits_{r_1'=1}^{r_1} \Delta x_{r_1'} \left/ \ell_o \right. ,
2 \sum\limits_{r_2'=1}^{r_2} \Delta x_{r_2'} \left/ \ell_o \right. \right) 
\;\;\;\;\;\;\;\;\;\;\;\;\;\;\;\;\; \\ \;\;\;\;\;\;\;\;\;\;\;\;\;\;\;\;\;\;\;\;\;\;
+ \frac{2\pi}{\ell_o} \, S_{n,n'}^{(2)} \left( 
2 \sum\limits_{r_1'=1}^{r_1} \Delta x_{r_1'} \left/ \ell_o \right. ,
2 \sum\limits_{r_2'=1}^{r_2} \Delta x_{r_2'} \left/ \ell_o \right. \right) \Bigg]
\Big( k + 2\pi n' \left/ \ell_o \right. \Big) 
\end{multline}
Substituting the summation evaluations given by Eqs. (\ref{appendixSum1}) and (\ref{appendixSum2}) we arrive at 
\begin{multline}
D_{n,n'}(k) = 
\frac{1}{ \mathcal{L}_{(R+1)/2} \mathcal{C}_{(R+1)/2} } 
{ \left( k + 2\pi n \left/ \ell_o \right. \right) }^2 \delta_{n,n'}
+ \frac{1}{ \pi \left( n - n' \right) } 
\\
\times\;
\sum\limits_{r=1}^{(R - 1)/2} 
\Bigg[
\frac{1}{\mathcal{L}_{(R+1)/2}} 
\Big( \frac{1}{\mathcal{C}_r} - \frac{1}{\mathcal{C}_{r+1}} \Big)
\Big( k + 2\pi n \left/ \ell_o \right. \Big) 
+ \Big( \frac{1}{\mathcal{L}_r} - \frac{1}{\mathcal{L}_{r+1}} \Big)
\frac{1}{\mathcal{C}_{(R+1)/2}} 
\Big( k + 2\pi n' \left/ \ell_o \right. \Big)
\Bigg]
\\
\times\;
\sin{ \left[ 2\pi \sum\limits_{r'=1}^r \Delta x_{r'} \, \left( n - n' \right) 
\left/ \ell_o \right. \right] }
\Big( k + 2\pi n' \left/ \ell_o \right. \Big)
+ \frac{1}{ \pi \left( n - n' \right) } 
\sum\limits_{r_1=1}^{(R - 1)/2} \sum\limits_{r_2=1}^{(R - 1)/2}
\Theta(r_1 - r_2) \\
\times\;
\Bigg[
\Big( \frac{1}{\mathcal{L}_{r_1}} - \frac{1}{\mathcal{L}_{r_1+1}} \Big)
\Big( \frac{1}{\mathcal{C}_{r_2}} - \frac{1}{\mathcal{C}_{r_2+1}} \Big)
\Big( k + 2\pi n \left/ \ell_o \right. \Big) 
+ \Big( \frac{1}{\mathcal{L}_{r_2}} - \frac{1}{\mathcal{L}_{r_2+1}} \Big)
\Big( \frac{1}{\mathcal{C}_{r_1}} - \frac{1}{\mathcal{C}_{r_1+1}} \Big)
\Big( k + 2\pi n' \left/ \ell_o \right. \Big) 
\Bigg]
\\
\times\;
\sin{ \left[
2\pi \sum\limits_{r'=1}^{r_2} \Delta x_{r'} \left( n - n' \right) \left/ \ell_o \right. 
\right] }  
\Big( k + 2\pi n' \left/ \ell_o \right. \Big) ,
\end{multline}
where we have used the fact that
\begin{equation}
\Theta \left( 2\sum\limits_{r=1}^{r_1} \Delta x_r \left/ \ell_o \right.
- 2\sum\limits_{r'=1}^{r_2} \Delta x_{r'} \left/ \ell_o \right. \right) = \Theta(r_1 - r_2) ,
\end{equation}
with $\Theta (r_1 - r_2)$ defined as in Eq. (\ref{stepFuncton}).

Explicitly evaluating $\Theta (r_1 - r_2)$ in the expression of $D_{n,n'}(k)$ above we obtain
\begin{multline}
D_{n,n'}(k) = 
\frac{1}{ \mathcal{L}_{(R+1)/2} \mathcal{C}_{(R+1)/2} } 
{ \left( k + 2\pi n \left/ \ell_o \right. \right) }^2 \delta_{n,n'}
\\
+ \frac{1}{ \pi \left( n - n' \right) } 
\sum\limits_{r=1}^{(R - 1)/2} 
\Bigg\{
\Big[
\frac{1}{\mathcal{L}_{(R+1)/2}} 
+ \frac{1}{2}
\Big( \frac{1}{\mathcal{L}_r} - \frac{1}{\mathcal{L}_{r+1}} \Big)
\Big]
\Big( \frac{1}{\mathcal{C}_r} - \frac{1}{\mathcal{C}_{r+1}} \Big)
\Big( k + 2\pi n \left/ \ell_o \right. \Big) \\
+ \Big( \frac{1}{\mathcal{L}_r} - \frac{1}{\mathcal{L}_{r+1}} \Big)
\Big[
\frac{1}{\mathcal{C}_{(R+1)/2}} 
+ \frac{1}{2}
\Big( \frac{1}{\mathcal{C}_r} - \frac{1}{\mathcal{C}_{r+1}} \Big)
\Big]
\Big( k + 2\pi n' \left/ \ell_o \right. \Big)
\Bigg\} 
\\
\times\; 
\sin{ \left[ 2\pi \sum\limits_{r'=1}^r \Delta x_{r'} \, \left( n - n' \right) 
\left/ \ell_o \right. \right] }
\Big( k + 2\pi n' \left/ \ell_o \right. \Big)
\\
+ \frac{1}{ \pi \left( n - n' \right) } 
\sum\limits_{r_1=2}^{(R - 1)/2} \; \sum\limits_{r_2=1}^{r_1-1} 
\Bigg[
\Big( \frac{1}{\mathcal{L}_{r_1}} - \frac{1}{\mathcal{L}_{r_1+1}} \Big)
\Big( \frac{1}{\mathcal{C}_{r_2}} - \frac{1}{\mathcal{C}_{r_2+1}} \Big)
\Big( k + 2\pi n \left/ \ell_o \right. \Big) \\
+ \Big( \frac{1}{\mathcal{L}_{r_2}} - \frac{1}{\mathcal{L}_{r_2+1}} \Big)
\Big( \frac{1}{\mathcal{C}_{r_1}} - \frac{1}{\mathcal{C}_{r_1+1}} \Big)
\Big( k + 2\pi n' \left/ \ell_o \right. \Big) 
\Bigg] 
\\
\times\;
\sin{ \left[
2\pi \sum\limits_{r'=1}^{r_2} \Delta x_{r'} \left( n - n' \right) \left/ \ell_o \right. 
\right] }  
\Big( k + 2\pi n' \left/ \ell_o \right. \Big) .
\end{multline}
After some manipulation of sums, the above expression of $D_{n,n'}(k)$ may be written more succinctly as
\begin{multline}	\label{arbitraryDispersionMatrix}
D_{n,n'}(k) = 
\frac{1}{ \mathcal{L}_{(R+1)/2} \mathcal{C}_{(R+1)/2} }
{ \Big( k + 2\pi n \left/ \ell_o \right. \Big) }^2 \delta_{n,n'} \\
+ \Bigg[ \mathcal{D}_{n-n'}(\left\{ \mathcal{L}_r \right\}, \left\{ \mathcal{C}_r \right\} ) \;
\Big( k + 2\pi n \left/ \ell_o \right. \Big) 
+ \mathcal{D}_{n-n'}(\left\{ \mathcal{C}_r \right\}, \left\{ \mathcal{L}_r \right\} ) \;
\Big( k + 2\pi n' \left/ \ell_o \right. \Big) \Bigg]
\Big( k + 2\pi n' \left/ \ell_o \right. \Big) ,
\end{multline}
where we have defined coefficients
\begin{equation}	\label{loadingCoefficient}
\mathcal{D}_n(\left\{ \mathcal{L}_r \right\}, \left\{ \mathcal{C}_r \right\} ) = 
\frac{1}{ 2\pi n } \sum\limits_{r=1}^{(R - 1)/2} 
\left( \frac{1}{\mathcal{L}_r} + \frac{1}{\mathcal{L}_{r+1}} \right)
\left( \frac{1}{\mathcal{C}_r} - \frac{1}{\mathcal{C}_{r+1}} \right)
\sin{ \left( 2\pi \sum\limits_{r'=1}^r \Delta x_{r'} \, n \left/ \ell_o \right. \right) } .
\end{equation}
Equations (\ref{arbitraryDispersionMatrix}) and (\ref{loadingCoefficient}) express an element of the dispersion matrix appropriate for an arbitrary loading design.

\section{Derivation of Nonlinear Traveling-Wave Equations}	\label{appendixNonlinearEquations}
Consider a solution of Eqs. (\ref{telegrapherEquation1}) and (\ref{telegrapherEquation2}) involving a mix of forward-traveling RF waves, where the $\mu$-th such traveling wave propagates with wave number $k_\mu$ and frequency $\omega_{\mu}$. Inside the waveguide the voltage and current of each wave are expressed in terms of Floquet-Bloch-like functions, similar to Eq. (\ref{linearRFTravelingWave}), but with coefficients of Fourier expansion slowly varying in $x$, in the manner of Eq. (\ref{slowlyVaryingApproximation}). Specifically, we write
\begin{equation}	\label{travelingWaves0}
\left\{
\begin{array}{c}
V^{(\mu)}_{k_\mu}(x,t) \\
I^{(\mu)}_{k_\mu}(x,t) 
\end{array}
\right\}
= \sum\limits_{n=-\infty}^{\infty} 
\left\{
\begin{array}{c}
V^{(\mu)}_n(x) \\
I^{(\mu)}_n(x)
\end{array}
\right\}
e^{i \left( k_\mu + 2\pi n \left/ \ell_o \right. \right) x - i\omega_\mu t} , 
\end{equation}
where in the sum over $\mu$ we also include complex-conjugate terms. That is, for every index $\mu$, for which we have $k_\mu$ and $\omega_\mu$, there is another complex-conjugate term $-\mu$ corresponding to $k_{-\mu}=-k_\mu$ and $\omega_{-\mu}=-\omega_\mu$, with amplitude coefficients $V^{(-\mu)}_n(x)={V^{(\mu)}_{-n}(x)}^*$ and $I^{(-\mu)}_n(x)={I^{(\mu)}_{-n}(x)}^*$. This nomenclature ensures voltage and current are real-valued, and permits simpler notation for nested summations. With DC given by $I_{DC}$, the full solution may be expressed as
\begin{gather}
V(x,t) = \sum\limits_{\mu} V^{(\mu)}_{k_\mu}(x,t) ,  \label{sol1} \\
I(x,t) = I_{DC} + \sum\limits_{\mu} I^{(\mu)}_{k_\mu}(x,t) ,  \label{sol2}
\end{gather}
which includes the complex-conjugate $\mu$ terms.

For example, substituting Eqs. (\ref{travelingWaves0}) through (\ref{sol2}) into Eq. (\ref{kineticInductancePerLength}) we obtain
\begin{multline}	\label{inductanceTransformed}
L(x,t) = L_o(x) \Bigg\{ 1 + 
2 \frac{I_{DC}}{I_*^2} \sum\limits_{\mu} \sum\limits_{n=-\infty}^{\infty} 
I^{(\mu)}_{n}(x) \, e^{ i \left( k_{\mu} + 2\pi n \left/ \ell_o\right. \right) x - i \omega_{\mu} t } \\
+ \frac{1}{I_*^2} \sum\limits_{\mu,\mu'} \sum\limits_{n,n'=-\infty}^{\infty} 
I^{(\mu)}_{n}(x) \, I^{(\mu')}_{n'}(x) \, 
e^{ i \left[ k_{\mu} + k_{\mu'} + 2\pi \left( n + n' \right) \left/ \ell_o\right. \right] x - i \left( \omega_{\mu} + \omega_{\mu'} \right) t } 
\Bigg\} ,
\end{multline}
where both the sums over $\mu$ and $\mu'$ include complex conjugates. Then substituting Eqs. (\ref{travelingWaves0}) through (\ref{inductanceTransformed}) into Eqs. (\ref{telegrapherEquation1}) and (\ref{telegrapherEquation2}) we obtain an expression of $I^{(\mu)}_{n}(x)$ and $V^{(\mu)}_{n}(x)$ decoupled from one another. As in the linear limit, we again use the discrete Fourier transforms of $C(x)$ and $L_o(x)$, as given in Eqs. (\ref{discreteFourierTransformPair1}) and (\ref{discreteFourierTransformPair2}), and we also apply Eqs. (\ref{matrixInverse1}) and (\ref{matrixInverse2}). Since $V^{(\mu)}_{n}(x)$ and $I^{(\mu)}_{n}(x)$ are slowly varying in $x$ we may treat the functions of discrete Fourier expansions as approximately orthogonal. In this way we obtain an intermediate expression of the current given by
\begin{multline}	\label{intermediateStep}
\sum\limits_{n'=-\infty}^{\infty} C_{n-n'}^{-1} \,
\left( k_\mu + 2\pi n \left/ \ell_o \right. - i \frac{\partial }{\partial x} \right)
\left( k_\mu + 2\pi n' \left/ \ell_o \right. - i \frac{\partial }{\partial x} \right) 
I^{(\mu)}_{n'}(x) \cong  \\
\omega_\mu^2 \, \left[ 1 + {\left( \frac{I_{DC}}{I_*} \right) }^2 \right] 
\sum\limits_{n'=-\infty}^{\infty} {L_o}_{n-n'} \, I^{(\mu)}_{n'}(x) \\
+ \frac{I_{DC}}{I_*^2} \, \omega_\mu^2
\sum\limits_{\substack{\mu', \mu'' \\ ( \omega_{\mu'} + \omega_{\mu''} = \omega_{\mu} ) }}
\sum\limits_{n', n''} {L_o}_{n - n' - n''} \, I^{(\mu')}_{n'}(x) \, I^{(\mu'')}_{n''}(x) \, e^{i \left( k_{\mu'} + k_{\mu''} - k_{\mu} \right) x} \\
+ \frac{\omega_\mu^2}{3 I_*^2} \,  
\sum\limits_{ \substack{\mu', \mu'', \mu''' \\ ( \omega_{\mu'} + \omega_{\mu''} + \omega_{\mu'''} = \omega_{\mu} ) } }
\sum\limits_{n', n'', n'''} {L_o}_{n - n' - n'' - n'''} \, I^{(\mu')}_{n'}(x) \, I^{(\mu'')}_{n''}(x) \, I^{(\mu''')}_{n'''}(x) 
\, e^{i \left( k_{\mu'} + k_{\mu''} + k_{\mu'''} - k_{\mu} \right) x} .
\end{multline}

Via Eq. (\ref{slowlyVaryingApproximation}) we keep only the first derivatives in $x$ on the left side of Eq. (\ref{intermediateStep}). Also, to both sides of this equation we multiple by ${L_o}_{n'' - n}^{-1}$, of Eq. (\ref{matrixInverse2}), and then sum over $n$. With change of arbitrary index names the result may be expressed as
\begin{multline}	\label{currentDiffEq}
- 2i \sum\limits_{n'} \Lambda_{n,n'} (k_\mu) \, \frac{\partial }{\partial x} I^{(\mu)}_{n'}(x) \cong 
\omega_\mu^2 \, \left[ 1 + {\left( \frac{I_{DC}}{I_*} \right) }^2 \right] I^{(\mu)}_{n}(x) 
- \sum\limits_{n'} D_{n,n'} (k_\mu) I^{(\mu)}_{n'}(x) \\
+ \frac{I_{DC}\,\omega_\mu^2}{I_*^2}
\sum\limits_{ \substack{ \mu', \mu'' \\ ( \omega_{\mu'} + \omega_{\mu''} = \omega_{\mu} ) } }
\sum\limits_{ \substack{ n, n', n'' \\ ( n' + n'' = n ) } } I^{(\mu')}_{n'}(x) \, I^{(\mu'')}_{n''}(x) \, 
e^{i \left( k_{\mu'} + k_{\mu''} - k_{\mu} \right) x} \\
+ \frac{1}{3} \, \frac{\omega_\mu^2}{I_*^2} \,
\sum\limits_{ \substack{ \mu', \mu'', \mu''' \\ 
( \omega_{\mu'} + \omega_{\mu''} + \omega_{\mu'''} = \omega_{\mu} ) } }
\sum\limits_{ \substack{ n, n', n'', n''' \\ ( n' + n'' + n''' = n ) } }
I^{(\mu')}_{n'}(x) \, I^{(\mu'')}_{n''}(x) \, I^{(\mu''')}_{n'''}(x) \, e^{i \left( k_{\mu'} + k_{\mu''} + k_{\mu'''} - k_{\mu} \right) x} ,
\end{multline}
where we have made use of (\ref{dispersionMatrix}) and introduced a new matrix of elements 
\begin{equation}	\label{lambda}
\Lambda_{n,n'}(k) =
\sum\limits_{n''} {L_o}_{n - n''}^{-1} \, C_{n'' - n'}^{-1} \, 
\left[ k + \pi \left( n'' + n' \right) \left/ \ell_o \right. \right] .
\end{equation}

Solving the eigenproblem of Eqs. (\ref{linearEigenproblem1}) and (\ref{linearEigenproblem2}) gives us the band structure of dispersion frequencies $\Omega_{\alpha_\mu}(k_\mu)$ of the KIT. The incident traveling wave $\mu$ of frequency $\omega_\mu$ is matched to a particular dispersion frequency such that $\omega_\mu=\Omega_{\alpha_\mu}(k_\mu)$, where $k_\mu$ and $\alpha_\mu$ are matching wave number and band index, respectively. Adiabatic propagation of the wave, with slowly-varying amplitudes $V^{(\mu)}_{n}(x)$ and $I^{(\mu)}_{n}(x)$, allows us to express the solution as a superpositions of band states corresponding to matching $k_\mu$, as in Eq. (\ref{currentExpansion}). The amplitude $A^{(\mu)}_\alpha(x)$ is introduced as dimensionless and slowly varying in $x$.

If we apply Eq. (\ref{currentExpansion}), along with the band-matching condition $\omega_\mu=\Omega_{\alpha_\mu}(k_\mu)$, to Eq. (\ref{currentDiffEq}), we may make use of Eqs. (\ref{linearEigenproblem1}) and (\ref{linearEigenproblem2}), and the orthonormality of left and right vectors in Eqs. (\ref{orthonormalityCompleteness1}) and (\ref{orthonormalityCompleteness2}) to find
\begin{multline}
- 2i \sum\limits_{\alpha'} \tilde{\Lambda}_{\alpha,\alpha'}(k_\mu) \frac{\partial }{\partial x} A^{(\mu)}_{\alpha'}(x) \cong 
\left[ {{\Omega_{\alpha_\mu}}(k_\mu)}^2 \, \left( 1 + A_{DC}^2 \right) - {{\Omega_\alpha}(k_\mu)}^2 \right] A^{(\mu)}_\alpha(x) \\
+ {{\Omega_{\alpha_\mu}}(k_\mu)}^2 \, A_{DC}
\sum\limits_{{\alpha}', {\alpha}''} \sum\limits_{ \substack{ \mu', \mu'' \\ ( \omega_{\mu'} + \omega_{\mu''} = \omega_{\mu} ) } }
\sum\limits_{ \substack{ n, n', n'' \\ ( n' + n'' = n ) } }
{{u_n^{(\alpha)}}(k_{\mu})} \, {e_{n'}^{(\alpha')}}(k_{\mu'}) \, {e_{n''}^{(\alpha'')}}(k_{\mu''}) \\
\times\; A^{(\mu')}_{\alpha'}(x) \, A^{(\mu'')}_{\alpha''}(x) \, e^{i \left( k_{\mu'} + k_{\mu''} - k_{\mu} \right) x} \\
+ \frac{1}{3} \, {{\Omega_{\alpha_\mu}}(k_\mu)}^2 
\sum\limits_{\alpha', \alpha'', \alpha'''} 
\sum\limits_{ \substack{ \mu', \mu'', \mu''' \\ ( \omega_{\mu'} + \omega_{\mu''} + \omega_{\mu'''} = \omega_{\mu} ) } }
\sum\limits_{ \substack{ n, n', n'', n''' \\ ( n' + n'' + n''' = n ) } }
{{u_n^{(\alpha)}}(k_{\mu})} \, {e_{n'}^{(\alpha')}}(k_{\mu'}) \, 
{e_{n''}^{(\alpha'')}}(k_{\mu''}) \, {e_{n'''}^{(\alpha''')}}(k_{\mu'''}) \\
\times\; A^{(\mu')}_{\alpha'}(x) \, A^{(\mu'')}_{\alpha''}(x) \, A^{(\mu''')}_{\alpha'''}(x) \, 
e^{i \left( k_{\mu'} + k_{\mu''} + k_{\mu'''} - k_{\mu} \right) x} ,
\end{multline}
where $A_{DC}=I_{DC}/I_*$ and
\begin{equation}	\label{lambdaTildeAppendix}
\tilde{\Lambda}_{\alpha,\alpha'}(k_\mu) =
\sum\limits_{n, n'} \Lambda_{n,n'}(k_\mu) \, 
{u_{n}^{(\alpha)}}(k_\mu) \, {e_{n'}^{(\alpha')}}(k_\mu) .
\end{equation}
Assuming the matrix of Eq. (\ref{lambdaTildeAppendix}) may be inverted we obtain the final form of the coupled amplitude equations governing multiwave mixing within the KIT, viz.
\begin{multline}	\label{mwmEquations}
-i \frac{\partial }{\partial x} A^{(\mu)}_{\alpha}(x) \cong
\sum\limits_{\alpha'} f^{(\mu)}_{\alpha, \alpha'}(A_{DC}) \, A^{(\mu)}_{\alpha'}(k_\mu,x) \\
+ A_{DC} \sum\limits_{{\alpha}', {\alpha}''}
\sum\limits_{ \substack{ \mu', \mu'' \\ ( \omega_{\mu'} + \omega_{\mu''} = \omega_{\mu} ) } }
f^{(\mu,\mu',\mu'')}_{\alpha, \alpha', \alpha''} \, A^{(\mu')}_{\alpha'}(x) \, A^{(\mu'')}_{\alpha''}(x) \, 
e^{i \left( k_{\mu'} + k_{\mu''} - k_{\mu} \right) x} \\
+ \frac{1}{3} 
\sum\limits_{\alpha',\alpha''\alpha'''}
\sum\limits_{ \substack{ \mu', \mu'', \mu''' \\ ( \omega_{\mu'} + \omega_{\mu''} + \omega_{\mu'''} = \omega_{\mu} ) } }
f^{(\mu,\mu',\mu'',\mu''')}_{\alpha, \alpha', \alpha'', \alpha'''} \, A^{(\mu')}_{\alpha'}(x) \, A^{(\mu'')}_{\alpha''}(x) \, A^{(\mu''')}_{\alpha'''}(x) \, 
e^{i \left( k_{\mu'} + k_{\mu''} + k_{\mu'''} - k_{\mu} \right) x} .
\end{multline}

In the above expression we have defined band-hybridization coefficients
\begin{equation}	\label{fHybridizationAppendix}
f^{(\mu)}_{\alpha, \alpha'}(A_{DC}) =
\frac{1}{2} \, \tilde{\Lambda}_{\alpha,\alpha'}^{-1}(k_\mu)
\left[ {{\Omega_{\alpha_\mu}}(k_\mu)}^2 \, 
\left( 1 + A_{DC}^2 \right) - {{\Omega_{\alpha'}}(k_\mu)}^2 \right] ,
\end{equation}
3WM coefficients
\begin{equation}	\label{f3WMAppendix}
f^{(\mu,\mu',\mu'')}_{\alpha, \alpha', \alpha''} =
\frac{1}{2} \, {{\Omega_{\alpha_\mu}}(k_\mu)}^2 \; \sum\limits_{ \substack{ n, n', n'' \\ ( n' + n'' = n ) } }
\sum\limits_{\beta} \tilde{\Lambda}_{\alpha,\beta}^{-1}(k_\mu) \,
{{u_n^{(\beta)}}(k_{\mu})} \; {e_{n'}^{(\alpha')}}(k_{\mu'}) \; {e_{n''}^{(\alpha'')}}(k_{\mu''}) ,
\end{equation}
and 4WM coefficients
\begin{equation}	\label{f4WMAppendix}
\begin{array}{l}
f^{(\mu,\mu',\mu'',\mu''')}_{\alpha, \alpha', \alpha'', \alpha'''} = 
\frac{1}{2} \, {{\Omega_{\alpha_\mu}}(k_\mu)}^2 \; \sum\limits_{ \substack{ n, n', n'', n''' \\ ( n' + n'' + n''' = n ) } }
\sum\limits_{\beta} \tilde{\Lambda}_{\alpha,\beta}^{-1}(k_\mu) \,
{{u_n^{(\beta)}}(k_{\mu})} \; {e_{n'}^{(\alpha')}}(k_{\mu'}) \; {e_{n''}^{(\alpha'')}}(k_{\mu''}) \; {e_{n'''}^{(\alpha''')}}(k_{\mu'''}) ,
\end{array}
\end{equation}
where one may use the relation $e_{n}^{(\alpha)}(k_{-\mu})=e_{n}^{(\alpha)}(-k_{\mu})={e_{-n}^{(\alpha)}(k_{\mu})}^*$, from Eq. (\ref{eigenRelations}), as applicable, to assist in evaluation. Equation (\ref{mwmEquations}) is a general description of the coupled amplitude equations of multiwave mixing. The mixing coefficients of Eqs. (\ref{fHybridizationAppendix}), (\ref{f3WMAppendix}), and (\ref{f4WMAppendix}) may be computed directly from the band theory, as obtained from the solution of Eqs. (\ref{linearEigenproblem1}) and (\ref{linearEigenproblem2}).

\section{Approximation of 3WM Signal Gain}
\label{appendixApproximation3wmGain}
We approximate 3WM signal gain in the presence of a strong undepleted pump. Via Eqs. (\ref{pump}) through (\ref{idler1}), the amplitude equations of 3WM become
\begin{equation}	\label{approxPump}
-i\frac{\partial }{\partial x} A^{(P)}_\alpha(x) \cong
\sum\limits_{\alpha'} f^{(P)}_{\alpha, \alpha'}(A_{DC}) \, A^{(P)}_{\alpha'}(x) ,
\end{equation}
\begin{equation}	\label{approxSignal}
-i \frac{\partial }{\partial x} A^{(S)}_\alpha(x) \cong
\sum\limits_{\alpha'} f^{(S)}_{\alpha, \alpha'}(A_{DC}) \, A^{(S)}_{\alpha'}(x)
+ 2A_{DC} \sum\limits_{{\alpha}', {\alpha}''} 
{f^{(S, \bar{1}, P)}_{\alpha, \alpha', \alpha''}} \, {A^{(1)}_{\alpha'}(x)}^* \, A^{(P)}_{\alpha''}(x) \, e^{-i \Delta\beta^{(1)}_{3wm} x} ,
\end{equation}
\begin{equation}	\label{approxIdler1}
i \frac{\partial }{\partial x} {A^{(1)}_\alpha(x)}^* \cong
\sum\limits_{\alpha'} f^{(1)}_{\alpha, \alpha'}(A_{DC}) \, {A^{(1)}_{\alpha'}(x)}^*
+ 2A_{DC} \sum\limits_{{\alpha}', {\alpha}''}
{f^{(1,\bar{S},P)}_{\alpha, \alpha', \alpha''}} \, A^{(S)}_{\alpha'}(x) \, {A^{(P)}_{\alpha''}(x)}^* \, e^{i \Delta\beta^{(1)}_{3wm} x} .
\end{equation}
From our numerical analysis, the lowest two dispersion-frequency bands are sufficient for the approximation of pump amplitudes, so we may estimate the solution of Eq. (\ref{approxPump}) as
\begin{equation}	\label{approxPump1}
A^{(P)}_0(x) \cong A^{(P)}_{0,+}(P_o,A_{DC}) \, e^{i\phi^{(P)}_+(A_{DC}) x} + A^{(P)}_{0,-}(P_o,A_{DC}) \,e^{i\phi^{(P)}_-(A_{DC}) x} ,
\end{equation}
\begin{equation}	\label{approxPump2}
A^{(P)}_{-1}(x) \cong A^{(P)}_{-1,+}(P_o,A_{DC}) \, e^{i\phi^{(P)}_+(A_{DC}) x} + A^{(P)}_{-1,-}(P_o,A_{DC}) \,e^{i\phi^{(P)}_-(A_{DC}) x} ,
\end{equation}
with two distinct phase coefficients
\begin{multline}	\label{coefficients}
\phi^{(P)}_\pm(A_{DC}) = 
\frac{1}{2} \, \left[ f^{(P)}_{0,0}(A_{DC}) + f^{(P)}_{-1,-1}(A_{DC}) \right] \\
\pm \frac{1}{2} \sqrt{ { \left[ f^{(P)}_{0,0}(A_{DC}) - f^{(P)}_{-1,-1}(A_{DC}) \right] }^2 + 4 f^{(P)}_{0,-1}(A_{DC}) \, f^{(P)}_{-1,0}(A_{DC}) } .
\end{multline}
The amplitudes may be expressed as
\begin{equation}
A^{(P)}_{\alpha,\pm}(P_o,A_{DC}) = 
\frac{ f^{(P)}_{\alpha,\alpha}(A_{DC}) - \phi^{(P)}_\mp(A_{DC}) }{ \phi^{(P)}_\pm(A_{DC}) - \phi^{(P)}_\mp(A_{DC}) } \, \sqrt{P_o} ,
\end{equation}
\begin{equation}
A^{(P)}_{\beta,\pm}(P_o,A_{DC}) = 
\frac{ f^{(P)}_{\beta,\alpha}(A_{DC}) } {\phi^{(P)}_\pm(A_{DC}) - \phi^{(P)}_\mp(A_{DC})} \, \sqrt{P_o} ,
\end{equation}
where $\alpha=0$, $\beta=-1$ ($\alpha=-1$, $\beta=0$) for a pump frequency matched to the first (second) band.

Applying the pump solution to the equations coupling signal to idler, and neglecting higher-lying bands, we may write for the amplitudes of the first band
\begin{equation}
-i \frac{\partial }{\partial x} A^{(S)}_0(x) \cong
f^{(S)}_{0,0}(A_{DC}) \, A^{(S)}_0(x)
+ \sum\limits_{m=\pm} \tilde{f}^{(S, \bar{1}, P)}_m(P_o,A_{DC}) \, e^{i \left[ \phi^{(P)}_m(A_{DC}) - \Delta\beta^{(1)}_{3wm} \right] x} \, {A^{(1)}_0(x)}^* ,
\end{equation}
\begin{equation}
i \frac{\partial }{\partial x} {A^{(1)}_0(x)}^* \cong
f^{(1)}_{0,0}(A_{DC}) \, {A^{(1)}_0(x)}^*
+ \sum\limits_{m=\pm} {\tilde{f}^{(1,\bar{S},P)}_m(P_o,A_{DC})}^* \, e^{-i \left[ \phi^{(P)}_m(A_{DC}) - \Delta\beta^{(1)}_{3wm} \right] x} \, A^{(S)}_0(x) ,
\end{equation}
where
\begin{equation}	\label{canonicalMixingCoefficients}
\tilde{f}^{(\mu,\bar{\nu},P)}_{m}(P_o,A_{DC}) = 2A_{DC} \sum\limits_{\alpha} {f^{(\mu,\bar{\nu},P)}_{0,0,\alpha}} \, A^{(P)}_{\alpha,m}(P_o,A_{DC}) .
\end{equation}
If we define a transformation $A^{(\mu)}_0(x)=B^{(\mu)}(x)\, \exp{[ i f^{(\mu)}_{0,0}(A_{DC})\, x ]}$, for $\mu\in\left\{ S, 1 \right\}$, and let 
\begin{equation}	\label{canonicalPhase}
\varphi_\pm(A_{DC})=\phi^{(P)}_\pm(A_{DC})-f^{(S)}_{0,0}(A_{DC})-f^{(1)}_{0,0}(A_{DC})-\Delta\beta^{(1)}_{3wm} ,
\end{equation}
then we have
\begin{equation}
-i \frac{\partial }{\partial x} B^{(S)}(x) \cong
\sum\limits_{m=\pm} \tilde{f}^{(S, \bar{1}, P)}_m(P_o,A_{DC}) \, e^{i \varphi_m(A_{DC})\, x} \, {B^{(1)}(x)}^* ,
\end{equation}
\begin{equation}
i \frac{\partial }{\partial x} {B^{(1)}(x)}^* \cong
\sum\limits_{m=\pm} {\tilde{f}^{(1,\bar{S},P)}_m(P_o,A_{DC})}^* \, e^{-i \varphi_m(A_{DC})\, x} \, B^{(S)}(x) .
\end{equation}

To solve the coupled differential equations of $B^{(S)}(x)$ and ${B^{(1)}(x)}^*$ we introduce another transformation
\begin{equation}
B^{(\mu)}(x) = C^{(\mu)}_+(x) \, e^{i \varphi_+(A_{DC})\, x / 2} + C^{(\mu)}_-(x) \, e^{i \varphi_-(A_{DC})\, x / 2} ,
\end{equation}
which yields
\begin{multline}	\label{appendixC1}
-i \left[ \frac{\partial }{\partial x} C^{(S)}_+(x) \right] e^{i \varphi_+(A_{DC})\, x / 2} 
-i \left[ \frac{\partial }{\partial x} C^{(S)}_-(x) \right] e^{i \varphi_-(A_{DC})\, x / 2} 
\cong \\
- \frac{1}{2} \varphi_+(A_{DC}) \, C^{(S)}_+(x) \, e^{i \varphi_+(A_{DC})\, x / 2} 
- \frac{1}{2} \varphi_-(A_{DC}) \, C^{(S)}_-(x) \, e^{i \varphi_-(A_{DC})\, x / 2} 
\\
+ \sum\limits_{m=\pm} \tilde{f}^{(S, \bar{1}, P)}_{m}(P_o,A_{DC}) \, e^{i \varphi_m(A_{DC})\, x} 
\left[ {C^{(1)}_+(x)}^* \, e^{-i \varphi_+(A_{DC})\, x / 2} + {C^{(1)}_-(x)}^* \, e^{-i \varphi_-(A_{DC})\, x / 2} \right] ,
\end{multline}
\begin{multline}	\label{appendixC2}
i \left[ \frac{\partial }{\partial x} {C^{(1)}_+(x)}^* \right] e^{-i \varphi_+(A_{DC})\, x / 2}
+ i \left[ \frac{\partial }{\partial x} {C^{(1)}_-(x)}^* \right] e^{-i \varphi_-(A_{DC})\, x / 2}
\cong \\
- \frac{1}{2} \varphi_+(A_{DC}) \, {C^{(1)}_+(x)}^* \, e^{-i \varphi_+(A_{DC})\, x / 2} 
- \frac{1}{2} \varphi_-(A_{DC}) \, {C^{(1)}_-(x)}^* \, e^{-i \varphi_-(A_{DC})\, x / 2} 
\\
+ \sum\limits_{m=\pm} {\tilde{f}^{(1,\bar{S},P)}_{m}(P_o,A_{DC})}^* \, e^{-i \varphi_m(A_{DC})\, x} 
\left[ C^{(S)}_+(x) \, e^{i \varphi_+(A_{DC})\, x / 2} + C^{(S)}_-(x) \, e^{i \varphi_-(A_{DC})\, x / 2} \right] .
\end{multline}
We then make an approximation by separately equating terms of the spatial harmonics $\exp{ [i \varphi_\pm(A_{DC})\, x / 2] }$, on either side of Eqs. (\ref{appendixC1}) and (\ref{appendixC2}). In so doing we neglect terms associated with beating of these harmonics. In this harmonic-balance approximation we arrive at two distinct criteria of phase matching, embodied in pairs of coupled equations
\begin{equation}
-i \frac{\partial }{\partial x} C^{(S)}_\pm(x) \cong
- \frac{1}{2} \varphi_\pm(A_{DC}) \, C^{(S)}_\pm(x) + \tilde{f}^{(S, \bar{1}, P)}_\pm(P_o,A_{DC}) \, {C^{(1)}_\pm(x)}^*  ,
\end{equation}
\begin{equation}
i \frac{\partial }{\partial x} {C^{(1)}_\pm(x)}^* \cong 
- \frac{1}{2} \varphi_\pm(A_{DC}) \, {C^{(1)}_\pm(x)}^* + {\tilde{f}^{(1,\bar{S},P)}_\pm(P_o,A_{DC})}^* \, C^{(S)}_\pm(x) .
\end{equation}
Thus, setting 
\begin{equation}
C^{(S)}_\pm(x) = \bar{C}^{(S)}_\pm(P_o,A_{DC})\, e^{g_\pm(P_o,A_{DC})\, x} ,
\end{equation}
\begin{equation}
{C^{(1)}_\pm (x)}^* = {\bar{C}^{(1)}_\pm(P_o,A_{DC})}^*\, e^{g_\pm(P_o,A_{DC})\, x} , 
\end{equation}
we obtain the gain coefficient of each criterion, viz.
\begin{equation}	\label{appendixGainCoefficients}
g_\pm(P_o,A_{DC}) = 
\sqrt{ \tilde{f}^{(S,\bar{1},P)}_\pm(P_o,A_{DC}) \, {\tilde{f}^{(1,\bar{S},P)}_\pm(P_o,A_{DC})}^* - \frac{1}{4} {\varphi_\pm(A_{DC})}^2 } ,
\end{equation}
such that $B^{(S)}(x)$ and ${B^{(1)}(x)}^*$ take the general form
\begin{multline}	\label{Bs}
B^{(S)}(x) \cong 
\left[ \bar{C}^{(S)}_{+,+}(P_o,A_{DC})\, e^{g_+(P_o,A_{DC})\, x} + \bar{C}^{(S)}_{+,-}(P_o,A_{DC})\, e^{-g_+(P_o,A_{DC})\, x} \right] 
e^{i \varphi_+(A_{DC})\, x / 2} \\
+ \left[ \bar{C}^{(S)}_{-,+}(P_o,A_{DC})\, e^{g_-(P_o,A_{DC})\, x} + \bar{C}^{(S)}_{-,-}(P_o,A_{DC})\, e^{-g_-(P_o,A_{DC})\, x} \right] 
e^{i \varphi_-(A_{DC})\, x / 2} ,
\end{multline}
\begin{multline}
{B^{(1)}(x)}^* \cong 
{\tilde{f}^{(1,\bar{S},P)}_+(P_o,A_{DC})}^* \Bigg[ 
\frac{ \bar{C}^{(S)}_{+,+}(P_o,A_{DC}) \, e^{g_+(P_o,A_{DC})\, x} }{ ig_+(P_o,A_{DC}) + \frac{1}{2} \varphi_+(A_{DC}) } \\
+ \frac{ \bar{C}^{(S)}_{+,-}(P_o,A_{DC}) \, e^{-g_+(P_o,A_{DC})\, x} }{ -ig_+(P_o,A_{DC}) + \frac{1}{2} \varphi_+(A_{DC}) } 
\Bigg] e^{-i \varphi_+(A_{DC})\, x / 2} \\
+ {\tilde{f}^{(1,\bar{S},P)}_-(P_o,A_{DC})}^* \Bigg[ 
\frac{ \bar{C}^{(S)}_{-,+}(P_o,A_{DC}) \, e^{g_-(A_{DC})\, x} }{ ig_-(P_o,A_{DC}) + \frac{1}{2} \varphi_-(A_{DC}) } \\
+ \frac{ \bar{C}^{(S)}_{-,-}(P_o,A_{DC}) \, e^{-g_-(P_o,A_{DC})\, x} }{ -ig_-(P_o,A_{DC}) + \frac{1}{2} \varphi_-(A_{DC}) } 
\Bigg] e^{-i \varphi_-(A_{DC})\, x / 2} .
\end{multline}

The four constants $\bar{C}^{(S)}_{\pm,\pm}(P_o,A_{DC})$ are determined from $B^{(S)}(0)=\bar{A}^{(S)}$, ${B^{(1)}(0)}^*=0$, ${|\partial B^{(S)}(x) / \partial x|}_{x=0}=0$, and ${|\partial {B^{(1)}(x)}^* / \partial x|}_{x=0}=\bar{A}^{(S)} \sum\limits_{m=\pm} {\tilde{f}^{(1,\bar{S},P)}_{m}(P_o,A_{DC})}^*$. After some algebra we obtain
\begin{equation}	\label{Cs}
\bar{C}^{(S)}_{m,n}(P_o,A_{DC}) = 
i \frac{m}{2} \, {\mathcal{A}_m(P_o,A_{DC})}^* \left[ 1 - i \frac{ n\, \varphi_m(A_{DC}) }{2 g_m(P_o,A_{DC})} \right] \bar{A}^{(S)} ,
\end{equation}
where $m,n=\pm$ and 
\begin{equation}	\label{psuedoAmplitudes}
\mathcal{A}_\pm(P_o,A_{DC}) = \frac{ \sum\limits_{m=\pm} \tilde{f}^{(1,\bar{S},P)}_{m}(P_o,A_{DC}) - i\tilde{f}^{(1,\bar{S},P)}_{ \mp }(P_o,A_{DC}) }
{ \tilde{f}^{(1,\bar{S},P)}_\pm(P_o,A_{DC}) - \tilde{f}^{(1,\bar{S},P)}_\pm(P_o,A_{DC}) } .
\end{equation}
From Eqs. (\ref{Bs}) and (\ref{Cs}), the 3WM signal gain, $G_S(P_o,A_{DC},\ell_T)\cong {| B^{(S)}(\ell_T) / \bar{A}^{(S)} |}^2$, may be expressed as
\begin{multline}	\label{appendixSignalGain}
G_S(P_o,A_{DC},\ell_T)\cong \\
\Bigg| {\mathcal{A}_+(P_o,A_{DC})}^* \Big[ \cosh{g_+(P_o,A_{DC})\, \ell_T} 
- \frac{i\varphi_+(A_{DC})}{2g_+(P_o,A_{DC})} \, \sinh{g_+(P_o,A_{DC})\, \ell_T} \Big] e^{i \varphi_+(A_{DC})\, \ell_T / 2} \\
+ {\mathcal{A}_-(P_o,A_{DC})}^* \Big[ \cosh{g_-(P_o,A_{DC})\, \ell_T} 
- \frac{i\varphi_-(A_{DC})}{2g_-(P_o,A_{DC})} \, \sinh{g_-(P_o,A_{DC})\, \ell_T} \Big] e^{i \varphi_-(A_{DC})\, \ell_T / 2} \Bigg| ^2 .
\end{multline}

\bibliography{bibliography}

\begin{thebibliography}{43}%
\makeatletter
\providecommand \@ifxundefined [1]{%
 \@ifx{#1\undefined}
}%
\providecommand \@ifnum [1]{%
 \ifnum #1\expandafter \@firstoftwo
 \else \expandafter \@secondoftwo
 \fi
}%
\providecommand \@ifx [1]{%
 \ifx #1\expandafter \@firstoftwo
 \else \expandafter \@secondoftwo
 \fi
}%
\providecommand \natexlab [1]{#1}%
\providecommand \enquote  [1]{``#1''}%
\providecommand \bibnamefont  [1]{#1}%
\providecommand \bibfnamefont [1]{#1}%
\providecommand \citenamefont [1]{#1}%
\providecommand \href@noop [0]{\@secondoftwo}%
\providecommand \href [0]{\begingroup \@sanitize@url \@href}%
\providecommand \@href[1]{\@@startlink{#1}\@@href}%
\providecommand \@@href[1]{\endgroup#1\@@endlink}%
\providecommand \@sanitize@url [0]{\catcode `\\12\catcode `\$12\catcode
  `\&12\catcode `\#12\catcode `\^12\catcode `\_12\catcode `\%12\relax}%
\providecommand \@@startlink[1]{}%
\providecommand \@@endlink[0]{}%
\providecommand \url  [0]{\begingroup\@sanitize@url \@url }%
\providecommand \@url [1]{\endgroup\@href {#1}{\urlprefix }}%
\providecommand \urlprefix  [0]{URL }%
\providecommand \Eprint [0]{\href }%
\providecommand \doibase [0]{http://dx.doi.org/}%
\providecommand \selectlanguage [0]{\@gobble}%
\providecommand \bibinfo  [0]{\@secondoftwo}%
\providecommand \bibfield  [0]{\@secondoftwo}%
\providecommand \translation [1]{[#1]}%
\providecommand \BibitemOpen [0]{}%
\providecommand \bibitemStop [0]{}%
\providecommand \bibitemNoStop [0]{.\EOS\space}%
\providecommand \EOS [0]{\spacefactor3000\relax}%
\providecommand \BibitemShut  [1]{\csname bibitem#1\endcsname}%
\let\auto@bib@innerbib\@empty
\bibitem [{\citenamefont {Wallraff}\ \emph {et~al.}(2004)\citenamefont
  {Wallraff}, \citenamefont {Schuster}, \citenamefont {Blais}, \citenamefont
  {Frunzio}, \citenamefont {Huang}, \citenamefont {Majer}, \citenamefont
  {Kumar}, \citenamefont {Girvin},\ and\ \citenamefont
  {Schoelkopf}}]{Wallraff2004}%
  \BibitemOpen
  \bibfield  {author} {\bibinfo {author} {\bibfnamefont {A.}~\bibnamefont
  {Wallraff}}, \bibinfo {author} {\bibfnamefont {D.~I.}\ \bibnamefont
  {Schuster}}, \bibinfo {author} {\bibfnamefont {A.}~\bibnamefont {Blais}},
  \bibinfo {author} {\bibfnamefont {L.}~\bibnamefont {Frunzio}}, \bibinfo
  {author} {\bibfnamefont {R.-S.}\ \bibnamefont {Huang}}, \bibinfo {author}
  {\bibfnamefont {J.}~\bibnamefont {Majer}}, \bibinfo {author} {\bibfnamefont
  {S.}~\bibnamefont {Kumar}}, \bibinfo {author} {\bibfnamefont {S.~M.}\
  \bibnamefont {Girvin}}, \ and\ \bibinfo {author} {\bibfnamefont {R.~J.}\
  \bibnamefont {Schoelkopf}},\ }\href {\doibase doi:10.1038/nature02851}
  {\bibfield  {journal} {\bibinfo  {journal} {Nature}\ }\textbf {\bibinfo
  {volume} {431}},\ \bibinfo {pages} {162} (\bibinfo {year}
  {2004})}\BibitemShut {NoStop}%
\bibitem [{\citenamefont {Day}\ \emph {et~al.}(2003)\citenamefont {Day},
  \citenamefont {Leduc}, \citenamefont {Mazin}, \citenamefont {Vayonakis},\
  and\ \citenamefont {Zmuidzinas}}]{DayNature2003}%
  \BibitemOpen
  \bibfield  {author} {\bibinfo {author} {\bibfnamefont {P.~K.}\ \bibnamefont
  {Day}}, \bibinfo {author} {\bibfnamefont {H.~G.}\ \bibnamefont {Leduc}},
  \bibinfo {author} {\bibfnamefont {B.~A.}\ \bibnamefont {Mazin}}, \bibinfo
  {author} {\bibfnamefont {A.}~\bibnamefont {Vayonakis}}, \ and\ \bibinfo
  {author} {\bibfnamefont {J.}~\bibnamefont {Zmuidzinas}},\ }\href {\doibase
  doi:10.1038/nature02037} {\bibfield  {journal} {\bibinfo  {journal} {Nature}\
  }\textbf {\bibinfo {volume} {425}},\ \bibinfo {pages} {817} (\bibinfo {year}
  {2003})}\BibitemShut {NoStop}%
\bibitem [{\citenamefont {Castellanos-Beltran}\ \emph
  {et~al.}(2008)\citenamefont {Castellanos-Beltran}, \citenamefont {Irwin},
  \citenamefont {Hilton}, \citenamefont {Vale},\ and\ \citenamefont
  {Lehnert}}]{CastellanosBeltran2008}%
  \BibitemOpen
  \bibfield  {author} {\bibinfo {author} {\bibfnamefont {M.~A.}\ \bibnamefont
  {Castellanos-Beltran}}, \bibinfo {author} {\bibfnamefont {K.~D.}\
  \bibnamefont {Irwin}}, \bibinfo {author} {\bibfnamefont {G.~C.}\ \bibnamefont
  {Hilton}}, \bibinfo {author} {\bibfnamefont {L.~R.}\ \bibnamefont {Vale}}, \
  and\ \bibinfo {author} {\bibfnamefont {K.~W.}\ \bibnamefont {Lehnert}},\
  }\href@noop {} {\bibfield  {journal} {\bibinfo  {journal} {Nature Physics}\
  }\textbf {\bibinfo {volume} {4}},\ \bibinfo {pages} {928} (\bibinfo {year}
  {2008})}\BibitemShut {NoStop}%
\bibitem [{\citenamefont {Bergeal}\ \emph {et~al.}(2010)\citenamefont
  {Bergeal}, \citenamefont {Schackert}, \citenamefont {Metcalfe}, \citenamefont
  {Vijay}, \citenamefont {Manucharyan}, \citenamefont {Frunzio}, \citenamefont
  {Prober}, \citenamefont {Schoelkopf}, \citenamefont {Girvin},\ and\
  \citenamefont {Devoret}}]{Bergreal2010}%
  \BibitemOpen
  \bibfield  {author} {\bibinfo {author} {\bibfnamefont {N.}~\bibnamefont
  {Bergeal}}, \bibinfo {author} {\bibfnamefont {F.~.}\ \bibnamefont
  {Schackert}}, \bibinfo {author} {\bibfnamefont {M.}~\bibnamefont {Metcalfe}},
  \bibinfo {author} {\bibfnamefont {R.}~\bibnamefont {Vijay}}, \bibinfo
  {author} {\bibfnamefont {V.~E.}\ \bibnamefont {Manucharyan}}, \bibinfo
  {author} {\bibfnamefont {L.}~\bibnamefont {Frunzio}}, \bibinfo {author}
  {\bibfnamefont {D.~E.}\ \bibnamefont {Prober}}, \bibinfo {author}
  {\bibfnamefont {R.~J.}\ \bibnamefont {Schoelkopf}}, \bibinfo {author}
  {\bibfnamefont {S.~M.}\ \bibnamefont {Girvin}}, \ and\ \bibinfo {author}
  {\bibfnamefont {M.~H.}\ \bibnamefont {Devoret}},\ }\href@noop {} {\bibfield
  {journal} {\bibinfo  {journal} {Nature}\ }\textbf {\bibinfo {volume} {465}},\
  \bibinfo {pages} {64} (\bibinfo {year} {2010})}\BibitemShut {NoStop}%
\bibitem [{\citenamefont {Spietz}\ \emph {et~al.}(2010)\citenamefont {Spietz},
  \citenamefont {Irwin}, \citenamefont {Lee},\ and\ \citenamefont
  {Aumentado}}]{Spietz2010}%
  \BibitemOpen
  \bibfield  {author} {\bibinfo {author} {\bibfnamefont {L.}~\bibnamefont
  {Spietz}}, \bibinfo {author} {\bibfnamefont {K.}~\bibnamefont {Irwin}},
  \bibinfo {author} {\bibfnamefont {M.}~\bibnamefont {Lee}}, \ and\ \bibinfo
  {author} {\bibfnamefont {J.}~\bibnamefont {Aumentado}},\ }\href@noop {}
  {\bibfield  {journal} {\bibinfo  {journal} {Appl. Phys, Lett.}\ }\textbf
  {\bibinfo {volume} {97}},\ \bibinfo {pages} {142502} (\bibinfo {year}
  {2010})}\BibitemShut {NoStop}%
\bibitem [{\citenamefont {Hatridge}\ \emph {et~al.}(2011)\citenamefont
  {Hatridge}, \citenamefont {Vijay}, \citenamefont {Slichter}, \citenamefont
  {Clarke},\ and\ \citenamefont {Siddiqi}}]{Hatridge2011}%
  \BibitemOpen
  \bibfield  {author} {\bibinfo {author} {\bibfnamefont {M.}~\bibnamefont
  {Hatridge}}, \bibinfo {author} {\bibfnamefont {R.}~\bibnamefont {Vijay}},
  \bibinfo {author} {\bibfnamefont {D.~H.}\ \bibnamefont {Slichter}}, \bibinfo
  {author} {\bibfnamefont {J.}~\bibnamefont {Clarke}}, \ and\ \bibinfo {author}
  {\bibfnamefont {I.}~\bibnamefont {Siddiqi}},\ }\href@noop {} {\bibfield
  {journal} {\bibinfo  {journal} {Phys. Rev. B}\ }\textbf {\bibinfo {volume}
  {83}},\ \bibinfo {pages} {134501} (\bibinfo {year} {2011})}\BibitemShut
  {NoStop}%
\bibitem [{\citenamefont {Hover}\ \emph {et~al.}(2012)\citenamefont {Hover},
  \citenamefont {Chen}, \citenamefont {Ribelli}, \citenamefont {Zhu},
  \citenamefont {Sendelbach},\ and\ \citenamefont {McDermott}}]{SLUG}%
  \BibitemOpen
  \bibfield  {author} {\bibinfo {author} {\bibfnamefont {D.}~\bibnamefont
  {Hover}}, \bibinfo {author} {\bibfnamefont {Y.~F.}\ \bibnamefont {Chen}},
  \bibinfo {author} {\bibfnamefont {G.~J.}\ \bibnamefont {Ribelli}}, \bibinfo
  {author} {\bibfnamefont {S.}~\bibnamefont {Zhu}}, \bibinfo {author}
  {\bibfnamefont {S.}~\bibnamefont {Sendelbach}}, \ and\ \bibinfo {author}
  {\bibfnamefont {R.}~\bibnamefont {McDermott}},\ }\href@noop {} {\bibfield
  {journal} {\bibinfo  {journal} {Appl. Phys. Lett.}\ }\textbf {\bibinfo
  {volume} {100}},\ \bibinfo {pages} {063503} (\bibinfo {year}
  {2012})}\BibitemShut {NoStop}%
\bibitem [{\citenamefont {Roch}\ \emph {et~al.}(2012)\citenamefont {Roch},
  \citenamefont {Flurin}, \citenamefont {Nguyen}, \citenamefont {Morfin},
  \citenamefont {Campagne-Ibarcq}, \citenamefont {Devoret},\ and\ \citenamefont
  {Huard}}]{Roch2012}%
  \BibitemOpen
  \bibfield  {author} {\bibinfo {author} {\bibfnamefont {N.}~\bibnamefont
  {Roch}}, \bibinfo {author} {\bibfnamefont {E.}~\bibnamefont {Flurin}},
  \bibinfo {author} {\bibfnamefont {F.}~\bibnamefont {Nguyen}}, \bibinfo
  {author} {\bibfnamefont {P.}~\bibnamefont {Morfin}}, \bibinfo {author}
  {\bibfnamefont {P.}~\bibnamefont {Campagne-Ibarcq}}, \bibinfo {author}
  {\bibfnamefont {M.~H.}\ \bibnamefont {Devoret}}, \ and\ \bibinfo {author}
  {\bibfnamefont {B.}~\bibnamefont {Huard}},\ }\href@noop {} {\bibfield
  {journal} {\bibinfo  {journal} {Phys. Phys, Lett.}\ }\textbf {\bibinfo
  {volume} {108}},\ \bibinfo {pages} {147701} (\bibinfo {year}
  {2012})}\BibitemShut {NoStop}%
\bibitem [{\citenamefont {Mutus}\ \emph {et~al.}(2013)\citenamefont {Mutus},
  \citenamefont {White}, \citenamefont {Jeffrey}, \citenamefont {Sank},
  \citenamefont {Barends}, \citenamefont {Bochmann}, \citenamefont {Chen},
  \citenamefont {Chen}, \citenamefont {Chiaro}, \citenamefont {Dunsworth},
  \citenamefont {Kelly}, \citenamefont {Megrant}, \citenamefont {Neill},
  \citenamefont {O’Malley}, \citenamefont {Roushan}, \citenamefont
  {Vainsencher}, \citenamefont {Wenner}, \citenamefont {Siddiqi}, \citenamefont
  {Vijay}, \citenamefont {Cleland},\ and\ \citenamefont
  {Martinis}}]{Mutas2013}%
  \BibitemOpen
  \bibfield  {author} {\bibinfo {author} {\bibfnamefont {J.~Y.}\ \bibnamefont
  {Mutus}}, \bibinfo {author} {\bibfnamefont {T.~C.}\ \bibnamefont {White}},
  \bibinfo {author} {\bibfnamefont {E.}~\bibnamefont {Jeffrey}}, \bibinfo
  {author} {\bibfnamefont {D.}~\bibnamefont {Sank}}, \bibinfo {author}
  {\bibfnamefont {R.}~\bibnamefont {Barends}}, \bibinfo {author} {\bibfnamefont
  {J.}~\bibnamefont {Bochmann}}, \bibinfo {author} {\bibfnamefont
  {Y.}~\bibnamefont {Chen}}, \bibinfo {author} {\bibfnamefont {Z.}~\bibnamefont
  {Chen}}, \bibinfo {author} {\bibfnamefont {B.}~\bibnamefont {Chiaro}},
  \bibinfo {author} {\bibfnamefont {A.}~\bibnamefont {Dunsworth}}, \bibinfo
  {author} {\bibfnamefont {J.}~\bibnamefont {Kelly}}, \bibinfo {author}
  {\bibfnamefont {A.}~\bibnamefont {Megrant}}, \bibinfo {author} {\bibfnamefont
  {C.}~\bibnamefont {Neill}}, \bibinfo {author} {\bibfnamefont {P.~J.~J.}\
  \bibnamefont {O’Malley}}, \bibinfo {author} {\bibfnamefont
  {P.}~\bibnamefont {Roushan}}, \bibinfo {author} {\bibfnamefont
  {A.}~\bibnamefont {Vainsencher}}, \bibinfo {author} {\bibfnamefont
  {J.}~\bibnamefont {Wenner}}, \bibinfo {author} {\bibfnamefont
  {I.}~\bibnamefont {Siddiqi}}, \bibinfo {author} {\bibfnamefont
  {R.}~\bibnamefont {Vijay}}, \bibinfo {author} {\bibfnamefont {A.~N.}\
  \bibnamefont {Cleland}}, \ and\ \bibinfo {author} {\bibfnamefont {J.~M.}\
  \bibnamefont {Martinis}},\ }\href@noop {} {\bibfield  {journal} {\bibinfo
  {journal} {Appl. Phys, Lett.}\ }\textbf {\bibinfo {volume} {103}},\ \bibinfo
  {pages} {122602} (\bibinfo {year} {2013})}\BibitemShut {NoStop}%
\bibitem [{\citenamefont {Macklin}\ \emph {et~al.}(2015)\citenamefont
  {Macklin}, \citenamefont {O’Brien}, \citenamefont {Hover}, \citenamefont
  {Schwartz}, \citenamefont {Bolkhovsky}, \citenamefont {Zhang}, \citenamefont
  {Oliver},\ and\ \citenamefont {Siddiqi}}]{Macklin2015}%
  \BibitemOpen
  \bibfield  {author} {\bibinfo {author} {\bibfnamefont {C.}~\bibnamefont
  {Macklin}}, \bibinfo {author} {\bibfnamefont {K.}~\bibnamefont {O’Brien}},
  \bibinfo {author} {\bibfnamefont {D.}~\bibnamefont {Hover}}, \bibinfo
  {author} {\bibfnamefont {M.~E.}\ \bibnamefont {Schwartz}}, \bibinfo {author}
  {\bibfnamefont {V.}~\bibnamefont {Bolkhovsky}}, \bibinfo {author}
  {\bibfnamefont {X.}~\bibnamefont {Zhang}}, \bibinfo {author} {\bibfnamefont
  {W.~D.}\ \bibnamefont {Oliver}}, \ and\ \bibinfo {author} {\bibfnamefont
  {I.}~\bibnamefont {Siddiqi}},\ }\href {\doibase 10.1126/science.aaa8525}
  {\bibfield  {journal} {\bibinfo  {journal} {Science}\ }\textbf {\bibinfo
  {volume} {350}},\ \bibinfo {pages} {307} (\bibinfo {year}
  {2015})}\BibitemShut {NoStop}%
\bibitem [{\citenamefont {Eom}\ \emph {et~al.}(2012)\citenamefont {Eom},
  \citenamefont {Day}, \citenamefont {LeDuc},\ and\ \citenamefont
  {Zmuidzinas}}]{EomNature2012amplifier}%
  \BibitemOpen
  \bibfield  {author} {\bibinfo {author} {\bibfnamefont {B.~H.}\ \bibnamefont
  {Eom}}, \bibinfo {author} {\bibfnamefont {P.~K.}\ \bibnamefont {Day}},
  \bibinfo {author} {\bibfnamefont {H.~G.}\ \bibnamefont {LeDuc}}, \ and\
  \bibinfo {author} {\bibfnamefont {J.}~\bibnamefont {Zmuidzinas}},\ }\href
  {\doibase 10.1038/nphys2356} {\bibfield  {journal} {\bibinfo  {journal}
  {Nature Physics}\ }\textbf {\bibinfo {volume} {8}},\ \bibinfo {pages} {623}
  (\bibinfo {year} {2012})}\BibitemShut {NoStop}%
\bibitem [{\citenamefont {Agrawal}(2001)}]{Agrawal2001}%
  \BibitemOpen
  \bibfield  {author} {\bibinfo {author} {\bibfnamefont {G.~P.}\ \bibnamefont
  {Agrawal}},\ }\href@noop {} {\emph {\bibinfo {title} {Nonlinear Fiber
  Optics}}},\ \bibinfo {edition} {3rd}\ ed.\ (\bibinfo  {publisher} {Academic
  Press},\ \bibinfo {address} {New York},\ \bibinfo {year} {2001})\ pp.\
  \bibinfo {pages} {389--436}\BibitemShut {NoStop}%
\bibitem [{\citenamefont {Agha}\ \emph {et~al.}(2009)\citenamefont {Agha},
  \citenamefont {Okawachi},\ and\ \citenamefont
  {Gaeta}}]{AghaOpticsExpress2009theoretical}%
  \BibitemOpen
  \bibfield  {author} {\bibinfo {author} {\bibfnamefont {I.~H.}\ \bibnamefont
  {Agha}}, \bibinfo {author} {\bibfnamefont {Y.}~\bibnamefont {Okawachi}}, \
  and\ \bibinfo {author} {\bibfnamefont {A.~L.}\ \bibnamefont {Gaeta}},\ }\href
  {\doibase 10.1364/OE.17.016209} {\bibfield  {journal} {\bibinfo  {journal}
  {Opt. Express}\ }\textbf {\bibinfo {volume} {17}},\ \bibinfo {pages} {16209}
  (\bibinfo {year} {2009})}\BibitemShut {NoStop}%
\bibitem [{\citenamefont {Chembo}\ and\ \citenamefont
  {Yu}(2010)}]{ChemboPRA2010modal}%
  \BibitemOpen
  \bibfield  {author} {\bibinfo {author} {\bibfnamefont {Y.~K.}\ \bibnamefont
  {Chembo}}\ and\ \bibinfo {author} {\bibfnamefont {N.}~\bibnamefont {Yu}},\
  }\href {\doibase 10.1103/PhysRevA.82.033801} {\bibfield  {journal} {\bibinfo
  {journal} {Phys. Rev. A}\ }\textbf {\bibinfo {volume} {82}},\ \bibinfo
  {pages} {033801} (\bibinfo {year} {2010})}\BibitemShut {NoStop}%
\bibitem [{\citenamefont {Chembo}\ \emph {et~al.}(2010)\citenamefont {Chembo},
  \citenamefont {Strekalov},\ and\ \citenamefont {Yu}}]{ChemboPRL2010spectrum}%
  \BibitemOpen
  \bibfield  {author} {\bibinfo {author} {\bibfnamefont {Y.~K.}\ \bibnamefont
  {Chembo}}, \bibinfo {author} {\bibfnamefont {D.~V.}\ \bibnamefont
  {Strekalov}}, \ and\ \bibinfo {author} {\bibfnamefont {N.}~\bibnamefont
  {Yu}},\ }\href {\doibase 10.1103/PhysRevLett.104.103902} {\bibfield
  {journal} {\bibinfo  {journal} {Phys. Rev. Lett.}\ }\textbf {\bibinfo
  {volume} {104}},\ \bibinfo {pages} {103902} (\bibinfo {year}
  {2010})}\BibitemShut {NoStop}%
\bibitem [{\citenamefont {Hansson}\ \emph {et~al.}(2013)\citenamefont
  {Hansson}, \citenamefont {Modotto},\ and\ \citenamefont
  {Wabnitz}}]{HanssonArXiv2013dynamics}%
  \BibitemOpen
  \bibfield  {author} {\bibinfo {author} {\bibfnamefont {T.}~\bibnamefont
  {Hansson}}, \bibinfo {author} {\bibfnamefont {D.}~\bibnamefont {Modotto}}, \
  and\ \bibinfo {author} {\bibfnamefont {S.}~\bibnamefont {Wabnitz}},\ }\href
  {\doibase http://dx.doi.org/10.1103/PhysRevA.88.023819} {\bibfield  {journal}
  {\bibinfo  {journal} {Phys. Rev. A}\ }\textbf {\bibinfo {volume} {88}},\
  \bibinfo {pages} {023819} (\bibinfo {year} {2013})}\BibitemShut {NoStop}%
\bibitem [{\citenamefont {Godey}\ \emph {et~al.}(2014)\citenamefont {Godey},
  \citenamefont {Balakireva}, \citenamefont {Coillet},\ and\ \citenamefont
  {Chembo}}]{GodeyArXiv2013stability}%
  \BibitemOpen
  \bibfield  {author} {\bibinfo {author} {\bibfnamefont {C.}~\bibnamefont
  {Godey}}, \bibinfo {author} {\bibfnamefont {I.}~\bibnamefont {Balakireva}},
  \bibinfo {author} {\bibfnamefont {A.}~\bibnamefont {Coillet}}, \ and\
  \bibinfo {author} {\bibfnamefont {Y.~K.}\ \bibnamefont {Chembo}},\ }\href
  {\doibase http://dx.doi.org/10.1103/PhysRevA.89.063814} {\bibfield  {journal}
  {\bibinfo  {journal} {Phys. Rev. A}\ }\textbf {\bibinfo {volume} {89}},\
  \bibinfo {pages} {063814} (\bibinfo {year} {2014})}\BibitemShut {NoStop}%
\bibitem [{\citenamefont {Kippenberg}(2004)}]{KippenbergPhD2008thesis}%
  \BibitemOpen
  \bibfield  {author} {\bibinfo {author} {\bibfnamefont {T.~J.~A.}\
  \bibnamefont {Kippenberg}},\ }\emph {\bibinfo {title} {Nonlinear Optics in
  Ultra-high-Q Whispering-Gallery Optical Microcavities}},\ \href@noop {}
  {Ph.D. thesis},\ \bibinfo  {school} {California Institue of Technology}
  (\bibinfo {year} {2004})\BibitemShut {NoStop}%
\bibitem [{\citenamefont {Del'Haye}\ \emph {et~al.}(2007)\citenamefont
  {Del'Haye}, \citenamefont {Schliesser}, \citenamefont {Arcizet},
  \citenamefont {Wilken}, \citenamefont {Holzwarth},\ and\ \citenamefont
  {Kippenberg}}]{DelHayeNature2007comb}%
  \BibitemOpen
  \bibfield  {author} {\bibinfo {author} {\bibfnamefont {P.}~\bibnamefont
  {Del'Haye}}, \bibinfo {author} {\bibfnamefont {A.}~\bibnamefont
  {Schliesser}}, \bibinfo {author} {\bibfnamefont {O.}~\bibnamefont {Arcizet}},
  \bibinfo {author} {\bibfnamefont {T.}~\bibnamefont {Wilken}}, \bibinfo
  {author} {\bibfnamefont {R.}~\bibnamefont {Holzwarth}}, \ and\ \bibinfo
  {author} {\bibfnamefont {T.~J.}\ \bibnamefont {Kippenberg}},\ }\href
  {\doibase 10.1038/nature06401} {\bibfield  {journal} {\bibinfo  {journal}
  {Nature}\ }\textbf {\bibinfo {volume} {450}},\ \bibinfo {pages} {1214}
  (\bibinfo {year} {2007})}\BibitemShut {NoStop}%
\bibitem [{\citenamefont {Del'Haye}\ \emph {et~al.}(2011)\citenamefont
  {Del'Haye}, \citenamefont {Herr}, \citenamefont {Gavartin}, \citenamefont
  {Gorodetsky}, \citenamefont {Holzwarth},\ and\ \citenamefont
  {Kippenberg}}]{DelhayPRL2011octavespanning}%
  \BibitemOpen
  \bibfield  {author} {\bibinfo {author} {\bibfnamefont {P.}~\bibnamefont
  {Del'Haye}}, \bibinfo {author} {\bibfnamefont {T.}~\bibnamefont {Herr}},
  \bibinfo {author} {\bibfnamefont {E.}~\bibnamefont {Gavartin}}, \bibinfo
  {author} {\bibfnamefont {M.~L.}\ \bibnamefont {Gorodetsky}}, \bibinfo
  {author} {\bibfnamefont {R.}~\bibnamefont {Holzwarth}}, \ and\ \bibinfo
  {author} {\bibfnamefont {T.~J.}\ \bibnamefont {Kippenberg}},\ }\href
  {\doibase 10.1103/PhysRevLett.107.063901} {\bibfield  {journal} {\bibinfo
  {journal} {Phys. Rev. Lett.}\ }\textbf {\bibinfo {volume} {107}},\ \bibinfo
  {pages} {063901} (\bibinfo {year} {2011})}\BibitemShut {NoStop}%
\bibitem [{\citenamefont {Foster}\ \emph {et~al.}(2011)\citenamefont {Foster},
  \citenamefont {Levy}, \citenamefont {Kuzucu}, \citenamefont {Saha},
  \citenamefont {Lipson},\ and\ \citenamefont
  {Gaeta}}]{FosterOpticsExpress2011SiBasedComb}%
  \BibitemOpen
  \bibfield  {author} {\bibinfo {author} {\bibfnamefont {M.~A.}\ \bibnamefont
  {Foster}}, \bibinfo {author} {\bibfnamefont {J.~S.}\ \bibnamefont {Levy}},
  \bibinfo {author} {\bibfnamefont {O.}~\bibnamefont {Kuzucu}}, \bibinfo
  {author} {\bibfnamefont {K.}~\bibnamefont {Saha}}, \bibinfo {author}
  {\bibfnamefont {M.}~\bibnamefont {Lipson}}, \ and\ \bibinfo {author}
  {\bibfnamefont {A.~L.}\ \bibnamefont {Gaeta}},\ }\href {\doibase
  http://dx.doi.org/10.1364/OE.19.014233} {\bibfield  {journal} {\bibinfo
  {journal} {Optics Express}\ }\textbf {\bibinfo {volume} {19}},\ \bibinfo
  {pages} {14233} (\bibinfo {year} {2011})}\BibitemShut {NoStop}%
\bibitem [{\citenamefont {Vissers}\ \emph {et~al.}(2016)\citenamefont
  {Vissers}, \citenamefont {Erickson}, \citenamefont {Ku}, \citenamefont
  {Vale}, \citenamefont {Xian}, \citenamefont {Hilton},\ and\ \citenamefont
  {Pappas}}]{Vissers2016}%
  \BibitemOpen
  \bibfield  {author} {\bibinfo {author} {\bibfnamefont {M.~R.}\ \bibnamefont
  {Vissers}}, \bibinfo {author} {\bibfnamefont {R.~P.}\ \bibnamefont
  {Erickson}}, \bibinfo {author} {\bibfnamefont {H.-S.}\ \bibnamefont {Ku}},
  \bibinfo {author} {\bibfnamefont {L.}~\bibnamefont {Vale}}, \bibinfo {author}
  {\bibfnamefont {W.}~\bibnamefont {Xian}}, \bibinfo {author} {\bibfnamefont
  {G.~C.}\ \bibnamefont {Hilton}}, \ and\ \bibinfo {author} {\bibfnamefont
  {D.~P.}\ \bibnamefont {Pappas}},\ }\href {\doibase 10.1063/1.4937922}
  {\bibfield  {journal} {\bibinfo  {journal} {Appl. Phys. Lett.}\ }\textbf
  {\bibinfo {volume} {108}},\ \bibinfo {pages} {012601} (\bibinfo {year}
  {2016})}\BibitemShut {NoStop}%
\bibitem [{\citenamefont {Mazin}\ \emph {et~al.}(2006)\citenamefont {Mazin},
  \citenamefont {Bumble}, \citenamefont {Day}, \citenamefont {Eckart},
  \citenamefont {Golwala}, \citenamefont {Zmuidzinas},\ and\ \citenamefont
  {Harris}}]{MazinAPL2006}%
  \BibitemOpen
  \bibfield  {author} {\bibinfo {author} {\bibfnamefont {B.~A.}\ \bibnamefont
  {Mazin}}, \bibinfo {author} {\bibfnamefont {B.}~\bibnamefont {Bumble}},
  \bibinfo {author} {\bibfnamefont {P.}~\bibnamefont {Day}}, \bibinfo {author}
  {\bibfnamefont {M.~E.}\ \bibnamefont {Eckart}}, \bibinfo {author}
  {\bibfnamefont {S.}~\bibnamefont {Golwala}}, \bibinfo {author} {\bibfnamefont
  {J.}~\bibnamefont {Zmuidzinas}}, \ and\ \bibinfo {author} {\bibfnamefont
  {F.~A.}\ \bibnamefont {Harris}},\ }\href@noop {} {\bibfield  {journal}
  {\bibinfo  {journal} {Appl. Phys. Lett.}\ }\textbf {\bibinfo {volume} {89}},\
  \bibinfo {pages} {222507} (\bibinfo {year} {2006})}\BibitemShut {NoStop}%
\bibitem [{\citenamefont {Barends}\ \emph {et~al.}(2010)\citenamefont
  {Barends}, \citenamefont {Vercruyssen}, \citenamefont {Endo}, \citenamefont
  {de~Visser}, \citenamefont {Zijlstra}, \citenamefont {Klapwijk},
  \citenamefont {Diener}, \citenamefont {Yates},\ and\ \citenamefont
  {Baselmans}}]{BarendsAPL2010}%
  \BibitemOpen
  \bibfield  {author} {\bibinfo {author} {\bibfnamefont {R.}~\bibnamefont
  {Barends}}, \bibinfo {author} {\bibfnamefont {N.}~\bibnamefont
  {Vercruyssen}}, \bibinfo {author} {\bibfnamefont {A.}~\bibnamefont {Endo}},
  \bibinfo {author} {\bibfnamefont {P.~J.}\ \bibnamefont {de~Visser}}, \bibinfo
  {author} {\bibfnamefont {T.}~\bibnamefont {Zijlstra}}, \bibinfo {author}
  {\bibfnamefont {T.~M.}\ \bibnamefont {Klapwijk}}, \bibinfo {author}
  {\bibfnamefont {P.}~\bibnamefont {Diener}}, \bibinfo {author} {\bibfnamefont
  {S.~J.~C.}\ \bibnamefont {Yates}}, \ and\ \bibinfo {author} {\bibfnamefont
  {J.~J.~A.}\ \bibnamefont {Baselmans}},\ }\href@noop {} {\bibfield  {journal}
  {\bibinfo  {journal} {Appl. Phys. Lett.}\ }\textbf {\bibinfo {volume} {97}},\
  \bibinfo {pages} {023508} (\bibinfo {year} {2010})}\BibitemShut {NoStop}%
\bibitem [{\citenamefont {Kittel}(1976)}]{Kittel1976}%
  \BibitemOpen
  \bibfield  {author} {\bibinfo {author} {\bibfnamefont {C.}~\bibnamefont
  {Kittel}},\ }\href@noop {} {\emph {\bibinfo {title} {Introduction to Solid
  State Physics}}},\ \bibinfo {edition} {5th}\ ed.\ (\bibinfo  {publisher}
  {John Wiley \& Sons},\ \bibinfo {address} {New York},\ \bibinfo {year}
  {1976})\ pp.\ \bibinfo {pages} {37--70,105--124,190,457--498}\BibitemShut
  {NoStop}%
\bibitem [{\citenamefont {Ashcroft}\ and\ \citenamefont
  {Mermin}(1976)}]{AshcroftMermin1976}%
  \BibitemOpen
  \bibfield  {author} {\bibinfo {author} {\bibfnamefont {N.~W.}\ \bibnamefont
  {Ashcroft}}\ and\ \bibinfo {author} {\bibfnamefont {N.~D.}\ \bibnamefont
  {Mermin}},\ }\href@noop {} {\emph {\bibinfo {title} {Solid State Physics}}},\
  \bibinfo {edition} {1st}\ ed.\ (\bibinfo  {publisher} {Holt, Rinehart and
  Winston},\ \bibinfo {address} {Philadelphia},\ \bibinfo {year} {1976})\ pp.\
  \bibinfo {pages} {85--129,133--141,430--447,671--689}\BibitemShut {NoStop}%
\bibitem [{\citenamefont {John}(1987)}]{John1987}%
  \BibitemOpen
  \bibfield  {author} {\bibinfo {author} {\bibfnamefont {S.}~\bibnamefont
  {John}},\ }\href {\doibase https://doi.org/10.1103/PhysRevLett.58.2486}
  {\bibfield  {journal} {\bibinfo  {journal} {Phys. Rev. Lett.}\ }\textbf
  {\bibinfo {volume} {58}},\ \bibinfo {pages} {2486} (\bibinfo {year}
  {1987})}\BibitemShut {NoStop}%
\bibitem [{\citenamefont {Yablonovitch}\ and\ \citenamefont
  {Gmitter}(1989)}]{Yablonovitch1989}%
  \BibitemOpen
  \bibfield  {author} {\bibinfo {author} {\bibfnamefont {E.}~\bibnamefont
  {Yablonovitch}}\ and\ \bibinfo {author} {\bibfnamefont {T.~J.}\ \bibnamefont
  {Gmitter}},\ }\href {\doibase https://doi.org/10.1103/PhysRevLett.63.1950}
  {\bibfield  {journal} {\bibinfo  {journal} {Phys. Rev. Lett.}\ }\textbf
  {\bibinfo {volume} {63}},\ \bibinfo {pages} {1950} (\bibinfo {year}
  {1989})}\BibitemShut {NoStop}%
\bibitem [{\citenamefont {Leung}\ and\ \citenamefont {Liu}(1990)}]{Leung1990}%
  \BibitemOpen
  \bibfield  {author} {\bibinfo {author} {\bibfnamefont {K.~M.}\ \bibnamefont
  {Leung}}\ and\ \bibinfo {author} {\bibfnamefont {Y.~F.}\ \bibnamefont
  {Liu}},\ }\href {\doibase https://doi.org/10.1103/PhysRevLett.65.2646}
  {\bibfield  {journal} {\bibinfo  {journal} {Phys. Rev. Lett.}\ }\textbf
  {\bibinfo {volume} {65}},\ \bibinfo {pages} {2646} (\bibinfo {year}
  {1990})}\BibitemShut {NoStop}%
\bibitem [{\citenamefont {Zhang}\ and\ \citenamefont
  {Satpathy}(1990)}]{Zhang1990}%
  \BibitemOpen
  \bibfield  {author} {\bibinfo {author} {\bibfnamefont {Z.}~\bibnamefont
  {Zhang}}\ and\ \bibinfo {author} {\bibfnamefont {S.}~\bibnamefont
  {Satpathy}},\ }\href {\doibase https://doi.org/10.1103/PhysRevLett.65.2650}
  {\bibfield  {journal} {\bibinfo  {journal} {Phys. Rev. Lett.}\ }\textbf
  {\bibinfo {volume} {65}},\ \bibinfo {pages} {2650} (\bibinfo {year}
  {1990})}\BibitemShut {NoStop}%
\bibitem [{\citenamefont {Ho}\ \emph {et~al.}(1990)\citenamefont {Ho},
  \citenamefont {Chan},\ and\ \citenamefont {Soukoulis}}]{Ho1990}%
  \BibitemOpen
  \bibfield  {author} {\bibinfo {author} {\bibfnamefont {K.~M.}\ \bibnamefont
  {Ho}}, \bibinfo {author} {\bibfnamefont {C.~T.}\ \bibnamefont {Chan}}, \ and\
  \bibinfo {author} {\bibfnamefont {C.~M.}\ \bibnamefont {Soukoulis}},\ }\href
  {\doibase https://doi.org/10.1103/PhysRevLett.65.3152} {\bibfield  {journal}
  {\bibinfo  {journal} {Phys. Rev. Lett.}\ }\textbf {\bibinfo {volume} {65}},\
  \bibinfo {pages} {3152} (\bibinfo {year} {1990})}\BibitemShut {NoStop}%
\bibitem [{\citenamefont {Lin}\ \emph {et~al.}(1998)\citenamefont {Lin},
  \citenamefont {Fleming}, \citenamefont {Hetherington}, \citenamefont {Smith},
  \citenamefont {Biswas}, \citenamefont {Ho}, \citenamefont {Sigalas},
  \citenamefont {Zubrzycki}, \citenamefont {Kurtz},\ and\ \citenamefont
  {Bur}}]{Lin1998}%
  \BibitemOpen
  \bibfield  {author} {\bibinfo {author} {\bibfnamefont {S.~Y.}\ \bibnamefont
  {Lin}}, \bibinfo {author} {\bibfnamefont {J.~G.}\ \bibnamefont {Fleming}},
  \bibinfo {author} {\bibfnamefont {D.~L.}\ \bibnamefont {Hetherington}},
  \bibinfo {author} {\bibfnamefont {B.~K.}\ \bibnamefont {Smith}}, \bibinfo
  {author} {\bibfnamefont {R.}~\bibnamefont {Biswas}}, \bibinfo {author}
  {\bibfnamefont {K.~M.}\ \bibnamefont {Ho}}, \bibinfo {author} {\bibfnamefont
  {M.~M.}\ \bibnamefont {Sigalas}}, \bibinfo {author} {\bibfnamefont
  {W.}~\bibnamefont {Zubrzycki}}, \bibinfo {author} {\bibfnamefont {S.~R.}\
  \bibnamefont {Kurtz}}, \ and\ \bibinfo {author} {\bibfnamefont
  {J.}~\bibnamefont {Bur}},\ }\href {\doibase https://doi.org/10.1038/28343}
  {\bibfield  {journal} {\bibinfo  {journal} {Nature}\ }\textbf {\bibinfo
  {volume} {394}},\ \bibinfo {pages} {251} (\bibinfo {year}
  {1998})}\BibitemShut {NoStop}%
\bibitem [{\citenamefont {O'Brien}\ \emph {et~al.}(2014)\citenamefont
  {O'Brien}, \citenamefont {Macklin}, \citenamefont {Siddiqi},\ and\
  \citenamefont {Zhang}}]{OBrien2014}%
  \BibitemOpen
  \bibfield  {author} {\bibinfo {author} {\bibfnamefont {K.}~\bibnamefont
  {O'Brien}}, \bibinfo {author} {\bibfnamefont {C.}~\bibnamefont {Macklin}},
  \bibinfo {author} {\bibfnamefont {I.}~\bibnamefont {Siddiqi}}, \ and\
  \bibinfo {author} {\bibfnamefont {X.}~\bibnamefont {Zhang}},\ }\href@noop {}
  {\bibfield  {journal} {\bibinfo  {journal} {Phys. Rev. Lett.}\ }\textbf
  {\bibinfo {volume} {113}},\ \bibinfo {pages} {157001} (\bibinfo {year}
  {2014})}\BibitemShut {NoStop}%
\bibitem [{\citenamefont {White}\ \emph {et~al.}(2015)\citenamefont {White},
  \citenamefont {Mutus}, \citenamefont {Hoi}, \citenamefont {Barends},
  \citenamefont {Campbell}, \citenamefont {Chen}, \citenamefont {Chen},
  \citenamefont {Chiaro}, \citenamefont {Dunsworth}, \citenamefont {Jeffrey},
  \citenamefont {Kelly}, \citenamefont {Megrant}, \citenamefont {Neill},
  \citenamefont {O'Malley}, \citenamefont {Roushan}, \citenamefont {Sank},
  \citenamefont {Wenner}, \citenamefont {Chaudhuri}, \citenamefont {Gao},\ and\
  \citenamefont {Martinis}}]{White2015}%
  \BibitemOpen
  \bibfield  {author} {\bibinfo {author} {\bibfnamefont {T.~C.}\ \bibnamefont
  {White}}, \bibinfo {author} {\bibfnamefont {J.~Y.}\ \bibnamefont {Mutus}},
  \bibinfo {author} {\bibfnamefont {I.-C.}\ \bibnamefont {Hoi}}, \bibinfo
  {author} {\bibfnamefont {R.}~\bibnamefont {Barends}}, \bibinfo {author}
  {\bibfnamefont {B.}~\bibnamefont {Campbell}}, \bibinfo {author}
  {\bibfnamefont {Y.}~\bibnamefont {Chen}}, \bibinfo {author} {\bibfnamefont
  {Z.}~\bibnamefont {Chen}}, \bibinfo {author} {\bibfnamefont {B.}~\bibnamefont
  {Chiaro}}, \bibinfo {author} {\bibfnamefont {A.}~\bibnamefont {Dunsworth}},
  \bibinfo {author} {\bibfnamefont {E.}~\bibnamefont {Jeffrey}}, \bibinfo
  {author} {\bibfnamefont {J.}~\bibnamefont {Kelly}}, \bibinfo {author}
  {\bibfnamefont {A.}~\bibnamefont {Megrant}}, \bibinfo {author} {\bibfnamefont
  {C.}~\bibnamefont {Neill}}, \bibinfo {author} {\bibfnamefont {P.~J.~J.}\
  \bibnamefont {O'Malley}}, \bibinfo {author} {\bibfnamefont {P.}~\bibnamefont
  {Roushan}}, \bibinfo {author} {\bibfnamefont {A.}~\bibnamefont {Sank},
  \bibfnamefont {D.~andVainsencher}}, \bibinfo {author} {\bibfnamefont
  {J.}~\bibnamefont {Wenner}}, \bibinfo {author} {\bibfnamefont
  {S.}~\bibnamefont {Chaudhuri}}, \bibinfo {author} {\bibfnamefont
  {J.}~\bibnamefont {Gao}}, \ and\ \bibinfo {author} {\bibfnamefont {J.~M.}\
  \bibnamefont {Martinis}},\ }\href@noop {} {\bibfield  {journal} {\bibinfo
  {journal} {Appl. Phys. Lett.}\ }\textbf {\bibinfo {volume} {106}},\ \bibinfo
  {pages} {242601} (\bibinfo {year} {2015})}\BibitemShut {NoStop}%
\bibitem [{\citenamefont {Mills}\ and\ \citenamefont
  {Trullinger}(1987)}]{Mills1987}%
  \BibitemOpen
  \bibfield  {author} {\bibinfo {author} {\bibfnamefont {D.~L.}\ \bibnamefont
  {Mills}}\ and\ \bibinfo {author} {\bibfnamefont {S.~E.}\ \bibnamefont
  {Trullinger}},\ }\href@noop {} {\bibfield  {journal} {\bibinfo  {journal}
  {Phys. Rev. B}\ }\textbf {\bibinfo {volume} {36}},\ \bibinfo {pages} {947}
  (\bibinfo {year} {1987})}\BibitemShut {NoStop}%
\bibitem [{\citenamefont {Sipe}\ and\ \citenamefont {Winful}(1988)}]{Sipe1988}%
  \BibitemOpen
  \bibfield  {author} {\bibinfo {author} {\bibfnamefont {J.~E.}\ \bibnamefont
  {Sipe}}\ and\ \bibinfo {author} {\bibfnamefont {H.~G.}\ \bibnamefont
  {Winful}},\ }\href@noop {} {\bibfield  {journal} {\bibinfo  {journal} {Optics
  Lett.}\ }\textbf {\bibinfo {volume} {13}},\ \bibinfo {pages} {132} (\bibinfo
  {year} {1988})}\BibitemShut {NoStop}%
\bibitem [{\citenamefont {Chen}\ and\ \citenamefont
  {Mills}(1987{\natexlab{a}})}]{Chen1987-1}%
  \BibitemOpen
  \bibfield  {author} {\bibinfo {author} {\bibfnamefont {W.}~\bibnamefont
  {Chen}}\ and\ \bibinfo {author} {\bibfnamefont {D.~L.}\ \bibnamefont
  {Mills}},\ }\href@noop {} {\bibfield  {journal} {\bibinfo  {journal} {Phys.
  Rev. Lett.}\ }\textbf {\bibinfo {volume} {58}},\ \bibinfo {pages} {160}
  (\bibinfo {year} {1987}{\natexlab{a}})}\BibitemShut {NoStop}%
\bibitem [{\citenamefont {Chen}\ and\ \citenamefont
  {Mills}(1987{\natexlab{b}})}]{Chen1987-2}%
  \BibitemOpen
  \bibfield  {author} {\bibinfo {author} {\bibfnamefont {W.}~\bibnamefont
  {Chen}}\ and\ \bibinfo {author} {\bibfnamefont {D.~L.}\ \bibnamefont
  {Mills}},\ }\href@noop {} {\bibfield  {journal} {\bibinfo  {journal} {Phys.
  Rev. B}\ }\textbf {\bibinfo {volume} {35}},\ \bibinfo {pages} {524} (\bibinfo
  {year} {1987}{\natexlab{b}})}\BibitemShut {NoStop}%
\bibitem [{\citenamefont {Christodoulides}\ and\ \citenamefont
  {Joeseph}(1988)}]{Christodoulides1988}%
  \BibitemOpen
  \bibfield  {author} {\bibinfo {author} {\bibfnamefont {D.~N.}\ \bibnamefont
  {Christodoulides}}\ and\ \bibinfo {author} {\bibfnamefont {R.~I.}\
  \bibnamefont {Joeseph}},\ }\href@noop {} {\bibfield  {journal} {\bibinfo
  {journal} {Optics Lett.}\ }\textbf {\bibinfo {volume} {13}},\ \bibinfo
  {pages} {794} (\bibinfo {year} {1988})}\BibitemShut {NoStop}%
\bibitem [{\citenamefont {Mandelik}\ \emph {et~al.}(2003)\citenamefont
  {Mandelik}, \citenamefont {Eisenberg}, \citenamefont {Silberberg},
  \citenamefont {Morandotti},\ and\ \citenamefont {Aitchison}}]{Mandelik2003}%
  \BibitemOpen
  \bibfield  {author} {\bibinfo {author} {\bibfnamefont {D.}~\bibnamefont
  {Mandelik}}, \bibinfo {author} {\bibfnamefont {H.~S.}\ \bibnamefont
  {Eisenberg}}, \bibinfo {author} {\bibfnamefont {Y.}~\bibnamefont
  {Silberberg}}, \bibinfo {author} {\bibfnamefont {R.}~\bibnamefont
  {Morandotti}}, \ and\ \bibinfo {author} {\bibfnamefont {J.~S.}\ \bibnamefont
  {Aitchison}},\ }\href@noop {} {\bibfield  {journal} {\bibinfo  {journal}
  {Phys. Rev. Lett.}\ }\textbf {\bibinfo {volume} {90}},\ \bibinfo {pages}
  {053902} (\bibinfo {year} {2003})}\BibitemShut {NoStop}%
\bibitem [{\citenamefont {Press}\ \emph {et~al.}(2007)\citenamefont {Press},
  \citenamefont {Teukolsky}, \citenamefont {Vetterling},\ and\ \citenamefont
  {Flannery}}]{NumericalRecipes2007}%
  \BibitemOpen
  \bibfield  {author} {\bibinfo {author} {\bibfnamefont {W.~H.}\ \bibnamefont
  {Press}}, \bibinfo {author} {\bibfnamefont {S.~A.}\ \bibnamefont
  {Teukolsky}}, \bibinfo {author} {\bibfnamefont {W.~T.}\ \bibnamefont
  {Vetterling}}, \ and\ \bibinfo {author} {\bibfnamefont {B.~P.}\ \bibnamefont
  {Flannery}},\ }\href@noop {} {\emph {\bibinfo {title} {Numerical Recipes: The
  Art of Scientific Computing}}},\ \bibinfo {edition} {3rd}\ ed.\ (\bibinfo
  {publisher} {Cambridge University Press},\ \bibinfo {address} {New York},\
  \bibinfo {year} {2007})\ p.\ \bibinfo {pages} {920}\BibitemShut {NoStop}%
\bibitem [{Note1()}]{Note1}%
  \BibitemOpen
  \bibinfo {note} {The numerical program that solves the coupled differential
  equations and obtains the signal gain is available from the authors upon
  request.}\BibitemShut {Stop}%
\bibitem [{\citenamefont {Liu}\ and\ \citenamefont {Houck}(2016)}]{Liu2016}%
  \BibitemOpen
  \bibfield  {author} {\bibinfo {author} {\bibfnamefont {Y.}~\bibnamefont
  {Liu}}\ and\ \bibinfo {author} {\bibfnamefont {A.~A.}\ \bibnamefont
  {Houck}},\ }\href {\doibase 10.1038/nphys3834} {\bibfield  {journal}
  {\bibinfo  {journal} {Nature Physics}\ } (\bibinfo {year} {2016}),\
  10.1038/nphys3834}\BibitemShut {NoStop}%
\end{thebibliography}%

\end{document}